%% file: ms.tex
\newif\ifdraft
\newif\ifARXIV
\draftfalse
\ARXIVtrue
\ifdraft
\documentclass[manuscript]{aastex62}
\else
\documentclass[singlecolumn]{aastex62}
\fi
\usepackage{amsmath,afterpage,multirow,color}

\usepackage{multirow,lineno}
\usepackage{url}      
\received{}
\revised{}
\accepted{}
\submitjournal{ApJS}

\shorttitle{HCN $J$=4--3, HNC $J$=1--0, $\mathrm{H^{13}CN}$ $J$=1--0, and $\mathrm{HC_3N}$ $J$=10--9 Maps of the Galactic Center Region I}
\shortauthors{Tanaka et al.}
\tighten

\input{defines.etex}
\ifdraft
\pagewiselinenumbers
\fi

\renewcommand\textcolor[2]{{#2}}
\begin{document}

\title{HCN $J$=4--3, HNC $J$=1--0, $\mathrm{H^{13}CN}$ $J$=1--0, and $\mathrm{HC_3N}$ $J$=10--9 Maps of Galactic Center Region I: Spatially-Resolved Measurements of Physical Conditions and Chemical Composition}

\correspondingauthor{Kunihiko Tanaka}
\email{ktanaka@phys.keio.ac.jp}

\author{Kunihiko Tanaka}
\affil{Department of Physics, Faculty of Science and Technology, Keio University, 3-14-1 Hiyoshi, Yokohama, Kanagawa 223--8522 Japan}
\author{Makoto Nagai}
\affil{Advanced Technology Center, National Astronomical Observatory Japan, 2-21-1 Osawa, Mitaka, Tokyo 181-8588, Japan}
\author{Kazuhisa Kamegai}
\affil{Astronomy Data Center, National Astronomical Observatory Japan, 2-21-1 Osawa, Mitaka, Tokyo 181-8588, Japan}
\author{Takahiro Iino}
\affil{Nature and Science Museum,Tokyo University of Agriculture and Technology, 2-24-6 Naka-machi, Koganei, Tokyo 184-8588, Japan}

\author{Takeshi Sakai}
\affil{Graduate School of Informatics and Engineering, The University of Electro-Communications, 1-5-1 Chofugaoka, Chofu, Tokyo 182-8585, Japan}


\begin{abstract}
\input{abstract.etex}

\end{abstract}


\section{INTRODUCTION}
\input{introduction.etex}
\section{OBSERVATIONS}
\input{observation.etex}

\section{DATA}\label{SECTION_DATA}
\input{data.etex}
\section{PHYSICAL CONDITIONS AND MOLECULAR ABUNDANCES: METHODS}\label{SECTION_ANALYSIS}
\input{analysis.etex}

\section{PHYSICAL CONDITIONS AND MOLECULAR ABUNDANCES: RESULTs}\label{SECTION_RESULTS}
\input{results.etex}

\section{DISCUSSION}
\input{discussion.etex}

\section{SUMMARY}
\input{summary.etex}

\acknowledgments
\input{acknowledgements.etex}

\bibliographystyle{apj}
\bibliography{mendeley,local}

\ifARXIV
\input{figures.etex}
\fi

\end{document}

%% file: abstract.etex
This {\it supplement} paper presents the maps of HCN $J$=4--3, HNC $J$=1--0, $\mathrm{H^{13}CN}$ $J$=1--0, and HC$_3$N $J$=10--9 for the Galactic central molecular zone (CMZ), which have been obtained using the Atacama Submillimeter Telescope Experiment and Nobeyama Radio Observatory 45-m telescope.
Three-dimensional maps (2-D in space and 1-D in velocity) of the gas kinetic temperature ($T_\mathrm{kin}$), hydrogen volume density ($n_\mathrm{H_2}$), and fractional abundances of eight molecules (HCN, HNC, $\mathrm{HC_3N}$, HCO$^+$, $\mathrm{H_2CO}$, SiO, CS, and $\mathrm{N_2H^+}$) have been constructed from our and archival data.
We have developed a method with hierarchical Bayesian inference for this analysis, which has successfully suppressed the artificial correlations among the parameters created by systematic errors due to the deficiency in the simple one-zone excitation analysis and the calibration uncertainty.
The typical values of $T_\mathrm{kin}$ and $n_\mathrm{H_2}$ are $10^{1.8}\ $K and $10^{4.2}\ \mathrm{cm}^{-3}$, respectively, and the presence of an additional cold/low-density component is also indicated.
The distribution of high-temperature regions is poorly correlated with known active star-forming regions, while a few of them coincide with shocked clouds.
Principal component analysis has identified two distinct groups in the eight analyzed molecules:   
one group with large PC1 and PC2 scores and the other with a large $T_\mathrm{kin}$ dependence, which could be explained using two regimes of shock chemistry with fast ($\gtrsim 20\ \mathrm{km\,s}^{-1}$) and slow ($\lesssim 20\ \mathrm{km\,s}^{-1}$) velocity shocks, respectively.
This supports the idea that the mechanical sputtering of dust grains and the mechanical heating play primary roles in the chemical and thermal processes in CMZ clouds.

%% file: introduction.etex
The innermost 200 pc of the Milky Way, or the central molecular zone (CMZ), is the Galactic version of the central molecular condensations observed in many spiral galaxies, which frequently act as molecular gas reservoirs that fuel starbursts and active galactic nuclei.
The observation of dense molecular gas is of fundamental importance in studying the Galactic CMZ, where the cold interstellar medium is primarily confined in giant molecular clouds (GMCs) with a volume density $\gtrsim 10^{3}\ \pcc$ \citep{Morris1996,Nagai2007}. 
The dense gas in the CMZ has been intensively studied in various molecular transitions with high critical densities (\ncrit) at millimeter \citep{Tsuboi1999,Jones2012,Jones2013} and submillimeter \citep{Oka2007,Ginsburg2016} wavelengths.
This supplement paper reports the results of the new CMZ surveys in HCN \JJ{4}{3}\ at 350 GHz and three high-density tracer lines in the 3-mm band (\HCNt\ \JJ{1}{0}, HNC \JJ{1}{0}, and \HCCCN\ \JJ{10}{9}), which were performed using the Atacama Submillimeter Telescope Experiment (ASTE) 10-m telescope and Nobeyama Radio Observatory (NRO) 45-m telescope.
The HCN \JJ{4}{3}\ transition, whose nominal \ncrit\ is $10^7\ \pcc$, is the highest-density tracer among those previously used in the CMZ survey, even though the effective \ncrit\ can be 1--2 orders of magnitude lower owing to the photon trapping effect.%
\footnote{
In this paper, we adopt the definition of \ncrit\ given in Equation 4 of \cite{Shirley2015}; $\ncrit \equiv A_{i,j} / \sum_{k \neq i} \gamma_{i,j}$ for the $i \rightarrow j$ level transition, where $A$ and $\gamma$ are the Einstein $A$ coefficient and collisional excitation rate, respectively.}

The measurement of the gas kinetic temperature ($\Tkin$) and hydrogen volume density (\nHH) is a fundamentally important issue in several fields of interstellar physics and chemistry.
In star formation studies, the volume density is a key parameter for understanding the dynamical evolution and the current physical status of molecular clouds. 
Spatially resolved measurements of the temperature and density may assist in determining the mechanisms that fuel the well-known local starburst in the Sgr B complex  and the formation of young massive clusters such as the Quintuplet and Arches clusters \citep{Nagata1995,Morris1996,Figer1999} while suppressing star formation in the remaining vast majority of the clouds \citep{Longmore2012b,Rathborne2014a,Kruijssen2014}.
In addition, the CMZ is  a unique target for studying molecular chemistry under the extreme environment of the innermost part of galaxies, where molecular clouds are exposed to intense ultraviolet (UV), X-ray, and cosmic-ray (CR) irradiation and placed in strong tidal and magnetic fields.
The Galactic CMZ is known for several interesting chemical characteristics such as widespread hot-core-like chemistry \citep{Requena-Torres2006}, a rich abundance of SiO \citep{Martin-Pintado1997}, and an anomalously high atomic-carbon-to-CO abundance ratio \citep{Tanaka2011}.
Information about the physical conditions is necessary for converting the molecular line {\it intensities} into their {\it abundances} and for investigating the dependence of the molecular abundances on environmental factors.

The immediate aim of this paper is to present 3-D (2-D in space and 1-D in velocity) maps of \Tkin\ and \nHH\ constructed using newly obtained data and published data taken from the literature.
There are multiple previous studies on spatially resolved \Tkin\ measurements, which use ammonia inversion lines \citep{Nagayama2007,Ott2014,Arai2016}, multi-$J$ CO and \COt\ lines \citep{Martin2004,Nagai2007}, and \formaldehyde\ lines \citep{Ginsburg2016}\ as probes.
For the measurement of \nHH, several studies have been conducted on the basis of observations of high-density tracers; however, they are primarily targeted at limited areas, and the large-scale distribution across the entire CMZ remains less studied.
The analysis by \cite{Nagai2007}\ using low-$J$ CO lines is the \nHH\ measurement with the best spatial resolution and coverage until now, but its scope is limited to the outer low-density part of the clouds because of the selection of tracer lines with $\ncrit \le 10^4\ \pcc$.
The new analysis we present in this paper uses the intensity ratios among multitransition HCN lines as the primary density probe and realizes sufficient parameter coverage to study dense regions of the CMZ.
In addition, the molecular line data of several popular species at the millimeter wavelength are incorporated, whereby we measure the spatial variations in their fractional abundances simultaneously with the physical conditions.

Our analysis was conducted using a method that we developed by employing the hierarchical Bayesian (HB) inference technique. 
In a few cases, the standard maximum likelihood (ML) method is unsuitable for analysis involving a large number of molecular lines because it is susceptible to unmeasurable systematic errors due to the calibration uncertainty and the approximations adopted in the model.   
The HB analysis realizes more robust parameter inference by introducing a prior probability function that applies stronger constraints on the behavior of the parameters. 

The outline of this paper is as follows:   In Section 2, we describe the details of the observations performed using the ASTE 10-m telescope and NRO 45-m telescope.    The data are presented in Section \ref{SECTION_DATA}.
Section 4 describes the HB model, and its results are presented in Section 5.
Section 6 describes the investigation of the mechanisms that govern the distribution of \Tkin\ and the molecular abundances.
The identification of high-density clumps in the HCN \JJ{4}{3}\ maps and the investigation of their star-forming activities will be provided in a separate forthcoming paper.

%% file: observation.etex
We observed the HCN \JJ{4}{3}\ line using the ASTE 10-m telescope and the \HCNt\ \JJ{1}{0}, HNC \JJ{1}{0}, and \HCCCN\ \JJ{10}{9}\ lines with the NRO 45-m telescope.
The frequencies, upper state energies ($\Eu$), and critical densities of the target lines are listed in Table \ref{TABLE_INPDATA}.
The observed regions are shown in Figure \ref{FIG1}.
The primary parts of the four major GMC complexes in the CMZ, i.e., Sgr A--D, are observed in all four lines, even though the gap region between the Sgr B and D complexes (Galactic longitude of $0.8^\circ$ to $1.1^\circ$) is not included in the ASTE observations.
The HCN \JJ{4}{3}\ and \HCNt\ \JJ{1}{0}\ maps of the Sgr C complex ($-0.6^\circ < \gl < -0.3^\circ$) have been published in a separated paper \citep{Tanaka2014}.

\subsection{ASTE Observations}
The ASTE HCN $\JJ{4}{3}$ observations were performed during the following four periods: the first from August to September 2010, the second from April to May 2011, the third in November 2011, and the last in August 2013.
The target regions were mapped using multiple on-the-fly \citep[OTF;][]{Sawada2008} scans, each covering a subregion of $20'$ square or less.
The fields mapped in the four periods are shown in Figure \ref{FIG1}.
The scans in the X and Y directions (i.e., in the directions of Galactic longitude and latitude, respectively) were obtained for each subregion to mitigate the scanning effect.
The off position was taken at ($l$, $b$) = ($0^\circ, -1^\circ$), where we confirmed the absence of significant HCN \JJ{4}{3}\ emission during the observations. 
The total observation time over the four periods was 76 hours including the integration time for off-position integration and intensity calibration, and the dead-time for antenna acceleration and the transient time between scan rows.
A sideband-separation-type receiver CATS345 was used as the receiver frontend.
The telescope beamsize at 350 GHz was $22''$. 
The backend was a 1024-channel auto-correlator system with a channel separation of 500 kHz, which provided a velocity separation of 0.44 $\kmps$ and a spectral coverage of 400 \kmps\ per sideband at 350 GHz.
We observed the CO $\JJ{3}{2}$ line toward V1427 Aql or RAFGL5379 approximately every 1 hour to maintain the pointing accuracy within $< 3''$ during observation runs.
The antenna temperatures were calibrated using the standard chopper wheel method.

We utilized the NOSTAR package developed by the NRO, for flagging bad data with unstable spectral baselines, the subtraction of spectral baselines, and the reduction of the data into $\gl$--$\gb$--\vlsr\ data cubes.  
A third-degree polynomial fitting was applied to spectral baseline subtraction to remove standing noise.  
Spectral data were spatially convolved using a Gaussian-tapered Bessel function and resampled onto an $8.5''\times8.5''\times 2\ \kmps$ grid.  
The final effective spatial resolution was $24''$.
The X- and Y-scan maps were combined into one map by employing the {\it PLAIT} algorithm to remove scanning noise \citep{Emerson1988}.
The typical noise per voxel is 0.14 K in the \Tmb\ scale.

\setcounter{figure}{0}
\begin{figure*}[ttt]
\epsscale{0.7}
\plotone{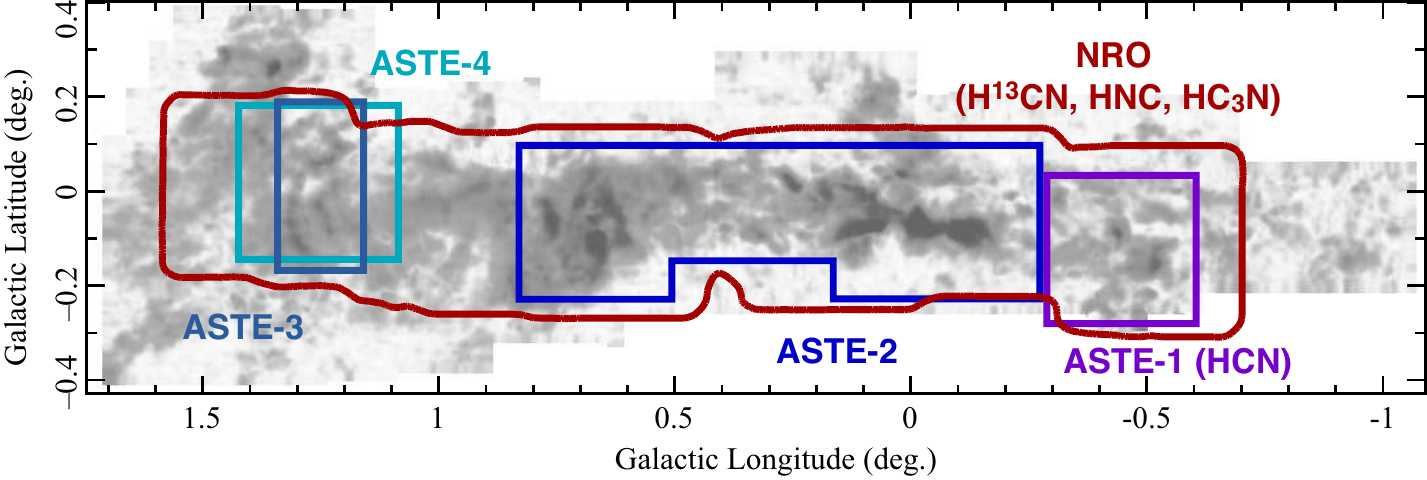}
\caption{Mapping regions of the ASTE HCN \JJ{4}{3}\ observations and the NRO HNC, \HCNt, and \HCCCN\ observations superposed on the grayscale map of the CS \JJ{1}{0}\ peak intensity \citep{Tsuboi1999}. The mapping regions for the four ASTE observation periods are indicated with different colors.}\label{FIG1}
\ifdraft
\vspace{5mm}
\else
\end{figure*}
\nobreak
\begin{figure*}[ttt]
\fi     
\epsscale{0.7}
\plotone{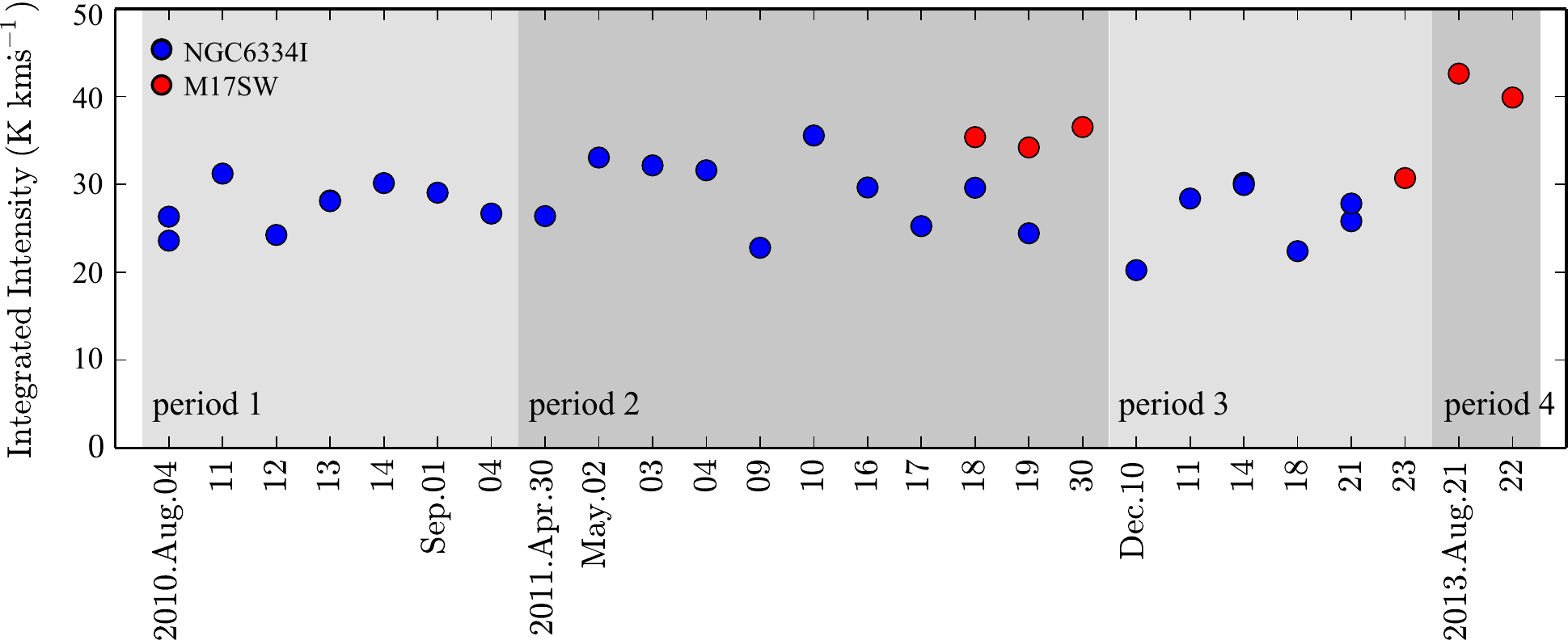}
\caption{Velocity-integrated HCN \JJ{4}{3}\ intensities of the calibrators in the scale of the chopper-wheel calibrated intensity, measured at different dates during observation periods 1--4. The blue and red points denote NGC6334 and M17SW, respectively.  }\label{FIG2}
\end{figure*}

The calibrator sources NGC6334I and M17SW were observed more than once in each observation period to measure the correction factor for the main-beam efficiency and image rejection ratio (IRR) and its instrumental variability among the four observation periods.
Figure \ref{FIG2} shows the measured chopper-wheel-calibrated HCN \JJ{4}{3}\ intensities of the calibrators.
As compared to the first two periods, there is a significant decline in the NGC6334\RN{1} intensity in period 3; the averaged intensities are $4.15\pm0.10\ \kelvin$, $4.25\pm0.15\ \kelvin$, and $3.63\pm0.25\ \kelvin$ for periods 1,2, and 3, respectively. 
Hence, we multiplied the data by a factor of 1.16 during  period 3 to correct this variation.
On the contrary, the M17SW intensity rises in period 4 as compared to the first three periods;  the averaged intensity for periods 2 and 3 after applying the above-mentioned correction is $5.86\pm0.12\ \kelvin$, while that for period 4 is $6.49\pm0.28\ \kelvin$.
We applied a correction factor of 0.90 for period 4.
After these corrections, we obtain the averaged intensities of NGC6334\RN{1} and M17SW as $4.12\pm0.10\ \kelvin$ and $5.86\pm0.09\ \kelvin$, respectively, with a relative daily variation of 10\%.
By comparing the M17SW intensity measured using the Caltech Submillimeter Observatory \citep[$9.8\pm0.3$ K;][]{Wang1994}, we estimate the scaling factor for the main-beam efficiency and IRR to be $1.67\pm0.06$.
The total systematic calibration errors are approximately estimated to be 0.1 from the daily variation in the calibrator intensities and the uncertainty in the scaling factor.

\subsection{NRO 45-m Observations}
The \HCNt\ \JJ{1}{0}, HNC \JJ{1}{0}, and \HCCCN\ \JJ{10}{9}\ lines were observed in January 2010 using the 25-beam array receiver system BEARS \citep{Sunada2000} installed on the NRO 45-m telescope.
The three lines were simultaneously observed in the double sideband (DSB) mode: \HCNt\ in the lower sideband and HNC and \HCCCN\ in the upper sideband.

We operated the digital backend in the wide-band mode with a channel separation of 0.5 MHz and a total bandwidth of 512 MHz, which correspond to a velocity channel separation of 1.7\ \kmps\ and a velocity coverage of 1748\ \kmps\ at 90 GHz.
Velocity ranges from $-200\ \kmps$ to $+200\ \kmps$ were covered by the spectrometer for the \HCNt\ and HNC lines; however, the velocities lower than $-150\ \kmps$ in the \HCCCN\ line were outside the correlator bandwidth.
The observation region was covered by performing multiple OTF scans for subregions whose sizes were $20'$ square or less.
The scans were performed in a manner such that each of the 25 beams of the BEARS multibeam receiver conducts full Nyquist sampling over the target region; then the resultant 25 maps were integrated into one map after the variations in the IRR and sensitivity among the beams were corrected.
X and Y scans were obtained for each subregion.
The antenna pointing accuracy was maintained within $5''$ by observing the SiO $\JJ{1}{0},\ v=1,2$ maser lines toward VX-Sgr.   

The data were converted into $\gl$--$\gb$--\vlsr\ data cubes with a $10.275''\times10.275''\times2\ \kmps$ grid using the same procedure as that for the ASTE data reduction, except for an additional correction for the variations in the sideband ratios among the 25 receiver units.
The relative correction factors were obtained by performing self-calibration for each sideband; we measured the ratios of the chopper-wheel-calibrated intensities among the 25 maps created for individual receiver channels.
The absolute values of the DSB correction factors were obtained by comparing the intensity toward Sgr B2 with the intensity measured using the S100 receiver.
The value of $\eta_{\mathrm{MB}}$ of the S100 receiver at 86 GHz is measured to be 0.45 by the observatory.
The effective beam size of the final maps was $\sim 20''$ after convolution is applied in the imaging process.  
The typical values of the noise per voxel are 0.19 K, 0.21 K, and 0.17 K for the \HCNt, HNC, and \HCCCN\ maps, respectively, in the \Tmb\ scale.         
The voxels in peripheral regions with RMS noise levels of the \HCNt\ spectra above 0.4 K in the $\Tmb$ scale were removed from the maps.



\begin{figure*}[ppp]
\epsscale{0.7}
\plotone{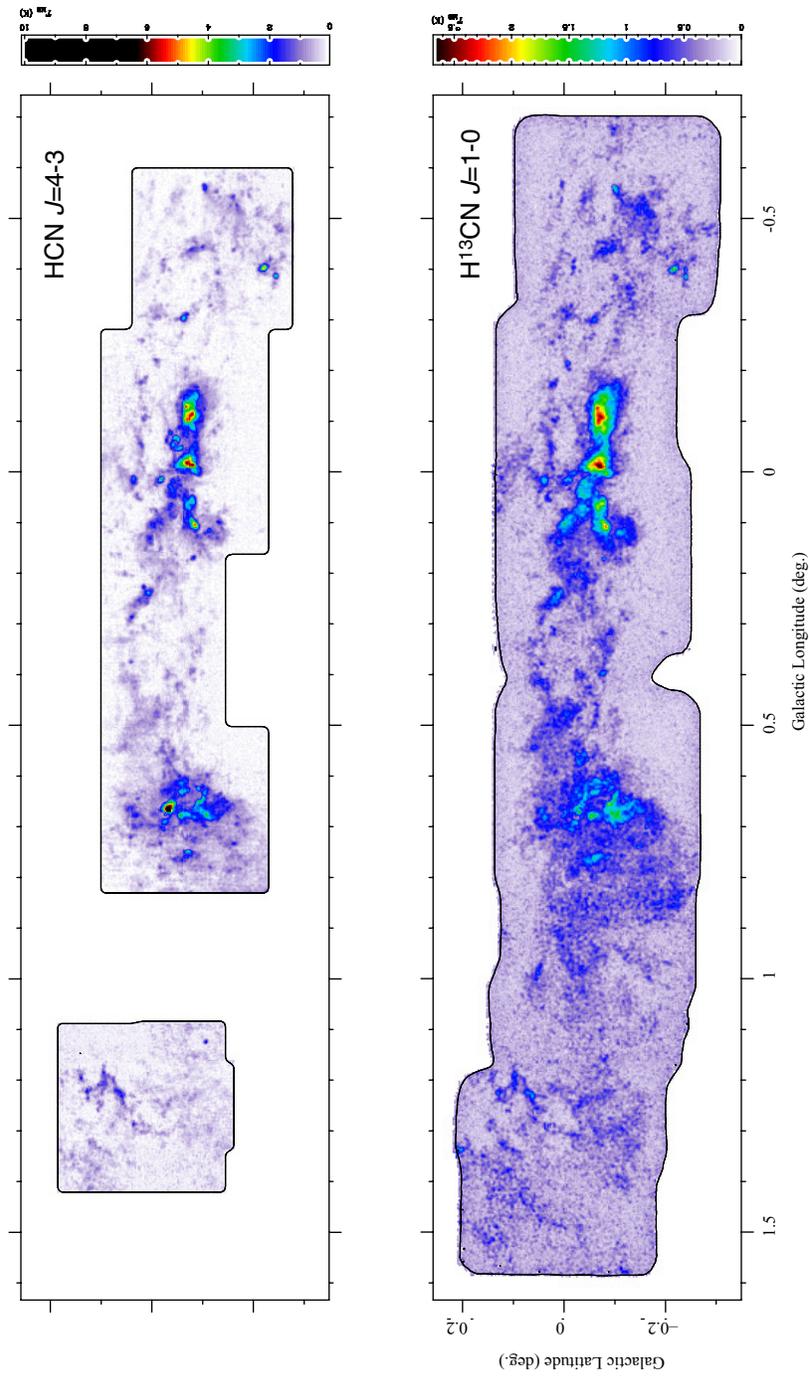}
\caption{{Peak intensity maps of the observed lines, calculated after smoothing along the velocity axis with 10-\kmps\ FWHM kernels.}}\label{FIG3}
\end{figure*}
\begin{figure*}[ppp]
\figurenum{3 ({\it Continued})}
\epsscale{0.7}
\plotone{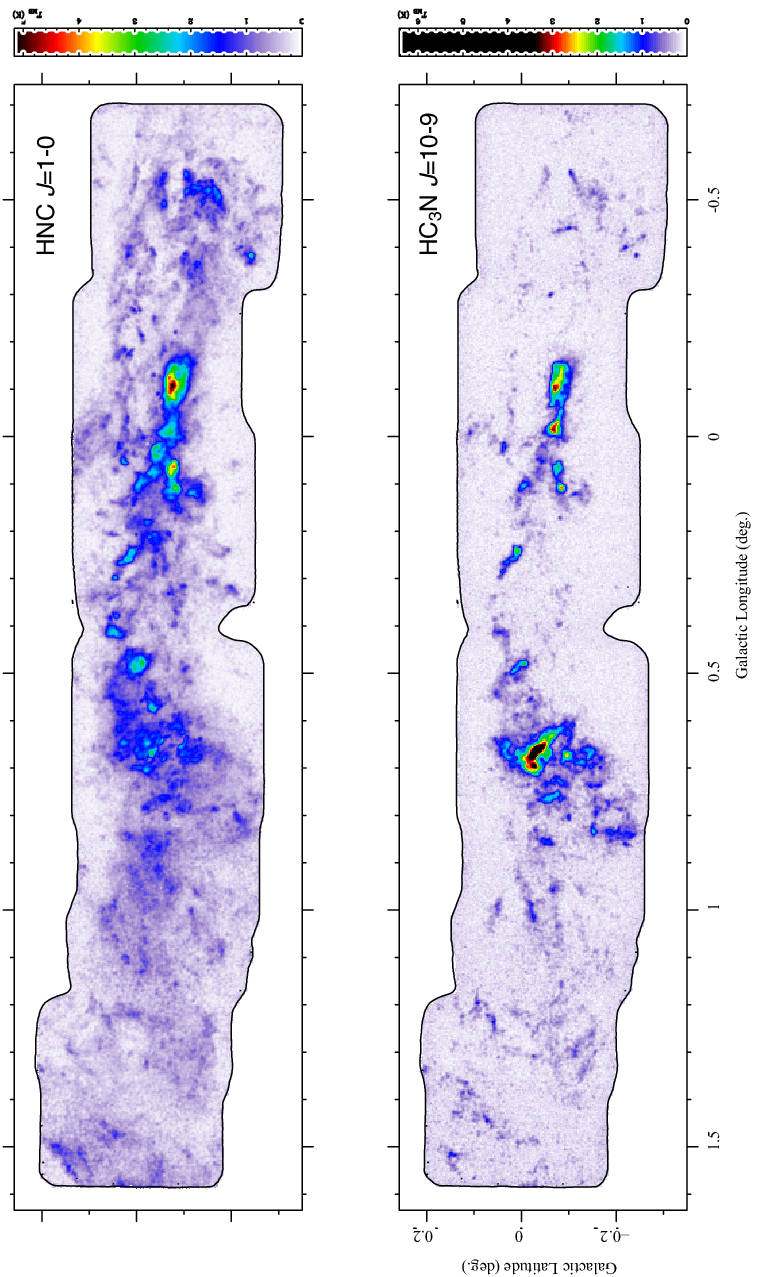}
\caption{{Peak intensity maps of the observed lines, calculated after smoothing along the velocity axis with 10-\kmps\ FWHM kernels.}}
\end{figure*}

\begin{figure*}[ttt]
\epsscale{1}
\plotone{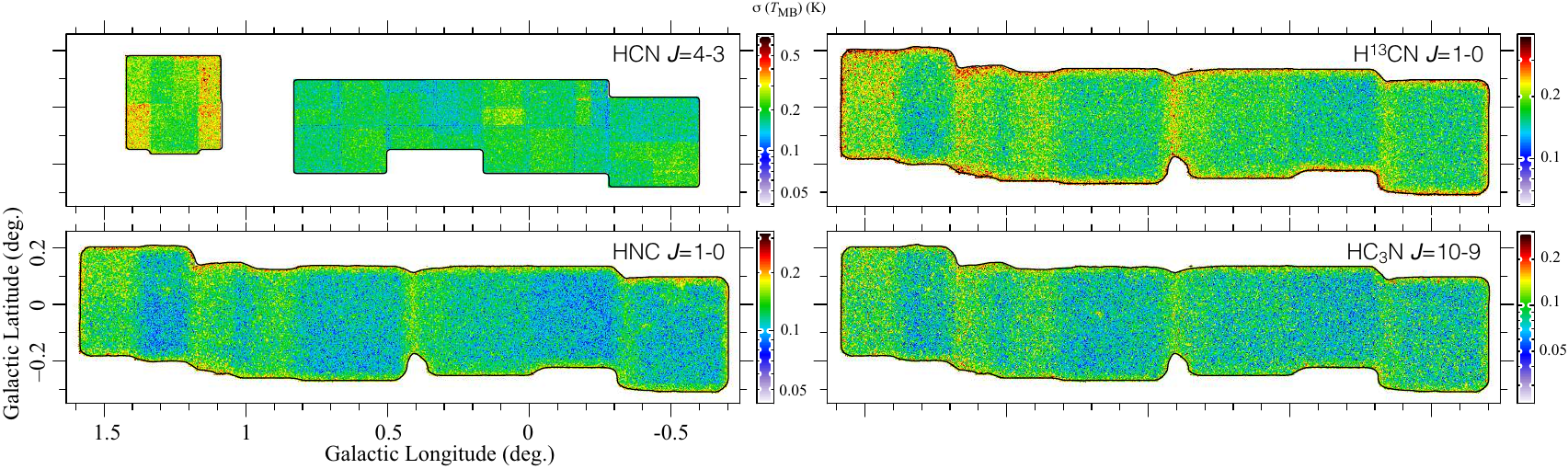}
\caption{Maps of the 1-$\sigma$ noise levels of the observed lines in logarithmic color scales.}\label{FIG4}
\end{figure*}


%% file: data.etex
Figure \ref{FIG3} shows the peak intensity maps of the observed lines smoothed with Gaussian kernels having an FWHM of 10 \kmps.
Note that low saturation levels are applied in the HCN and \HCCCN\ maps to soften the image contrast due to the extremely bright emission from the Sgr B2 cluster-forming region.  
The 1-$\sigma$ noise maps calculated from the emission-free spectral channels are shown in Figure \ref{FIG4}.
\ifARXIV
The velocity channel maps in bins with widths of 20\ \kmps\ and the longitude--velocity diagrams in latitude bins of $1.5'$ are presented in Figures \ref{FIG5} and \ref{FIG6}, respectively, which are attached at the end of the paper. 
\else
The velocity channel maps in bins with widths of 20\ \kmps\ and the longitude--velocity diagrams in latitude bins of $1.5'$ are presented in Figures \ref{FIG5} and \ref{FIG6}, respectively.

\begin{figure*}[ppp]
\epsscale{0.9}
\plottwox{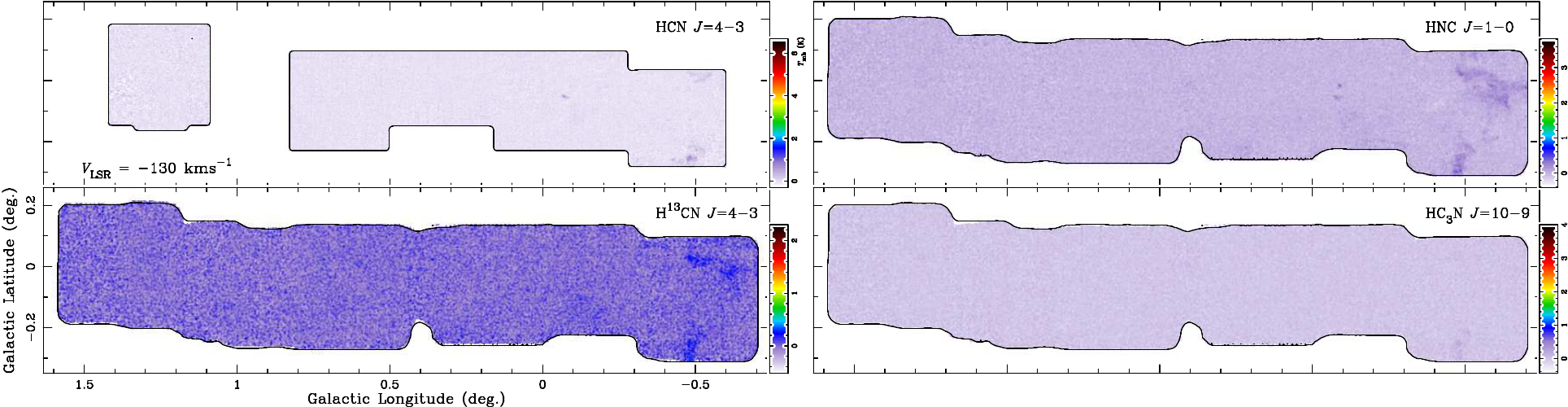}{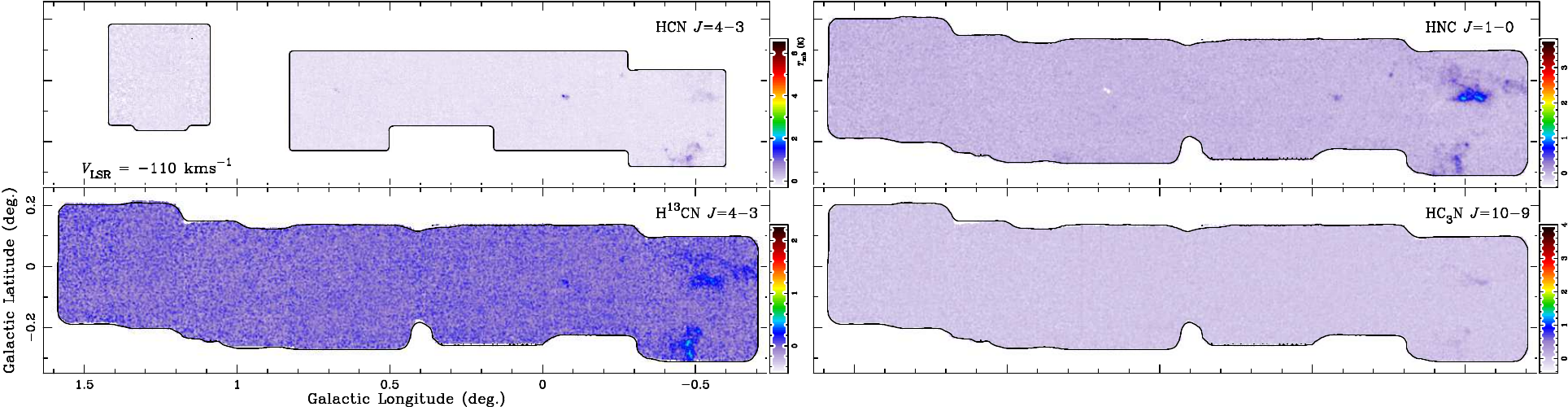}
\textcolor{red}{
\caption{Velocity channel maps of HCN \JJ{4}{3} (top left), HNC \JJ{1}{0} (top right), \HCNt\ \JJ{1}{0} (bottom left), and \HCCCN\ \JJ{10}{9} (bottom right) for a \vlsr\ range of {$-130$}\ \kmps\ to {+170}\ \kmps\ at an interval of 20 \kmps.  The channel center velocity is given on the HCN $\JJ{4}{3}$\ panel for each channel.    Full channel maps are available in the Figure Set \ref{FIGSET5}. }\label{FIG5}
}
\end{figure*}

\textcolor{red}{
\input{figset5.etex}
}

\begin{figure*}[ppp]
\epsscale{0.9}
\plottwoy{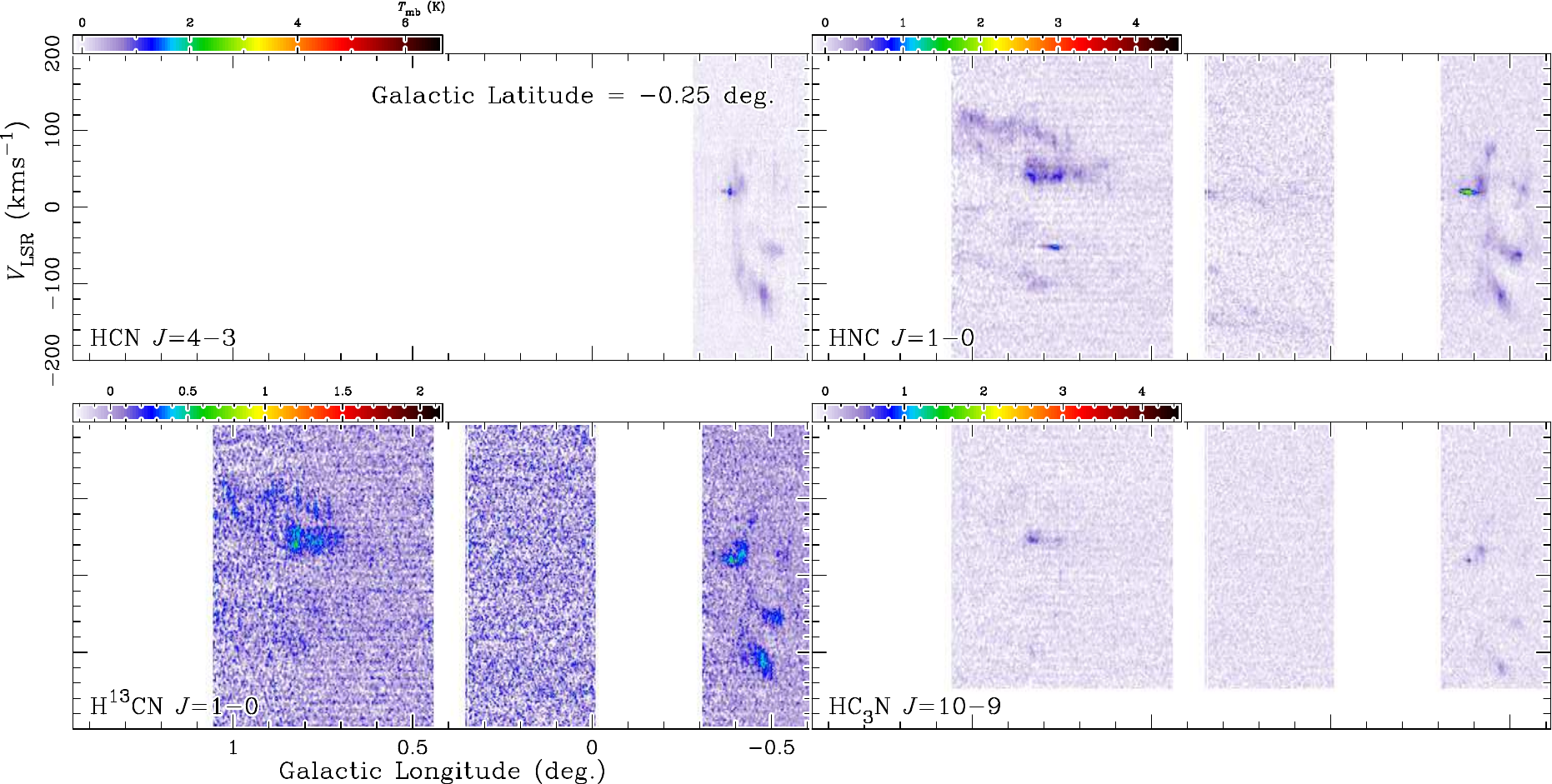}{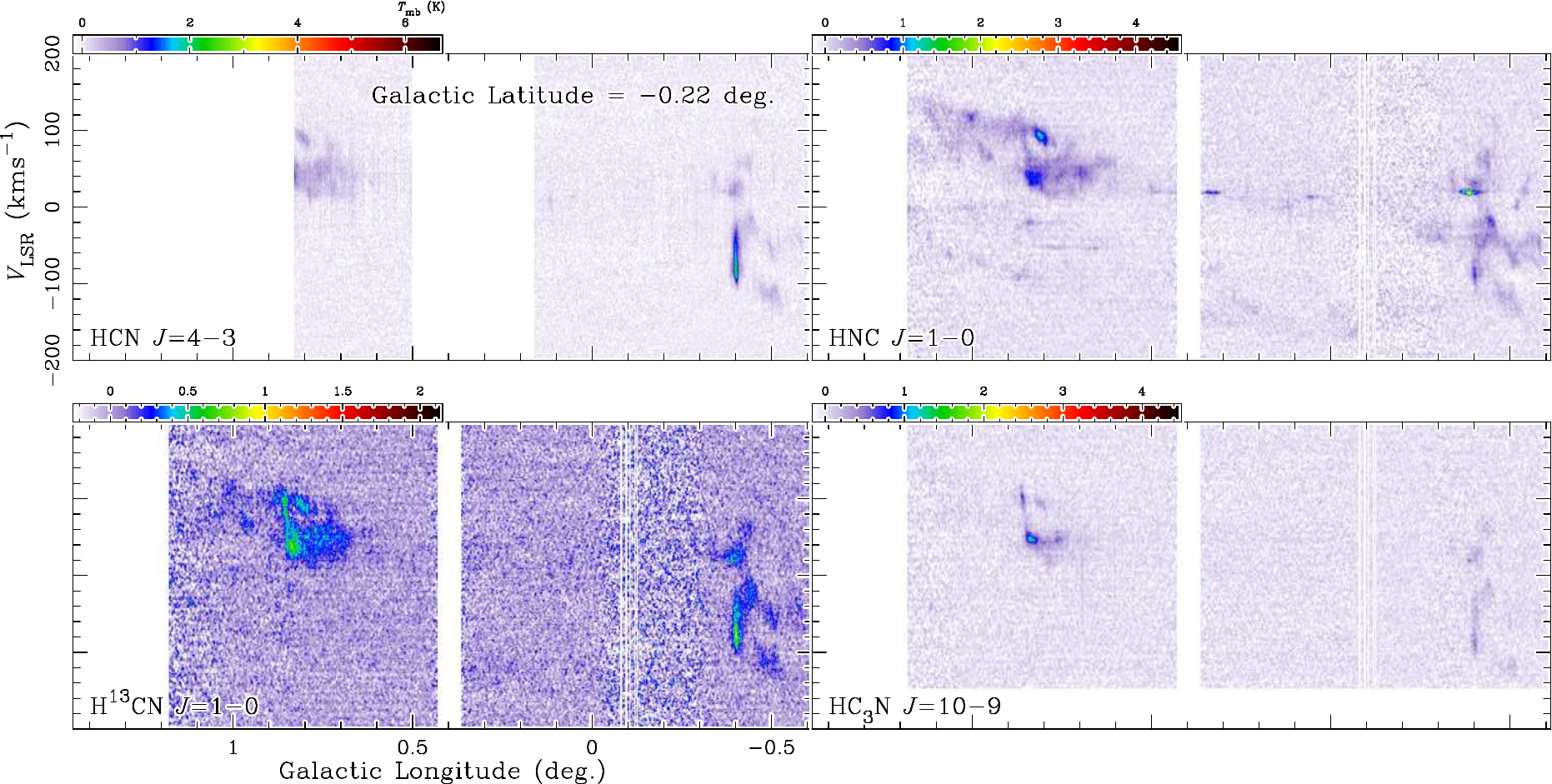}
\textcolor{red}{
\caption{Galactic longitude--\vlsr\ maps of HCN \JJ{4}{3} (top left), HNC \JJ{1}{0} (top right), \HCNt\ \JJ{1}{0} (bottom left), and \HCCCN\ \JJ{10}{9} (bottom right) for a latitude range of $-0^\circ.25$ to $+0^\circ.12$ at an interval of $90''$.  The channel center latitude is given on the HCN \JJ{4}{3}\ panel for each latitude channel.  Diagrams for the full latitude range are available in the Figure Set  \ref{FIGSET6}. }\label{FIG6}
}
\end{figure*}

\textcolor{red}{
\input{figset6.etex}
}
\fi

The overall spatial extension of molecular gas is best traced by the HNC \JJ{1}{0}\ map, where both intense emission from the central regions of the GMC complexes and low-level emission surrounding them are visible owing to the relatively low $\ncrit$ and $\Eu$ compared with the other lines. 
The \HCNt\ map is similar to the HNC map, except that the low-level emission is less clear owing to the low S/N ratio.
The HCN \JJ{4}{3}\ and \HCCCN\ maps are more dominated by compact features compared with the other two maps; 
their bright emissions are limited to the narrow areas close to the centers of the GMC complexes of the Sgr A, Sgr B, and M$0.11$$-0.08$ regions and several compact clumps such as the brick cloud (G0.253+0.016), CO$-0.40$, and CO$-0.30$ \citep{Tanaka2014}.  
Low-level emissions primarily have a clumpy or filamentary shape, and diffuse extended emissions are absent or considerably weak except for the envelopes of the Sgr B and Sgr A complexes. 
The most conspicuous feature in the HCN \JJ{4}{3}\ and \HCCCN\ maps is the bright compact emission from the Sgr B2 cluster-forming region, which is observed as absorption holes in the HNC and \HCNt\ maps against a strong continuum that is subtracted in the data reduction process. 

Considering that the upper state energies of HCN \JJ{4}{3}\ and \HCCCN\ are relatively high and their critical densities differ by two orders of magnitude, their morphological similarity to each other and difference from the other two lines suggest that the ratios between the intensities of different lines primarily reflect the temperature variation.
Therefore, the conventional assumption of constant \Tkin\ cannot be used for measuring \nHH\ through the excitation analysis; the distributions of \Tkin\ and \nHH\ must be determined simultaneously.




%% file: analysis.etex
The purpose of this study is to construct 3-D maps (2-D in space and 1-D in velocity) of the physical conditions and molecular abundances from our molecular line maps.
This section (\S\ref{SECTION_ANALYSIS}) is devoted to a description of the non local thermodynamical equilibrium (non-LTE) method and the HB inference used for the analysis.
The first subsection (\S\ref{SUBSECTION_NONLTE}) describes the parameters calculated with the non-LTE method.
We overview the framework for the parameter inference using the HB method in Subsection \ref{SUBSECTION_METHOD}.
Full details of the probability density functions (PDFs) used in the HB method are given in Subsection \ref{SUBSECTION_PDFS}.
We also present the numerical method to calculate the posterior distribution (\S\ref{SUBSECTION_MCMC}), a comparison with a simple nonhierarchical method (\S\ref{SUBSECTION_NONH}), and the details of the input data (\S\ref{SUBSECTION_DATA}). 
The results of the analysis are presented in the next section (\S\ref{SECTION_RESULTS}).

\subsection{Parameters}\label{SUBSECTION_NONLTE}
This analysis requires a larger dataset including other transitions of the molecules we observed, as the number of the observed lines (four) is less than the minimum number of parameters to be determined, i.e., six (the gas kinetic temperature \Tkin, the hydrogen volume density \nHH, the hydrogen column density per unit velocity width \NHH, the ratios between the fractional abundances \xmol\ of the three molecules, and the beam filling factor \ff).
We use the additional data of the nine lines listed in Table \ref{TABLE_INPDATA}, which are taken from the literature \citep{Tsuboi1999, Jones2012, Tsuboi2015, Ginsburg2016}, and determine 13 values, i.e., \Tkin, \nHH, \NHH, \ff, the C/\Ct\ isotopic ratio (\Rt), and \xmol\ of eight molecules (HCN, \HCOp, \NNHp, HNC, CS, SiO, \HCCCN, and \pformaldehyde). 

The model parameters are described in Table \ref{TABLE_MODELPARAM}. 
We use the logarithms of the parameters to constrain their values to be positive.
We introduce the parameter \fff, which defines the beam filling factor as $\ff = 1 - e^{-\fff}$, instead of directly including \ff\ in the model parameter set; this formulation limits \ff\ within a finite range of [0, 1], whereby we can assume infinite variable ranges for all model parameters.
We fix the \COn\ fractional abundance at $10^{-4}$ and use the molecule as a proxy for $\mathrm{H_2}$.
The \Ct\ isotopic abundance is assumed to be common for all C-bearing species.

\begin{figure*}
\setcounter{figure}{6}
\epsscale{1}
\plotone{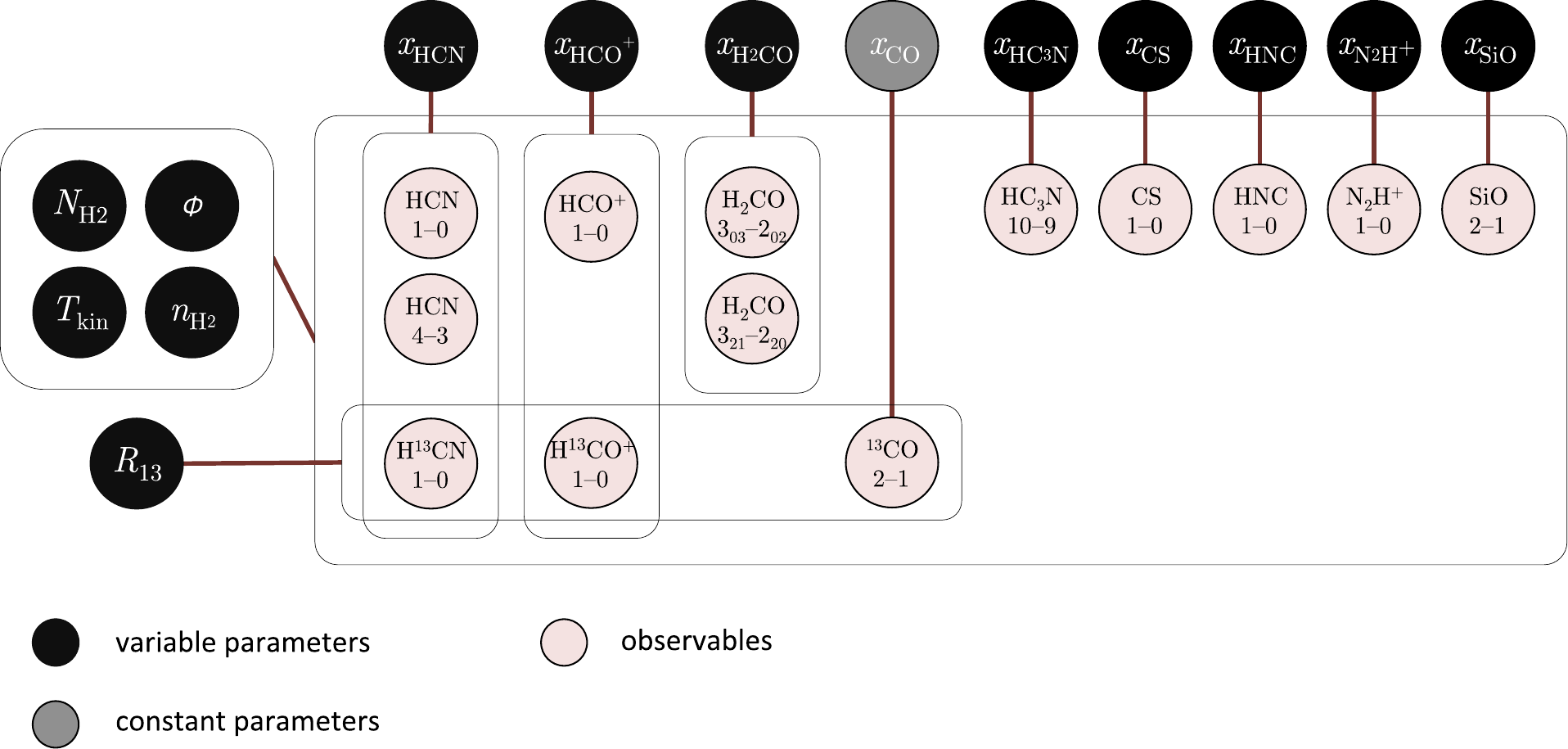}
\caption{Schematic diagram showing the dependencies of the observed line intensities on the model parameters.  Four parameters, \NHH, \Tkin, \nHH, and \ff\ affect all line intensities.  These parameters and molecular abundances \xmol\ determine the individual line intensities. The \Ct\ isotopologue lines (\HCNt \JJ{1}{0}, \HCOpt \JJ{1}{0}, and \COt \JJ{2}{1}) have additional dependencies on \Rt. }\label{FIG_RAWLVG}
\end{figure*}

\begin{figure*}
\epsscale{0.7}
\plotone{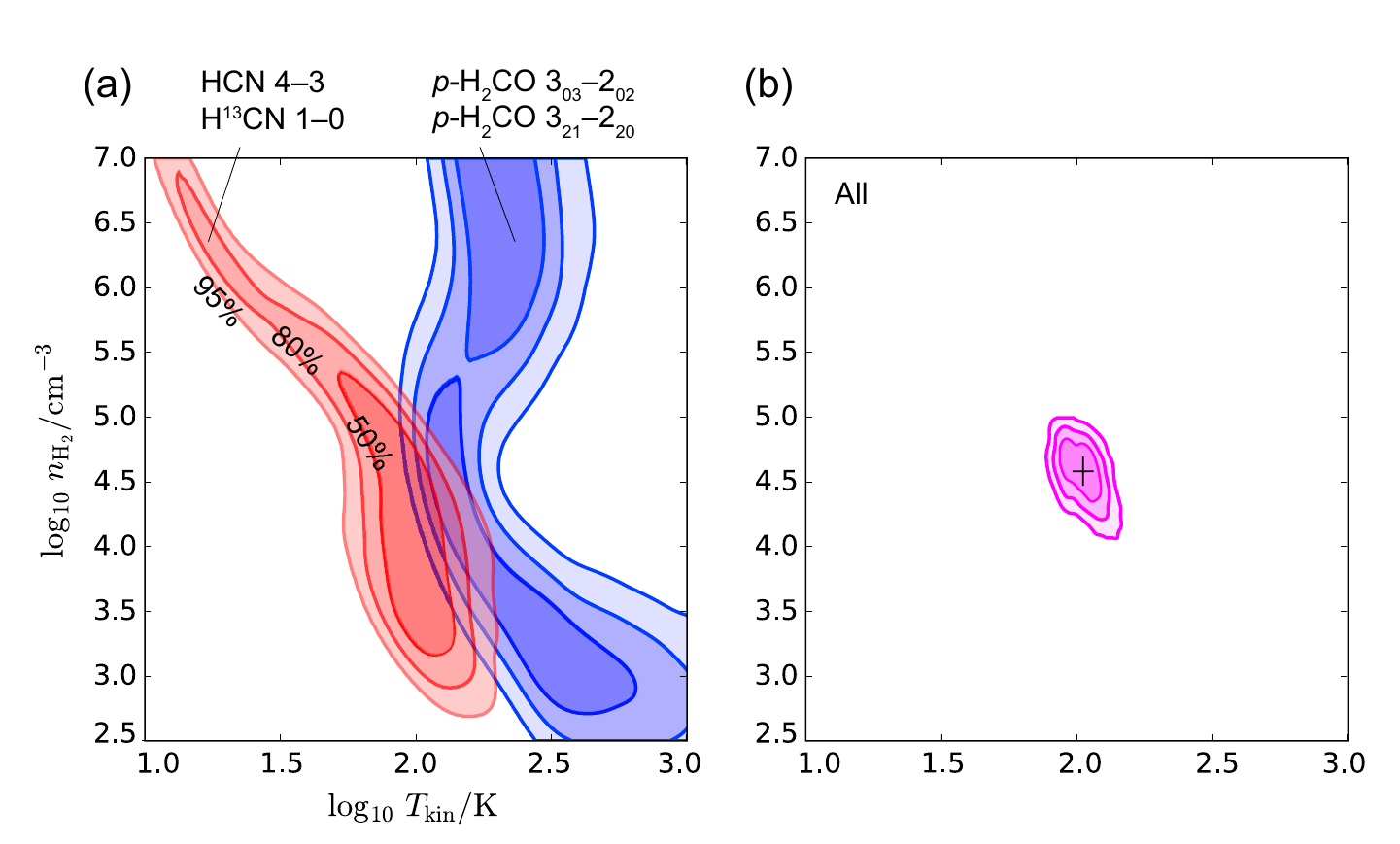}
\caption{(a) \Tkin--\nHH\ ranges calculated for the center position of the 50-\kmps\ cloud using the LVG analysis, where \ff = 0.25 and \Rt = 25\ are assumed.  The red  contours are for the simultaneous credible intervals calculated from the HCN \JJ{4}{3}\ and \HCNt\ \JJ{1}{0}\ intensities, and the blue contours are those for the two \formaldehyde\ lines (\pfx\ and \pfy).  The credible intervals are calculated using the nonhierarchical Bayesian analysis with a flat prior function.   (b) The credible intervals calculated using the four lines simultaneously.  The cross mark denotes the median of the simultaneous probability.}\label{LVG_SINGLE}

\epsscale{0.7}
\plotone{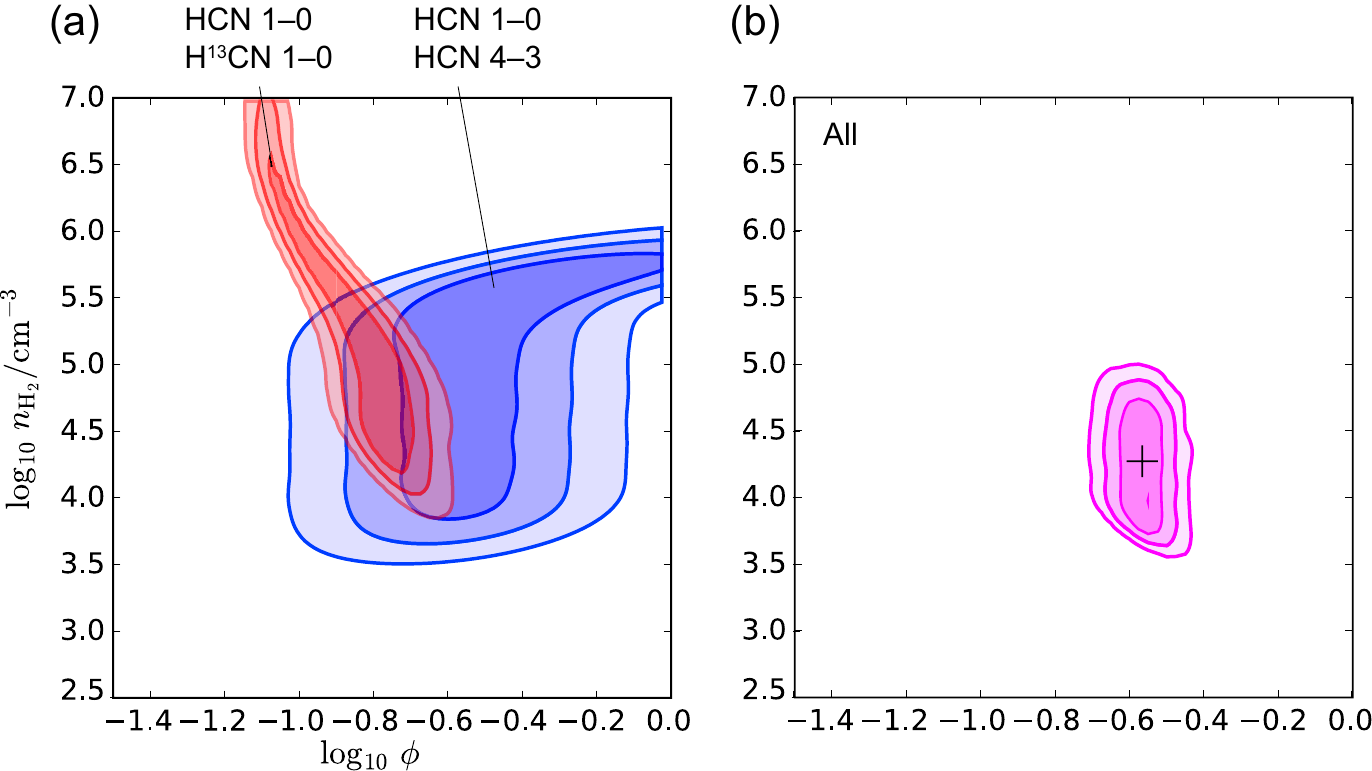}
\caption{(a) \fff--\nHH\ ranges calculated for the center position of the 50-\kmps\ cloud using the LVG analysis, where \Tkin = $10^2$ and \Rt = 25 are assumed.  The red  contours are for the simultaneous credible intervals calculated from the HCN \JJ{1}{0}\ and \HCNt\ \JJ{1}{0}\ intensities, and the blue contours are those for the HCN \JJ{1}{0}\ and \JJ{4}{3}\ lines.  The credible intervals are calculated using the nonhierarchical Bayesian analysis with a flat prior function.   (b) The credible intervals calculated using the three lines simultaneously. The cross mark denotes the median of the simultaneous probability. }\label{LVG_SINGLE2}
\end{figure*}

Figure \ref{FIG_RAWLVG} shows a schematic diagram of the dependencies of the molecular line intensities on the model parameters. 
In the non-LTE scheme the line intensities are a function of three parameters: the column density per unit velocity width ($N_X = \xmolp{X}\cdot\NHH$), \nHH, and \Tkin.   
The column density of \Ct\ isotopologues is given as a product of \Rt\ and the column density of the main $^{12}\mathrm{C}$ isotopologues.
Intensities multiplied by a beam filling factor \ff\ are the intensities that are actually observed with telescopes.   
Hence, as $\xmolp{CO}$ is constant in our model, $(N_{\rm mol} + 4)$ parameters are necessary to calculate the observed line intensity, where $N_{\rm mol}$ is the number of molecular species not including \Ct\ isotopologues.

The emissivity per unit \NHH\ is dependent on the population distribution among the internal energy levels in the molecule, which is determined by the equilibrium between radiative transitions, excitation by photon trapping, and excitation/de-excitation through collisions with molecular hydrogens.
The rates of these processes depend on the transition dipole moments and the collisional cross sections; hence, different lines have different dependencies on \Tkin\ and \nHH.
Figure \ref{LVG_SINGLE}a\ shows the \Tkin--\nHH\ parameter range calculated using the large velocity gradient (LVG) analysis \citep{Goldreich1974} from the HCN \JJ{4}{3}\ and \HCNt\ \JJ{1}{0}\ intensities toward the 50-\kmps\ cloud on the assumption that \Rt = 25 and \ff = 0.25. 
The rectangular-prior Bayesian analysis (\S\ref{SUBSECTION_NONH}) is used for parameter inference.
The highest probability parameter set is not uniquely determined since the number of input data (two) is less than the number of the free parameters (three; note that \xmolp{HCN} and \NHH\ degenerate in the parameter space).
Figure \ref{LVG_SINGLE}a\ also shows the parameter range calculated using the two \pformaldehyde\ lines;  their intensity ratio is a good probe of \Tkin\ as it is insensitive to \nHH \citep{Mangum1993,Ginsburg2016}. 
Thus the unique solution  $\left(\Tkin, \nHH\right) = (10^2\ \kelvin, 10^{4.5}\ \pcc) $ is obtained by combining the HCN and \formaldehyde\ results when \Rt\ and \ff\ are given.

The parameters $N_X$ and \ff\ degenerate in optically thin lines, but as the optical depth increases the intensity depends nonlinearly on $N_X$;  hence, the optically thick lines,  HCN \JJ{1}{0}\ and HCN \JJ{4}{3}\ in our dataset, contain information on \ff. 
Figure \ref{LVG_SINGLE2}\ shows the \ff--\nHH\ parameter ranges calculated from the three HCN lines toward the 50-\kmps\ cloud, assuming $\Tkin = 10^2\ \kelvin$ and $\Rt = 25$.    
\ff\ is determined well as $\log_{10}\ \ff = -0.56\pm0.06$ in this example.  
\Rt\ is mainly measured from the intensity ratio of the \HCOp \JJ{1}{0}\ line to its \Ct\ isotopologue line, as they are generally not too optically thick in the CMZ \citep{Riquelme2013}.
Therefore, the seven lines from HCN, \formaldehyde, and \HCOp\ and their \Ct\ isotopologues are sufficient to determine the physical condition parameters.

The HNC, \NNHp, HNC, CS, SiO, and \HCCCN\ data are primarily used to estimate their abundances; since multitransition data of these species are not included in our dataset, they do not contain significant information of the physical condition parameters. 
The molecular abundances are measured in the unit of fractional abundance to \NHH, whose column density is obtained from the \COt\ data.

\subsection{Hierarchical Bayesian Analysis}\label{SUBSECTION_METHOD}
We determine the optimal parameter distribution, $\pv = \left\{\pvi\right\}$, from the observed line intensities $\Iobsv=\left\{\Iobsv_i\right\}$ using the above non-LTE framework. 
The subscript index $i$ ($\in \left\{1,2, ..., N\right\}$) specifies a voxel in the 3-D space.
The individual line intensities and parameters at the \thup{i}\ voxel are represented by the vector components ${\Iobs_{i,j}}$ ($j \in \left\{1,2, ..., \Nl\right\}$) and ${p_{i,k}}$ ($k \in \left\{1,2, ..., \Np\right\}$), respectively, where  \Np\ and \Nl\ are the numbers of parameters and lines, respectively, both of which are 13 in our analysis.

The ML method is commonly used for parameter optimization problems; 
it is designed to define \pv\ that maximizes the likelihood function 
\begin{eqnarray}  
\PDF{\Iobsv|\pv}  & = & \prod_{i,j}\frac{1}{\delta_{i,j}}\exp\left[- \frac{1}{2} w_j \left(\frac{\Iobs_{i,j} - \Icalci_{j}}{\delta_{i,j}}\right)^2\right], \label{EQ1}
\end{eqnarray}
where \myvector{F\left(\pvi\right)}\ denotes the model intensities calculated with the non-LTE analysis.   
The factor $w_j$ is the weight of the $j^{\mathrm{th}}$ line in the fitting.
The notation \PDF{\myvector{x}|\myvector{a}}\ represents a conditional probability density function (PDF) for the variable \myvector{x}\ when the parameter \myvector{a}\ is given.
A normal error distribution with a standard deviation of $\delta_{i,j}$ is assumed for each $\Iobs_{i,j}$. 
Here and henceforth, we omit the normalization constants in the expressions of PDFs wherever possible.  
The rectangular-prior Bayesian analysis used in the previous subsection is essentially equivalent to the ML method, as we will show in a later section (\S\ref{SUBSECTION_NONH}).

The ML method frequently does not provide a satisfactory solution for the excitation analysis due to insufficient information about the errors present in the observations and models.
In addition to random noise $\delta$, which explicitly appears in Equation \ref{EQ1}, the input data may contain systematic errors due to calibration errors and uncertainty in spectral baseline subtraction, which are difficult to measure.
Moreover, systematic errors could be also introduced by the simplifications used in the model, such as the assumption of spatial uniformity along the line of sight and within the beam and the LVG approximation;
in general, different lines could probe different depths of molecular clouds according to their upper state energies and critical densities, and this discrepancy from the ideal one-zone situation is regarded as a systematic error.  
Furthermore, the lack of resolution for separating multiphase gas is non-negligible; interferometric observations have shown that cold quiescent clumps and warm turbulent filaments with a size of $\sim 0.1$ pc ($\sim 1''.5$)  are frequently entangled in typical single-dish beam sizes ($10''$--$20''$).
Another potential source of systematic errors in our analysis is the omission of the isotopic fractionation in C-bearing species. 
These systematic errors can seriously affect the results of the excitation analysis. 
As we will show in Subsection \ref{SECT_COMP_NONH}, nonhierarchical Bayesian analysis that is equivalent to the ML method creates an anticorrelation between \nHH\ and \NHH, \textcolor{red}{which is likely to be unphysical} since the column and volume densities should be positively correlated in realistic clouds.
This nonphysical anticorrelation is caused by the degeneracy of \nHH\ and \NHH\ in the LVG model; the standard ML method fails to resolve the degeneracy owing to the insufficient information about the systematic errors.
A severe degeneracy also exists between \NHH\ and \ff, as both parameters scale the line intensities in the model.

We adopt a Bayesian inference method similar to that used for the spectral energy distribution (SED) fittings of thermal dust emission \citep{Kelly2012}.
The basic idea of the Bayesian analysis is to apply additional constraints on the parameters by introducing the prior probability, thereby making it possible to assume reasonable values of the systematic errors and to forbid the parameters from exhibiting unphysical behaviors.
First, we modify the likelihood function (Equation \ref{EQ1}) to a form that includes a factor that represents the systematic errors, $\Emv = \left\{\Em_{i,j}\right\}$:
\begin{eqnarray}
\PDF{\Iobsv|\pv,\Emv} & = &
 \prod_{i,j}\frac{1}{\delta_{i,j}}\exp\left[- \frac{1}{2} {w_j} \left(\frac{\Iobs_{i,j} - \Em_{i,j}\cdot\Icalci_{j}}{\delta_{i,j}}\right)^2\right] . \label{EQ2}
\end{eqnarray}
The systematic errors are assumed to be multiplicative because we consider calibration errors and the deviation from the one-zone LVG model for error sources, which are likely to affect the intensities in multiplicative manners. 
The ML method cannot determine the unique optimal parameters for this modified likelihood function because the degree of freedom is less than zero.
Instead, the Bayesian method evaluates the posterior probability given by Bayes' theorem: 
\begin{eqnarray}
\PDF{\pv,\Emv|\Iobsv} & = & 
\frac{
\PDF{ \Iobsv|\pv,\Emv} \cdot \PDF{\pv,\Emv}
}
{
\PDF{\Iobsv} 
}. \label{EQ3_0}
\end{eqnarray}
The posterior probability \PDF{\pv,\Emv|\Iobsv}\ is the conditional PDF of the explanatory parameters (\pv, \Emv) where the observable \Iobsv\ is given.
$\PDF{\pv,\Emv}$ is the prior function, which is the simultaneous PDF for \pv\ and \Emv\ given {\it a priori} by assumptions and information other than \Iobsv. 
The denominator, $\PDF{\myvector{I}} = \int \PDF{I|\pv,\Emv}\cdot\PDF{\pv,\Emv}\cdot\mathrm{d}\pv\cdot\mathrm{d}\Emv$, is the normalization constant of the posterior, which can be omitted in the analysis.

The specific form of the prior is usually determined using statistical or physical modeling, as prior functions are not directly measurable.
One approach to model the prior probability is to use a function that explicitly limits the parameter variation within physically allowed ranges, such as rectangular or logistic functions \citep[e.g.][]{Kamenetzky2012}.    
In this paper, we adopt another approach using the HB modeling, which uses a prior that is further parameterized by the hyperparameter \myvector{\theta}:
\begin{eqnarray}
\PDF{\pv,\Emv,\myvector{\theta}|\Iobsv} & = & 
\frac{
\PDF{ \Iobsv|\pv,\Emv} \cdot \PDF{\pv,\Emv|\myvector{\theta}} \cdot \PDF{\myvector{\theta}}
}
{
\PDF{\Iobsv} 
}, \label{EQ3}
\end{eqnarray}
where $\PDF{\myvector{\theta}}$ is the hyperprior function, which defines the prior distribution on \myvector{\theta}.
In this analysis we adopt log-normal and multivariate Student functions for \Emv\ and \pv, respectively, whose mean values and scale factors are treated as hyperparameters;
namely, we assume that the {\it a priori} probability of the parameters has a unimodal distribution with unknown means and widths.
This is one of the simplest ways to model a prior distribution that is more realistic than a flat or rectangular distribution;  for example, \nHH\ and \NHH\ are known to have log-normal PDFs in turbulent-dominated molecular gas from  theories and observations \citep[e.g.][]{Nordlund1999,Pineda2010}.
The usage of unimodal priors ensures that the posterior distribution has at least one peak even when the likelihood function does not have a unique maximum value.

\begin{figure*}
\epsscale{1}
\plotone{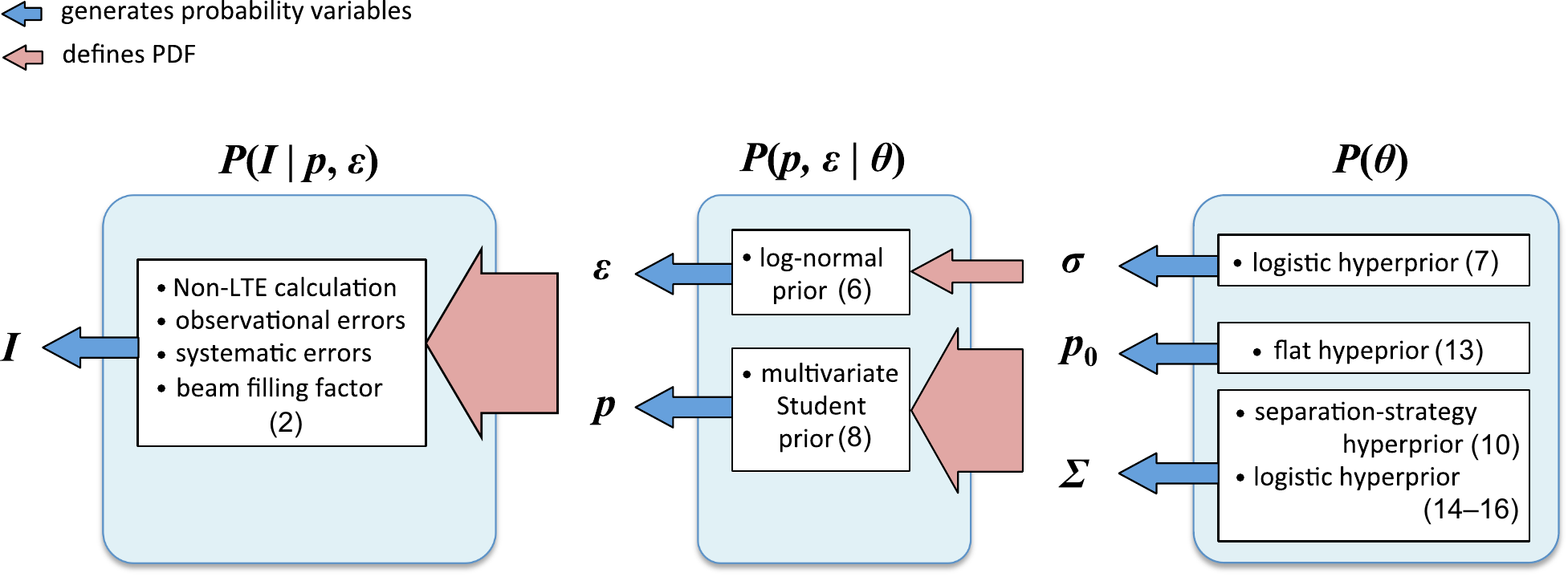}
\caption{Schematic diagram showing the hierarchical structure of the variables and probability functions in the HB analysis.  The numbers in parentheses are the equation numbers where the functions are defined in the text.  In the first level (rightmost in the figure),  the hyperparameter $\theta = \left(\pv_0, \myvector{\Sigma}, \smv \right)$ is a probability variable generated according to the probability given by the hyperprior function, $\PDF{\theta}$.  In the next level of the hierarchy, $\theta$ is an input parameter of the prior function $\PDF{\pv,\Emv|\theta}$\ that defines the PDF for the parameter $\pv$ and systematic errors $\Emv$.  The likelihood function $\PDF{\Iobsv|\pv,\Emv}$ defined by $\pv$ and $\Emv$ gives the PDF for the observables.   Note that the prior and hyperprior limit the statistical properties of the parameters, and the observational constraints from the physical processes are first considered in the likelihood function. }\label{FIG_SCHEMATIC_HB}
\end{figure*}

In this analysis we basically use a flat hyperprior distribution for $\theta$, but use logistic functions for a few hyperparameter elements whose variable ranges have to be explicitly controlled.
We summarize the hierarchical structure of the variables and PDFs in the HB analysis in the schematic diagram in Figure \ref{FIG_SCHEMATIC_HB}.   
The specific functional forms of the prior and hyperprior are given in the next subsection (\S\ref{SUBSECTION_PDFS}).
Examples of the function shapes used as the prior and hyperprior functions, the multivariate Student function, the log-normal function, and the logistic function, are presented in 2-D or 1-D space in Figure \ref{FIG_FUNCEXAMPLES}.

Note that the explanatory parameter \pv\ is a probability variable in the posterior function, whereas the ML method treats it as a constant parameter that defines the likelihood function.
Hence, the goal of the analysis is to calculate the numerical values of the posterior probability, and the determination of the parameter for the maximum probability is not a necessary step. 
The probability distribution of an individual parameter is given by the marginal posterior PDF, which are obtained by integrating the posterior \PDF{\pv,\Emv,\myvector{\theta}|\Iobsv}\ with the parameters that are not of interest.   
The marginal posterior of an arbitrary value $X$ is  
\begin{eqnarray}
\PDF{X|\Iobsv} & \equiv & \int \delta\left(X - y\left(\pv, \Emv\right)\right)\cdot\PDF{\pv,\Emv,\myvector{\theta}|\Iobsv} \cdot{\mathrm{d}}\pv\cdot{\mathrm{d}}\Emv\cdot{\mathrm{d}\myvector{\theta}},
\end{eqnarray}
where $y\left(\pv,\Emv\right)$ is the function that relates the variable $X$ and the model parameters.
We use the median $\tilde{X}$ and the \thn{25}--\thn{75}-percentile interval $\Delta_X$ of \PDF{X|\Iobsv}\ as the representative value and the uncertainty in the inferred parameter, respectively.

\subsection{Priors and Hyperpriors}\label{SUBSECTION_PDFS}
\subsubsection{Systematic Errors}
\begin{figure*}
\epsscale{1}
\plotone{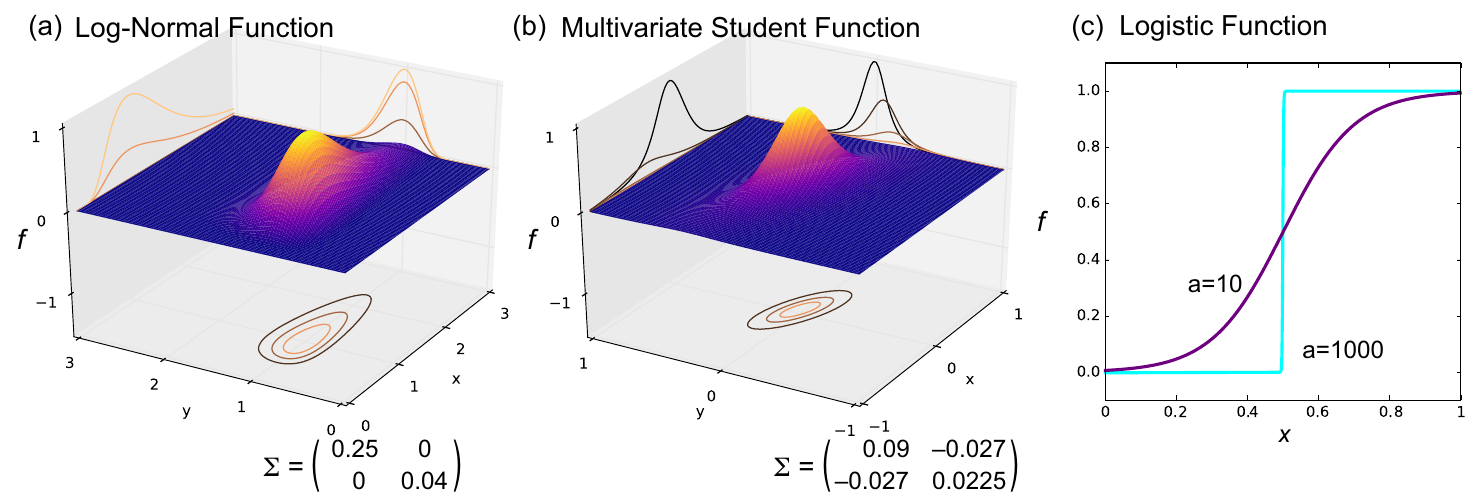}
\caption{Examples of the functions used for the prior and hyperprior PDFs:  (a)  Two-dimensional log-normal function $f(x,y) = \exp\left[-\frac{1}{2}\cdot\transverse{\myvector{x}}\cdot\Sigma^{-1}\cdot\myvector{x}\right]$,  (b) multivariate two-dimensional Student function $f(x, y) = \left[1 + \frac{1}{\nu}\cdot\transverse{\myvector{x}}\cdot\Sigma^{-1}\cdot\myvector{x}\right]^{-\frac{\nu+2}{2}}$ with $\nu = 2$,   and (c) logistic function $f(x) = \frac{1}{1 + \exp\left[\textcolor{red}{-a}\cdot\left(x - x_0\right)\right]}$ with $x_0 = 0.5$.  Contours are drawn at every 0.25. }\label{FIG_FUNCEXAMPLES}
\end{figure*}

The hyperparameter $\myvector{\theta}$ consists of three independent variables, i.e., $\myvector{\sigma}$, $\myvector{p_0}$, and $\myvector{\Sigma}$.   
The parameter $\myvector{\sigma}$ defines the prior for systematic errors, \Emv, for which we assume a log-normal form (Figure \ref{FIG_FUNCEXAMPLES}a) as follows:
\begin{eqnarray}
\PDF{\Emv|\myvector{\sigma}} & = & \prod_{i,j} \frac{1}{\sm_j\cdot\Em_{i,j}}\cdot \exp\left[-\frac{1}{2}\left(\frac{\ln{\Em_{i,j}}}{\sm_j}\right)^2\right]. \label{EQ4}
\end{eqnarray}
By adopting this formulation, which omits the cross terms of $\Em_{i,j}$, we implicitly assume that they are not correlated.
\PDF{\Emv|\myvector{\sigma}}\ is maximum at $\ln \Em_{i,j} = 0$ for all $i$ and $j$; therefore, the HB analysis will choose solutions with $\Em_{i,j}$ closer to unity if the other conditions are equal.

A logistic hyperprior is assumed for $\myvector{\sigma}$:
\begin{eqnarray}
\PDF{\myvector{\sigma}} & = & \prod_j  \frac{1}{ 1 + \exp\left[-a\cdot \left(\sigma_j - \sigma_{\mathrm{min},j}\right) \right]} , \label{EQ4_2}
\end{eqnarray}
 which gives 0 for $\sigma_j < \sigma_{\mathrm{min},j}$ and 1 otherwise for sufficiently large positive $a$ values (Figure \ref{FIG_FUNCEXAMPLES}c).   We find $a = 1000$ works for our problem. 
The lower limit values $\smv_{\mathrm{min}}$ are manually given parameters, which approximately correspond to the calibration uncertainties in the observations.   
We fix $\sm_{\mathrm{min},j}=  0.1$ for all $j$.

\begin{deluxetable*}{llccccl}
\tabletypesize{\scriptsize}
\tablecaption{Input Data\label{TABLE_INPDATA}}
\input{table1.etex}
\end{deluxetable*}

\begin{deluxetable*}{crll}
\tabletypesize{\scriptsize}
\tablecaption{Model Parameters\label{TABLE_MODELPARAM}}
\input{table2.etex}
\end{deluxetable*}

\subsubsection{Physical Conditions and Molecular Abundances}
The other two hyperparameters, $\myvector{p_0}$ and $\myvector{\Sigma}$, define the prior for \pv.
We adopt a multivariate Student prior (Figure \ref{FIG_FUNCEXAMPLES}b), which is one of the simplest ways to model a unimodal PDF for parameters that are potentially correlated with each other:
\begin{eqnarray}
\PDF{\pv|\myvector{p_0}, \myvector{\Sigma}} &=& |\myvector{\myvector{\Sigma}}|^{-\frac{N}{2}}\cdot
 {\mathop{\prod}\limits_i^{N}} \left[ 1 + \frac{1}{\nu} 
                                   \left( \pv_{i}- \pv_0 \right)^{\rm T}\cdot \myvector{\myvector{\Sigma}^{-1}}\cdot\left(\pv_{i} - \pv_0\right)
                                   \right]^{-\frac{\nu + \Np}{2}} ,  \label{EQ5}
\end{eqnarray}
where \myvector{\Sigma}\ and \myvector{p_0}\ denote the scale  matrix and  central value vector, respectively.
We fix the shape parameter $\nu$ as 2.
It is theoretically and observationally known that the PDFs of \NHH\ and \nHH\ are log-normal-shaped in turbulence-dominated gas, frequently with a power-law tail to the high-density end in self-gravitating clouds.  
We aim to emulate multicomponent turbulence and self-gravitating portions using the Student prior with a small $\nu$, as it allows more outliers than the normal prior with the same scale matrix.
The hyperparameter $\myvector{\Sigma}$ defines the {\it prior} distribution of $\pv$, which is not equal to the {\it posterior} variance--covariance matrix calculated using $\transverse{\pv}\cdot\pv$. 

As the scale matrix \myvector{\Sigma}\ is a symmetric non-negative definite matrix by definition, we cannot apply flat priors independently for its elements.
We adopt the separation strategy proposed by \cite{Barnard2000}, which decomposes \myvector{\Sigma}\ into the product of the prior correlation coefficient matrix $R = \{R_{i,j}\}$ and scaling diagonal matrix $S=\{S_i\cdot \delta_{i,j}\}$:
\begin{eqnarray} 
\Sigma & = & 
\left(
\begin{array}{cccc}
S_1 &     &     & 0 \\
    & S_2 &     &  \\
    &     & \ddots &  \\
0   &     &     & S_{\Np}
 \end{array}
\right) \cdot 
\left(
\begin{array}{cccc}
1      & R_{1,2} & \hdots & R_{1,\Np} \\
R_{1,2} & 1      & \hdots & R_{2,\Np} \\
\vdots & \vdots & \ddots & \vdots \\
R_{1,\Np} & R_{2,\Np} & \hdots & 1
\end{array}
\right) \cdot
\left(
\begin{array}{cccc}
S_1 &     &     & 0 \\
    & S_2 &     &  \\
    &     & \ddots &  \\
0   &     &     & S_{\Np} 
\end{array} 
\right). \label{EQ_SEPARATION}
\end{eqnarray}
Independent hyperpriors are assumed for $R$ and $S$:
\begin{eqnarray}
\PDF{\Sigma} &= &|S|^{-\Np}\cdot \PDF{S}\cdot\PDF{R} \label{EQ6_1},
\end{eqnarray}
where
\begin{eqnarray}
\PDF{R} & = & 
\left\{
\begin{array}{ll}
|R|^{-(\Np+1)}\cdot\mathop{\prod}\limits_k^\Np \left[\left({R^{-1}}_{k,k}\right)^{-\frac{N}{2}}\right] & (for\ |R| > 0 ) \\
0 &  (otherwise)
\end{array}
\right. \label{EQ6_2} \\
\PDF{S} & = & 1
. \label{EQ6_3}
\end{eqnarray}
${R^{-1}}_{k,k}$ denotes the diagonal elements of the inverse matrix of $R$.  
The factor $|S|^{-\Np}$ in Equation (\ref{EQ6_1}) originates from the Jacobian for variable conversion (Equation \ref{EQ_SEPARATION}); this factor does not appear if we adopt $R$ and $S$ as hyperparameters instead of $\myvector{\Sigma}$.
This PDF provides a flat distribution for each element of $S_k$ and $R_{k_1,k_2}$  over infinite and $[-1,1]$ ranges, respectively.

A flat hyperprior is assumed for the central value vector \myvector{p_0}:
\begin{eqnarray}
\PDF{\myvector{p_0}} = 1 .  \label{EQ7_2}
\end{eqnarray}

\subsection{Additional Hyperpriors}
In addition to the hyperpriors described in the above subsections (Equations \ref{EQ4_2}\ and \ref{EQ6_1}--\ref{EQ7_2}), we introduce additional hyperpriors for a few individual elements of $R$ and $S$ to explicitly prohibit unphysical behavior of the parameter \pv.

\paragraph{Prior Correlation Coefficient between \NHH\ and \nHH}
As mentioned in Subsection \ref{SUBSECTION_METHOD}, the degeneracy between \NHH\ and \nHH\ in the non-LTE excitation equations often creates an unphysical anticorrelation.
To prevent this artifact, we introduce a logistic hyperprior for $R_{\NHH, \nHH}$:
\begin{eqnarray}
\PDF{R_{\NHH, \nHH}} & = & \frac{1}{ 1 + \exp\left(\textcolor{red}{-a}\cdot R_{\NHH,\nHH}\right)},   \label{EQ7}
\end{eqnarray}
which constrains $R_{\NHH,\nHH}$ to be non-negative.
\paragraph{Prior Correlation Coefficient between \NHH\ and \fff}
Similarly, the severe degeneracy between \NHH\ and \fff\ may create an artificial anticorrelation between them, although it is reasonably assumed that voxels containing high-density regions are more likely to have higher volume filling factors, and hence higher beam-filling factors.   
Therefore, we impose a logistic hyperprior on  $R_{\NHH, \fff}$:
\begin{eqnarray}
\PDF{R_{\NHH, \fff}} & = & \frac{1}{ 1 + \exp\left(\textcolor{red}{-a}\cdot R_{\NHH,\fff}\right)} .   \label{EQ7_1}
\end{eqnarray}
\paragraph{Prior Scaling Factor for \Rt}
\cite{Riquelme2013} measured that the variation of \Rt\ in the inner 100-pc region is relatively small, approximately within a factor of 2. 
Therefore, we limit the variable range of \Rt\ by imposing a logistic hyperprior on $S_\Rt$:
\begin{eqnarray}
\PDF{S_{\Rt}} & = & \frac{1}{ 1 + \exp\left[\textcolor{red}{a}\cdot \left(S_\Rt - S_{\mathrm{max},\Rt}\right) \right]}. \label{EQ7_3}
\end{eqnarray}
We fix the upper limit value $S_{\mathrm{max},\Rt} = 0.03$ for this analysis; this corresponds to a relative variation of 30\%\ (\thn{5}--\thn{95}\ percentile) from the central value for the Student function with $\nu = 2$.
By limiting the variation of \Rt\ to this reasonably small value, we are able to prevent the variation of \Rt\ from creating an artificial correlation among \Rt, \xmolp{\HCN}, and \xmolp{\HCOp}; 
as these abundances are determined primarily by the intensities of the \Ct\ isotopologue lines, they are scaled by \Rt.

\bigskip
In total, the final prior \PDF{\pv,\Emv|\myvector{\theta}}\ and hyperprior \PDF{\myvector{\theta}}\ are given by the product of the functions for the individual parameters and hyperparameters:
\begin{eqnarray}
\PDF{\pv,\Emv|\myvector{\theta}} & = & \PDF{{\pv|\myvector{p_0},\myvector{\Sigma}}} \cdot \PDF{\Emv|\myvector{\sigma}} \\
\PDF{\myvector{\theta}} &=& \PDF{\myvector{p_0}} \cdot \PDF{\myvector{\Sigma}} \cdot  \PDF{\myvector{\sigma}} \cdot \PDF{R_{\NHH, \nHH}} \cdot \PDF{R_{\NHH, \fff}} \cdot \PDF{S_{\Rt}}.
\end{eqnarray}
Then, the posterior function $\PDF{\pv,\Emv,\myvector{\theta}|\Iobsv}$ is obtained by multiplying  Equations \ref{EQ2}, \ref{EQ4}--\ref{EQ5}, \ref{EQ6_1}, and \ref{EQ7_2}--\ref{EQ7_3}. 

\subsection{Numerical Solution Using the Markov-chain Monte Carlo Method}\label{SUBSECTION_MCMC}
The posterior function \PDF{\pv,\Emv,\myvector{\theta}|\Iobsv}\ is numerically calculated using the Markov-chain Monte Carlo (MCMC) method.
For the parameter vectors \pv\ and \Emv, the hybrid Monte Carlo (HM) method \citep{Duane1987} is employed to simultaneously sample all of their elements at each voxel.
The HM method is used also for the hyperparameter $\myvector{p_0}$.
The random-walk Metropolis--Hastings (MH) algorithm is adopted for the $R$ component of the hyperparameter $\myvector{\Sigma}$, where the Wishart proposal function is used to draw the proposal value, $\myvector{\Sigma}^{(K+1)}$, from the current value of $\myvector{\Sigma}^{(k)}$ at the \thup{K}\ step:
\begin{eqnarray}
\PDF{\myvector{\Sigma}^{(K+1)}|\myvector{\Sigma}^{(K)}} & = & 
|\Sigma^{(K+1)}|^{\frac{b -N_p  - 1}{2}}\cdot
|\Sigma|^{-\frac{N_p}{2}}\cdot
\exp\left[-\mathrm{tr}\left(\frac{{\Sigma^{(K)}}^{-1}\cdot\Sigma^{(K+1)}}{2}\right)\right] , 
\end{eqnarray}
so that $\myvector{\Sigma}$ performs a random walk while remaining symmetric positive definite. 
The shape parameter $b$ is manually adjusted so that the acceptance rate is not too high or low.
The $S$ component of $\myvector{\Sigma}$ is sampled using the HM method.
The random-walk MH algorithm with the Gaussian proposal function is used for the hyperparameter $\myvector{\sigma}$.

The model intensities, \Icalc, are calculated by solving the equilibrium between radiative transitions and the collisional excitation and de-excitation by molecular hydrogen.
The photon-trapping effect and cosmic microwave background are considered using the LVG approximation.   
The transition rates are taken from the Leiden Atomic and Molecular Database \citep[LAMDA;][]{Schoier2005}. 
We neglect the radiative excitation and infrared pumping by local continuum sources for simplicity, as they are insignificant compared with collisional excitation for  most of our data.
The exceptions are positions near the cluster-forming regions in the Sgr A and Sgr B2 complexes, for which our estimates are likely to include large errors. 

We ignore the hyperfine splitting of the HCN \JJ{1}{0}\ and \NNHp\ \JJ{1}{0}\ transitions since the absorption among the hyperfine components cannot be precisely treated with the scheme of the LVG approximation for the CMZ, where the intrinsic line width is greater than the width of the hyperfine splitting.  
This simplification may introduce a large uncertainty in the HCN \JJ{1}{0}\ transition, which is generally optically thick.

\subsection{Nonhierarchical Analysis}\label{SUBSECTION_NONH}
We perform another Bayesian analysis using a simpler nonhierarchical prior function for the purpose of comparison with the HB analysis.
In this analysis, the original form of the likelihood function (Equation \ref{EQ1}) without \Em\ is adopted.
A rectangular prior PDF is assumed to constrain \pv\  within reasonable ranges:
\begin{eqnarray}
\PDF{\pv} & = & \prod_{i,j} H\left(p_{i,j} - p_{\mathrm{min},j}\right)\cdot H\left( p_{\mathrm{max},j} - p_{i,j}\right) \label{EQ_NH1} 
\end{eqnarray} 
using the step function $H(x)$. 
The parameter ranges, $p_{\mathrm{min,max}}$, are fixed at the values given in Table \ref{TABLE_NONH}.
The final nonhierarchical posterior function is given as
\begin{eqnarray}
\PDF{\pv|\Iobsv} &=& \frac{\PDF{\Iobsv|\pv}\cdot\PDF{\pv}}{\PDF{\Iobsv}} . \label{EQ_NH2}
\end{eqnarray}
This nonhierarchical analysis is equivalent to the ML method with a finite parameter range, as the expression of the posterior function $\PDF{\pv|\Iobsv}$ is identical to the likelihood function $\PDF{\Iobsv|\pv}$, except for the normalization constant, as long as \pv\ remains within the range defined by the prior function.

\begin{deluxetable}{lcc}
\tabletypesize{\scriptsize}
\tablecaption{Parameter Ranges Applied in the Nonhierarchical Method\label{TABLE_NONH}}
\input{table3.etex}
\end{deluxetable}

\subsection{Input Data and Weights}\label{SUBSECTION_DATA}
The data from the 3-mm-band Mopra 22-m telescope survey are given in the scale of $\Ta$\ the in the literature \citep{Jones2012}. 
We applied a scale factor of 2.65 for the Mopra data so that the \textcolor{red}{HNC,} \HCCCN\ and \HCNt\ intensities included in the Mopra survey data become consistent with \Tmb\ of the same lines in our NRO 45-m data;   
the factor of 2.65 is consistent with the main-beam efficiency of the Mopra 22-m telescope, $0.4$--$0.5$ \citep{Ladd2005,Jones2012}. 
The differences between the spatial resolutions in the data obtained from five different instruments are corrected by resampling the images to a grid spacing of $30''\times30''\times10\ \kmps$  with a resolution of $60''\times60''\times10\ \kmps$.   
The $60''$ angular size translates into a linear size of 2.4 pc at the distance of the Galactic Center.
The Sgr D region ($l > 1^\circ$) is not used owing to the lack of \HCOpt\ data in this region.
The velocity ranges for the foreground spiral arm regions are manually filtered to prevent deep self-absorption dips in the HCN, \HCOp, and HNC lines in the arms.
We approximate the RMS noise levels, $\delta$, to be spatially uniform and measure them using the emission-less regions of the smoothed maps.  The measured values of $\delta$ are given in Table \ref{TABLE_INPDATA}.

The input data are classified into two categories; the first comprises the lines from the molecules that have multiple transitions in our dataset (HCN, \HCOp, \pformaldehyde, and their \Ct\ isotopologues), which are necessary to determine the physical conditions, \Rt, and \fff, while the second comprises molecules with single transitions, which are required only for calculating \xmol.
We apply a 3-$\sigma$ threshold for the first group lines; as a result of this filtering, almost the entire part of the CND and 180-pc ring are excluded from the analysis.
We apply different weights ($w_j$ in Equations \ref{EQ1} and \ref{EQ2}) for these two groups;  $w_j = 1$ for the first group, and $w_j = 0.1$ for the second group.
This ensures that the physical condition parameters are determined primarily by the intensity ratios among the first group lines.   
The effect of these weights on the final results are discussed in a later section (\S\ref{SECTION_CAVEATS}).

As mentioned earlier, the \COt\ \JJ{2}{1}\ data \citep{Ginsburg2016} are used to trace the hydrogen column densities by assuming a constant fractional abundance of $10^{-4}$ for \COn.
The assumption of constant \COn\ abundance may not be true, particularly for the Sgr A complex where a considerably high atomic carbon ($\Cn$)-to-\COn\ abundance ratio of $\sim 1$ is reported \citep{Tanaka2011}.
In addition, the low critical density of the line ($6\times 10^3\ \pcc$) may violate the one-zone assumption, which states that all lines originate from the same volume.
This effect is not negligible, as we will show in Subsection \ref{SECT_MULTCOMP}. 
Nevertheless, the impact of this effect on our analysis is limited, as these uncertainties only affect the \COn\ fractional abundance; changes in the \COt\ \JJ{2}{1}\ intensity or \xmolp{\COn}\ would systematically increase or decrease the fractional abundances of the other molecules, but \Tkin, \nHH, and the relative abundances among the molecules except for \COn\ are almost unaffected.

%% file: table1.etex


\tablecolumns{7}
\small
\tablehead{\\ \colhead{molecule} & \colhead{transition} & \colhead{frequency$^{\rm a}$} & \colhead{${\Eu}/{\kB}$$^{\rm a}$} & \colhead{$\log_{10}{\ncrit}$$^{\rm a,b}$}& \colhead{$\delta^{\rm c}$} & \colhead{reference} \\
  & & \colhead{(GHz)} & \colhead{(K)} & \colhead{($\pcc$)} & \colhead{(mK)} &  }
\tabletypesize{\scriptsize}
\tablewidth{0pt}
\startdata
HCN               & \JJ{4}{3}              & 354.505\phn & 42.5\phn & 7.3\phn & 20 &\multirow{4}{*}{this work}      \\
\HCNt             & \JJ{1}{0}              &  86.340\phn & 4.1\phn  & 6.5\phn & 27 &\\
HNC               & \JJ{1}{0}              &  90.664\phn & 4.4\phn  & 5.6\phn & 33 &\\
\HCCCN            & \JJ{10}{9}             &  90.979\phn & 24.0\phn & 5.2\phn & 26 &\\
\hline
HCN               & \JJ{1}{0}              &  88.632\phn &  4.3\phn & 6.4\phn & 50 &\multirow{3}{*}{\cite{Jones2012}} \\
\HCOp             & \JJ{1}{0}              &  89.189\phn &  4.3\phn & 5.3\phn & 76 &\\
\NNHp             & \JJ{1}{0}              &  93.174\phn &  4.5\phn & 5.3\phn & 37 &\\
\hline
\HCOpt            & \JJ{1}{0}              &  86.754\phn &  4.2\phn & 5.3\phn & 24 &\multirow{2}{*}{\cite{Tsuboi2015}}\\
SiO               & \JJ{2}{1}              &  86.847\phn &  6.3\phn & 5.4\phn & 24 &\\
\hline
\COt              & \JJ{2}{1}              & 220.399\phn & 15.9\phn & 3.8\phn & 93 &\multirow{3}{*}{\cite{Ginsburg2016}}\\
\pformaldehyde    & $3_{03}\mbox{--}2_{02}$ & 218.222\phn & 21.0\phn & 6.1\phn & 9  &\\
                  & $3_{21}\mbox{--}2_{20}$ & 218.760\phn & 68.1\phn & 5.6\phn & 13 &\\
\hline
CS                & \JJ{1}{0}              &  48.991\phn &  7.1\phn & 5.3\phn & 42 &\cite{Tsuboi1999} 
\enddata
\tablenotetext{a}{Taken from the from the Leiden Molecular and Atomic Database \citep{Schoier2005}.}
\tablenotetext{b}{Values at $T = 50\ \kelvin$, calculated as $\ncrit = A_{i,i-1} / \mathop{\sum}\limits_{j <  i} C_{i,j}$, where $A$ and $C$ denote the Einstein A coefficient and collision rate coefficient, respectively. }
\tablenotetext{c}{RMS noise level, measured in the data cube smoothed to a $60''\times60''\times 10\ \kmps$ resolution.}

%% file: table2.etex


\tablecolumns{4}
\small
\tablehead{\\ \colhead{name} & & & \colhead{description} }
\tabletypesize{\scriptsize}
\startdata
\sidehead{\parbox[l]{60pt}{Local\\Parameters}}
&$\pvi$ & & logarithms of the parameters at the \thup{i}\ voxel     \\
&& $\log_{10}\ \frac{\mathrm{d}\NHH}{\dv}$/$\psc(\kmps)^{-1}$  & {hydrogen column density per unit velocity} \\
&& $\log_{10}\ \nHH/\pcc$      & hydrogen volume density \\                    
&& $\log_{10}\ \Tkin/\kelvin $ & gas kinetic temperature \\                  
&& $\log_{10}\ \Rt$ & $^{12}$C/$^{13}$C isotopic abundance          \\
&& $\log_{10}\ \fff$ & parameter defining the beam filling factor $\ff$ \\
&& & ($\ff \equiv 1 - e^{-\fff}$) \\
&& $\log_{10}\ \xmolp{\HCN}$ & fractional abundances \\
&& $\log_{10}\ \xmolp{\HCOp}$ & \\
&& $\log_{10}\ \xmolp{\HNC}$ &  \\
&& $\log_{10}\ \xmolp{\HCCCN}$ &  \\
&& $\log_{10}\ \xmolp{\pformaldehyde}$ &  \\
&& $\log_{10}\ \xmolp{\CS}$ & \\
&& $\log_{10}\ \xmolp{\SiO}$ &  \\
&& $\log_{10}\ \xmolp{\NNHp}$ &  \\
&$\Em_{i,j}$ & & multiplicative error of the \thup{j}\ line at the \thup{i}\ voxel \\
\sidehead{\parbox[l]{60pt}{Hyper-\\parameters}}
&\myvector{p_0} & & central value of the multivariate Student prior for \pv  \\
&$\Sigma$       & & scale matrix of the multivariate Student prior for \pv   \\
&$\sm_{j}$      & & standard deviation of the log-lormal prior $\Em_{i,j}$\\ 
\sidehead{Constants} 
&$\nu$ & = 2 & scale parameter of the Student prior for \pv \\
&\textcolor{black}{\xmolp{\COn}}  & = $10^{-4}$ & fractional abundance of \textcolor{black}{\COn} \\
& {\textcolor{black}{$\smv_{\mathrm{min}}$}} & \textcolor{black}{= 0.1} & {\textcolor{black}{minimum $\smv$}} \\
& {\textcolor{black}{$S_{\mathrm{max}, \Rt}$}} & \textcolor{black}{= 0.03} & {\textcolor{black}{maximum $S_{\Rt}$}} \\
& {\textcolor{black}{$w_j$}} & \textcolor{black}{\parbox[c]{60pt}{\begin{eqnarray} = \left\{ \begin{array}{c}1 \\ 0.1\end{array}\right.\nonumber\end{eqnarray}}} & \parbox[l]{300pt}{\textcolor{black}{weight of \HCN, \HCOp, \formaldehyde, and \Ct\ isotopomar lines \\weight of the other lines}}\\
\enddata


%% file: table3.etex


\tablecolumns{3}
\small
\tablehead{\\ \colhead{parameter} & \colhead{$p_{\mathrm{min}}$} & \colhead{$p_{\mathrm{max}}$} }
\tabletypesize{\scriptsize}
\tablewidth{0pt}
\startdata
$\log_{10}\ \frac{\mathrm{d}\NHH}{\dv}$ / $\psc(\kmps)^{-1}$  & \textcolor{black}{20} & 24 \\
$\log_{10}\ \nHH/\pcc$ & \textcolor{black}{2} & \textcolor{black}{5} \\
$\log_{10}\ \Tkin/\kelvin $& 1 & \textcolor{black}{3} \\
$\log_{10}\ \Rt$ & \textcolor{black}{1.2} & \textcolor{black}{1.6} \\
$\log_{10}\ \fff$ & $\textcolor{black}{-2}$ &  1 \\
$\log_{10}\ \xmolp{\cdot}$ & \textcolor{black}{$-9$} & \textcolor{black}{$-6$}
\enddata
 

%% file: results.etex
\newcommand\nclumps{206}

\begin{figure}
\ifdraft
\epsscale{0.6}
\else
\epsscale{0.6}
\fi
\plotone{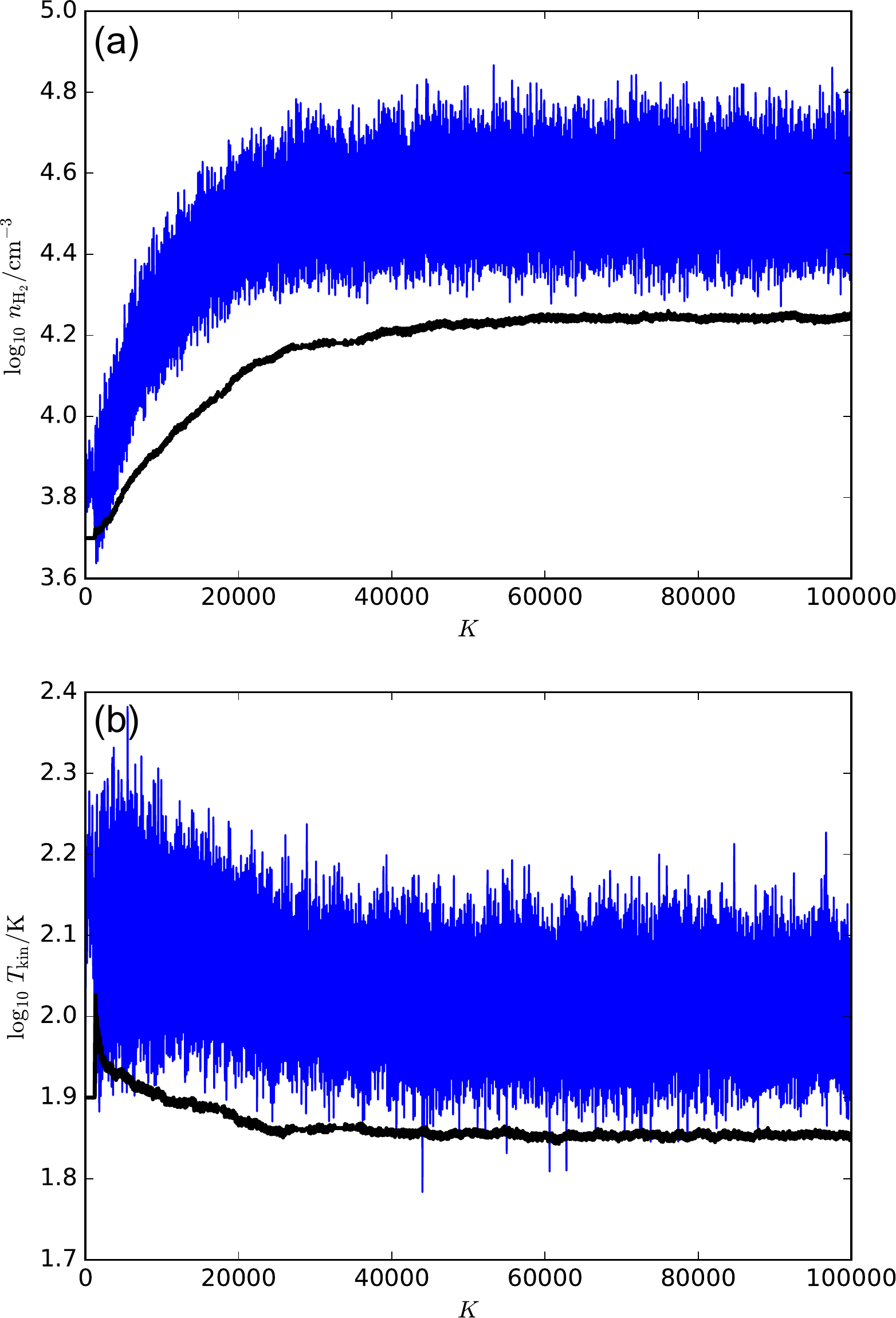}
\caption{Trace plots of \nHH\ and \Tkin\ in the HB analysis (panels a and b, respectively) against the MCMC step $K$.  The blue and black lines denote the plots for the local parameter $\pv$ at the center position of the 50-\kmps\ cloud and for the hyperparameter $\pv_0$, respectively. We kept $\pv_0$ constant and the step sizes small until $K$=1000 in this MCMC run.}\label{FIG_RAWPOSTERIOR}
\end{figure}

\begin{figure}
\ifdraft
\epsscale{0.6}
\else
\epsscale{0.65}
\fi
\plotone{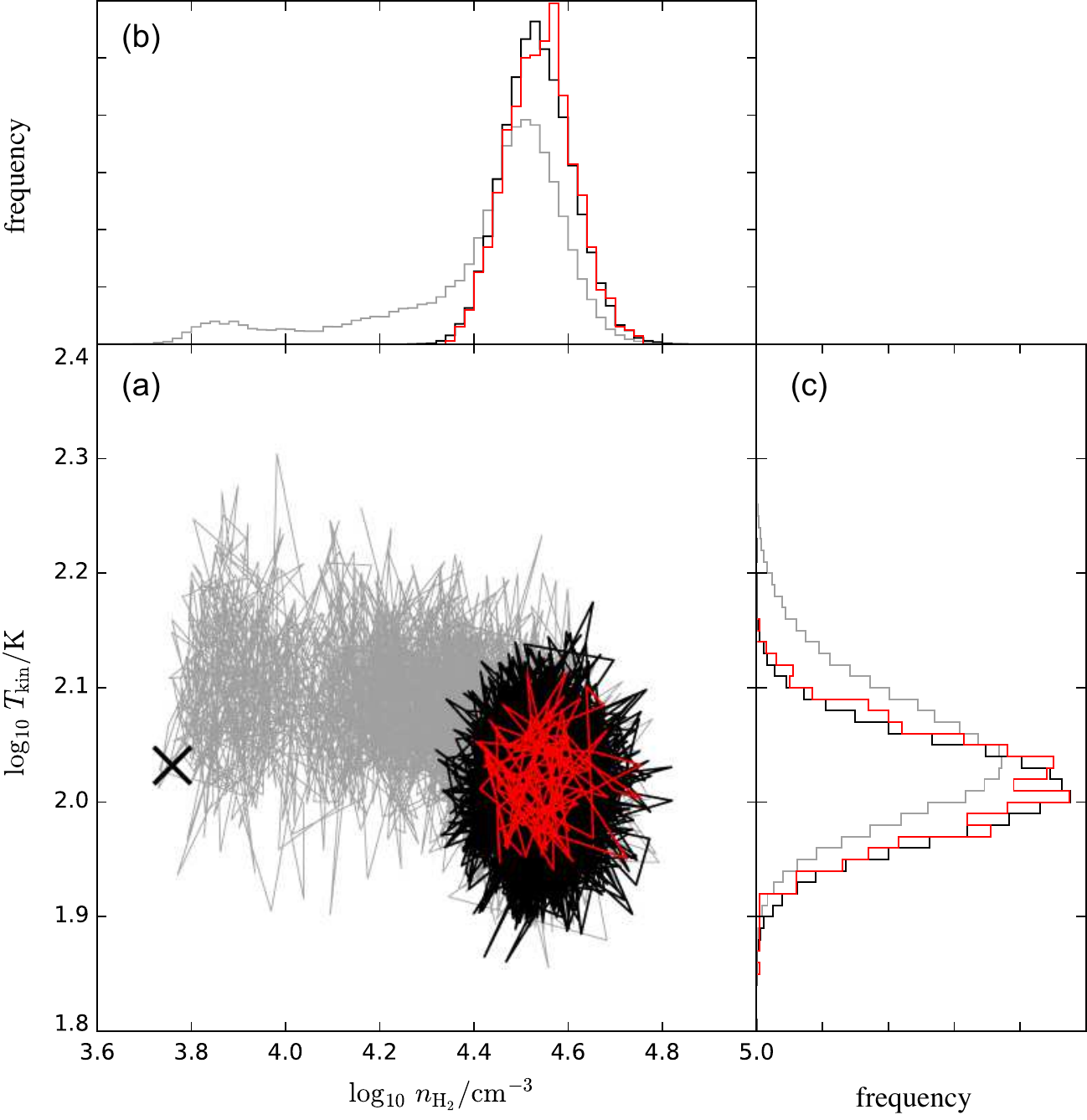}
\caption{(a) Trace plot of (\nHH, \Tkin) at the center position of the 50-\kmps\ cloud for the HB analysis on the two-dimensional plane at every 10 MCMC steps.  The gray and black lines are for the burn-in steps ($K < 60000$) and the steps after convergence ($K > 60000$), respectively.  The initial parameter is indicated by the cross mark.   The plot points for $K$ = 70000--71000 are highlighted in red, showing that the parameter space is sampled sufficiently randomly within this short interval. (b) and (c) Frequency histograms of \nHH\ and \Tkin, respectively. The gray, black, and red lines correspond to the plot points with the respective colors in panel (a).   }\label{FIG_RAWPOSTERIOR2}
\end{figure}

\begin{figure*}
\epsscale{1}
\plotone{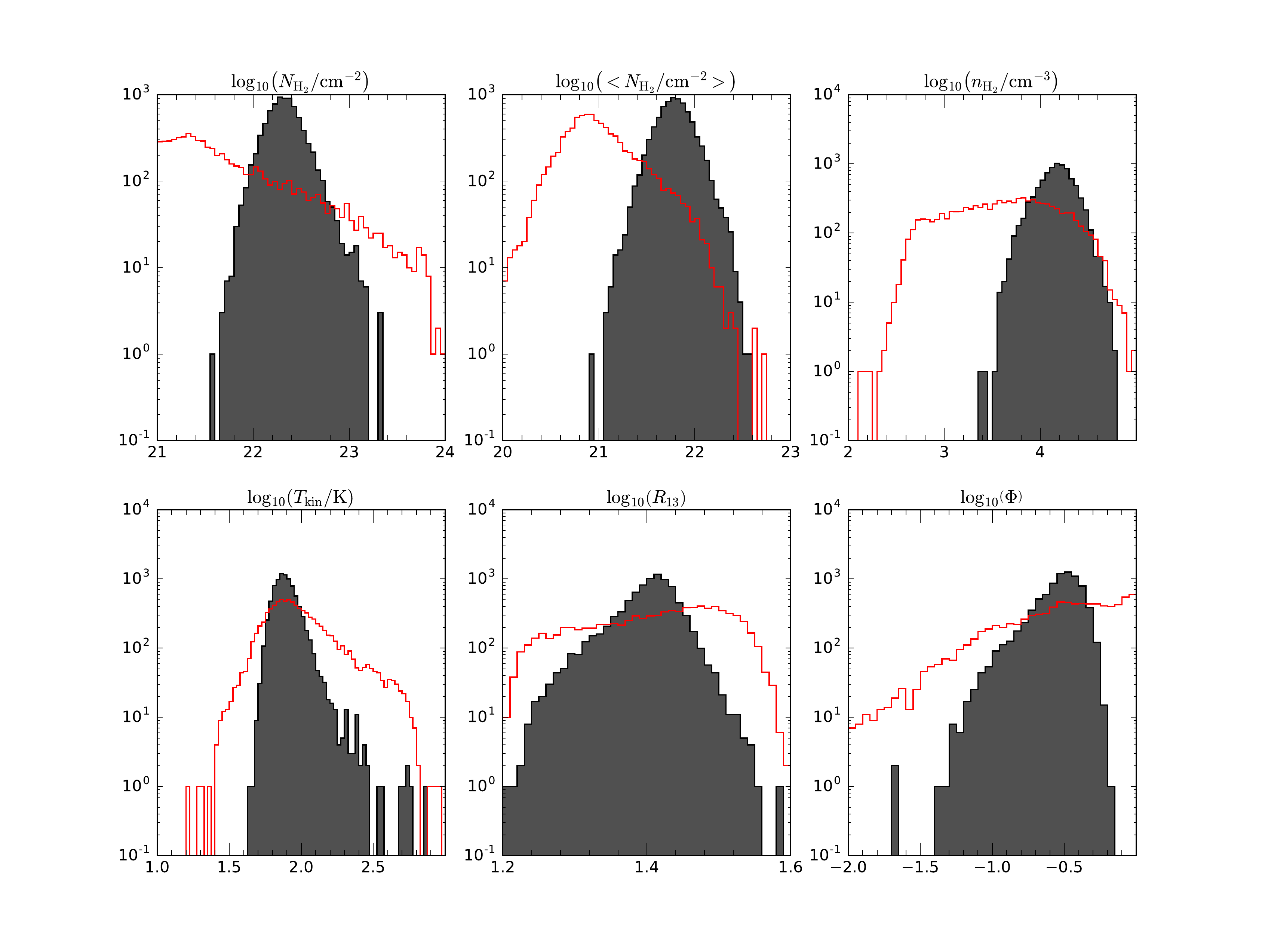}
\caption{Frequency histograms of the median values of the hydrogen column density (\NHH), beam-averaged \NHH (\NHHavg), hydrogen volume density (\nHH), gas kinetic temperature (\Tkin), C/\Ct\ isotopic abundance ratio ($\Rt$), and beam filling factor ($\Phi$) for the HB analysis (gray filled histograms) and the nonhierarchical analysis (red open histograms). } \label{FIG_HIST1}
\end{figure*}

\begin{figure*}
\plotone{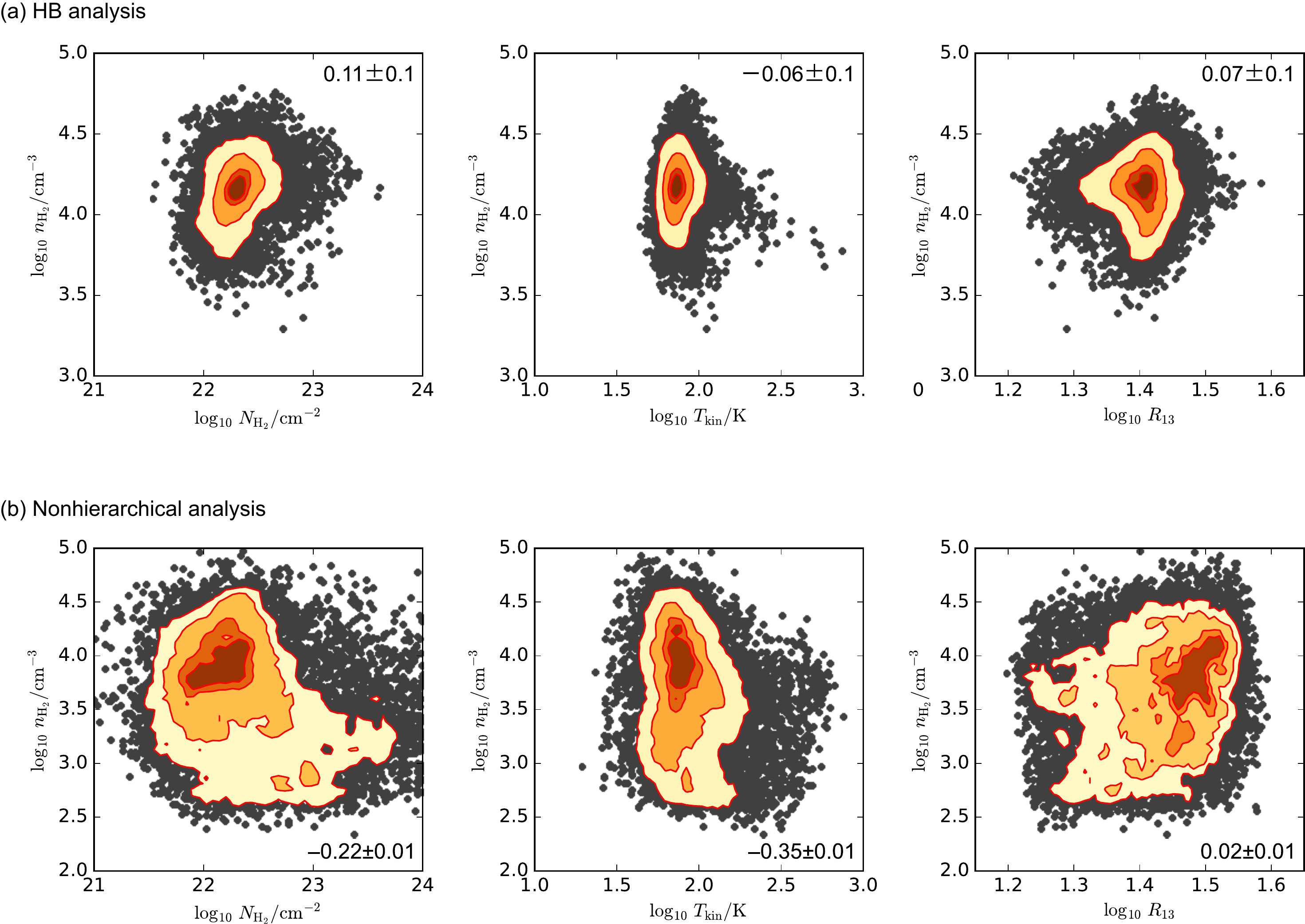}
\caption{(a) Scatter plots of \nHH\ versus \NHH\ per 10-\kmps\ velocity bin, \Tkin, and \Rt\ from left to right for the results of the HB analysis.   The overlaid colored contours show the 2-D frequency histograms drawn at the $10^\mathrm{th}$--$90^\mathrm{th}$,  $25^\mathrm{th}$--$75^\mathrm{th}$,  $40^\mathrm{th}$--$60^\mathrm{th}$, and  $45^\mathrm{th}$--$55^\mathrm{th}$-percentile intervals. 
The medians and the \thup{50}-percentile credible intervals of the correlation coefficients are shown on the panels.  (b) Same as (a) but for the nonhierarchical analysis. }\label{FIG_CORR1}
\end{figure*}

\subsection{Raw MCMC Results}\label{SECTION_PARAMSEARCH}
For the HB analysis, we repeated several MCMC runs starting from different sets of initial parameters.  
Figure \ref{FIG_RAWPOSTERIOR}\ shows the trace plot of the hyperparameters $p_{0,\Tkin}$ and $p_{0,\nHH}$, along with that of the local parameters $\Tkin$ and $\nHH$ at the center position of the 50-\kmps\ cloud, for one MCMC run.  
For both of the hyperparameters and the local parameters,  \nHH\ and \Tkin\ exhibit steady increasing and decreasing trends, respectively, until convergence has been reached at $K \sim 60000$;  these burn-in steps will be discarded when calculating the posterior and marginal posterior distributions. 
We confirmed that the converged values of the hyperparameters and local parameters were consistent among the different MCMC runs.

Figure \ref{FIG_RAWPOSTERIOR2}a shows the trace plot of the local parameters in the \nHH--\Tkin\ plane.
The steps $K$ = 70000--71000 are highlighted in the figure (red lines) to exemplify that the autocorrelation and correlation among the parameters are insignificant so that the parameter space is sampled sufficiently randomly.  
Frequency histograms of the parameters are shown in panels b and c.  
The partial frequency distributions for $K$ = 70000--71000 are consistent with the entire distributions for $K > 60000$;
this confirms that the MCMC run converged after $K$ = 60000 and that the parameter space is fully covered by this MCMC run.

\subsection{Maps}
After discarding the early burn-in steps before reaching convergence, the marginal posterior is calculated for each parameter at each voxel:
\begin{eqnarray}
\PDF{p_{i,j}|\Iobsv} & = & \int \PDF{\pv,\Emv,\myvector{\theta}|\Iobsv} \cdot \left(\prod_{k\ne i,l\ne j}{\mathrm{d}}{\pv_{k,l}}\right)\cdot{\mathrm{d}}\Emv\cdot{\mathrm{d}\myvector{\theta}}
\end{eqnarray}
from the frequency distribution of the parameter of interest in the MCMC-sampled variables.
The 3-D maps of the $i^{\mathrm{th}}$ parameter and its uncertainty are constructed from the median, $\tilde{p_{i,j}}$ and \textcolor{red}{$25^\mathrm{th}$--$75^\mathrm{th}$}-percentile interval, $\Delta\left(p_{i,j}\right)$, of their marginal posteriors.
For each parameter, the voxels with uncertainties greater than 0.2 are discarded from the final map;  as the parameter \pv\ is defined as a base-10 logarithm, the uncertainty of 0.2 translates into to a $\pm 29\%$ relative uncertainty in the linear scale.

Figure \ref{FIG_HIST1} shows the frequency histograms for the PDF-median values of \NHH (per 10-\kmps\ velocity interval), \nHH, \Tkin, \Rt, and $\ff = 1-e^{-\fff}$, which are obtained using the HB and nonhierarchical analyses.  
In addition to them, the histogram of the beam-averaged column density $\langle{\NHH}\rangle \equiv \ff\cdot\NHH$ is also presented.
Through the HB analysis, \Tkin\ was derived to be $10^{1.5}$ K to $10^{2.5}$ K, with the most frequent value being $10^{1.8}$\ K.  
This result is consistent with \textcolor{red}{the non-LTE analysis of the same \pformaldehyde\ data assuming constant \nHH}\ \citep{Ginsburg2016}, as expected from the weak \nHH\ dependence of their intensity ratio. 
The values of \nHH\ are in a range of $10^{3.5}$--$10^{5}$\ \pcc\ with the highest frequency at $10^{4.2}\ \pcc$, which is consistent with the previous non-LTE analysis performed using tracers of a similar \ncrit\ range \citep[$\sim 10^5\ \pcc$;][]{Tsuboi2011,Tsuboi2015}, while being approximately an order of magnitude higher than the analysis by \cite{Nagai2007} using low-$J$ CO transitions. 
The median \Rt\ is 25, which is in a good agreement with the canonical value for the CMZ, i.e., 24, measured through $\mathrm{^{12}C^{18}O}$ and $\mathrm{^{13}C^{18}O}$ observations \citep[][]{Lis1989, Lis1990}.

Figure \ref{FIG_CORR1}a shows the scatter plots and 2-D frequency histograms of \nHH\ versus \NHH\ (per 10-\kmps\ bin), \Tkin, and \Rt.  
The posterior correlation coefficients, $r_{i_1, i_2}  = \frac{\transverse{\pv_{i_1}}\cdot\transverse{\pv_{i_2}}}{|\pv_{i_1}|\cdot|\pv_{i_2}|}$, are shown in the figure.
Through the HB analysis, a small but significant positive correlation was obtained between \nHH\ and \NHH, whose correlation coefficient is $0.11\pm0.01$. 
The correlation coefficient is slightly negative for the \nHH--\Tkin\ plot, but this is mostly due to the outlying points with $\Tkin > 10^{2.5}\ \kelvin$ and $\nHH < 10^4\ \pcc$, which correspond to the Sgr B2 cluster-forming region.   No noticeable correlation is present in the main component.

\subsection{Comparison with the Nonhierarchical Method}\label{SECT_COMP_NONH}
A comparison of the results obtained with the nonhierarchical and HB methods is shown in Figures \ref{FIG_HIST1} and \ref{FIG_CORR1}.  
Figure \ref{FIG_HIST1}b shows that the histograms for the nonhierarchical analysis have broader profiles than those for the HB analysis for all of the six parameters displayed in the figure. 
The difference from the HB analysis is most remarkable for \NHH\ and \ff; their frequency distributions for the nonhierarchical analyses are spread over almost the full variable ranges and truncated at the boundary values imposed by the rectangular priors.
In particular, the histogram of \ff\ has a frequency peak at the upper limit value, i.e., $1$.
\remove{In addition, the high frequency of $\NHH = 10^{24}\ \psc$ is obviously unrealistic, because regions with such extremely high column density should be highly gravitationally unstable.}
This indicates that the nonhierarchical analysis failed to determine the optimal parameter set that maximizes \PDF{\Iobsv|\pv}\ in a physically reasonable range for considerable fractions of the voxels.
Meanwhile, the frequency profile of \Tkin\ does not differ greatly between the HB and nonhierarchical analysis, except that the width of the histogram in the latter is approximately twice that in the former.
This reflects that \Tkin\ is primarily determined by the intensity ratio between the two \formaldehyde\ lines and is relatively insensitive to other conditions.

Figure \ref{FIG_CORR1}b\ shows scatter plots and 2-D histograms of \nHH\ versus \NHH\ (per 10-\kmps\ bin), \Tkin, and \Rt\ for the nonhierarchical analysis.
The \NHH--\nHH\ plot exhibits a weak anticorrelation, whose correlation coefficient is $r_{\NHH, \nHH} = -0.22 \pm 0.01$; as mentioned earlier (\S\ref{SUBSECTION_METHOD}), \textcolor{red}{this anticorrelation is likely to be unphysical}.
The plot also indicates that the data points within the \thn{40}--\thn{60}-percentile interval of the 2-D histogram show a positive correlation, and their parameter range is approximately consistent with the result of the HB analysis. 
The unphysical anticorrelation is created by the tail extending to the high-\NHH and low-\nHH\ region in the 2-D histogram.

On the basis of the above comparison, we conclude that the results of the nonhierarchical analysis are less reliable than those of the HB analysis; 
although the nonhierarchical analysis was successful in measuring \Tkin\ and \nHH\ in high-density regions,  approximately 80\% of the total number of voxels are severely affected by the artificial correlation created through the analysis.
Conversely, the HB analysis successfully suppressed these unphysical behaviors of the physical condition parameters without introducing  complicated physical modeling.

\begin{deluxetable}{crl}
\tabletypesize{\scriptsize}
\tablecaption{Systematic Multiplicative Errors\label{TABLE_ERR}}
\input{table4.etex}
\end{deluxetable}

\subsection{Validity of the HB Analysis}
The posterior distribution of \Emv, which represents the multiplicative systematic errors in the measured line intensities, is determined simultaneously with that for \pv.
The parameter $\Em$ provides a measure of the discrepancy in the calculated line intensities from the observed intensities; thus, they can be used for checking the validity of the HB analysis.
Table \ref{TABLE_ERR}\ summarizes the medians and the \thn{25}--\thn{75}-percentile intervals of the voxel values of $\Em$.
The HB analysis could be judged to be successful as a whole; the errors are less than $\sim 20$\%\ for all of the lines, which would be sufficiently small to be interpreted as calibration errors or  minor deviations from the ideal model.
The relatively large dispersions in \Em\ for HCN \JJ{1}{0}\ and \HCOp\ \JJ{1}{0}\ are presumably because these lines are closely related to the inference of the parameter \Rt\ (see Figure \ref{FIG_RAWLVG}) and hence are strongly affected by the hyperprior $\PDF{S_\Rt}$ that suppresses the variation in \Rt. 
In conclusion, the HB analysis was able to determine the parameters that consistently explain the observed line intensities without introducing large systematic errors in addition to the measured random noise.

We note that the median values of $\Em$\ close to unity do not indicate that the intensity calibration is accurate, but instead they are simply the immediate consequences of the log-normal prior distribution assumed for \Em\ (Equation \ref{EQ4}), in which the mean value is fixed at 1.
Therefore, our analysis is not specifically designed to measure the calibration accuracy;  
$\Emv$ is a nuisance parameter introduced for convenience of calculation, which is to be integrated out in the final step of the analysis.
The prior for \Em\ is given so that $\Em = 1$ if a parameter set that simultaneously reproduces all of the observed intensities exists, even when \Iobs\ systematically includes large calibration errors. 
This explains the reason for the small dispersions in \Em\ for the lines of HNC, \HCCCN, SiO, \NNHp, and CS (the second group lines defined in \S{\ref{SUBSECTION_DATA}});  as these molecules have only one transition in the dataset, their intensities can be always fitted by modifying the abundances regardless of the other parameters. 
The inclusion of new parameters or hyperparameters might enable us to measure the systematic calibration offsets among different observations, but such an analysis would require an extended dataset.

\subsection{Spatial Variations in the Physical Condition Parameters and $^{13}$C Isotopic Abundance} \label{SECT_MAPS}
 
\begin{figure*}
\epsscale{0.9}
\plotone{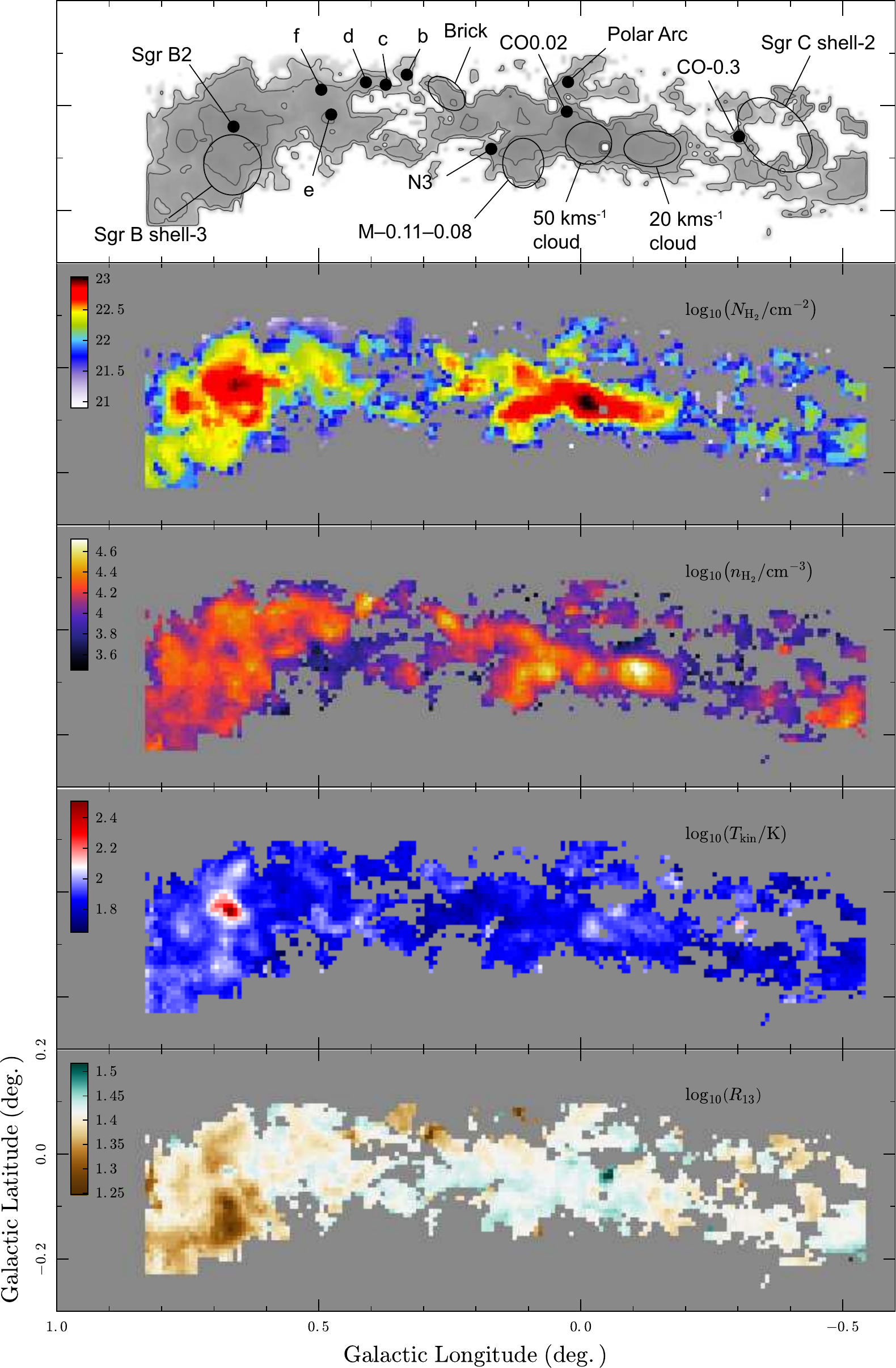}
\caption{Results of the HB analysis for the beam-averaged hydrogen column density (\NHHavg), hydrogen volume density (\nHH), gas kinetic temperature (\Tkin), and [$^{12}\mathrm{C}$]/[\Ct]\ isotopic ratio (\Rt) in the projection onto the $l$--$b$ plane.    The \NHHavg\ map shows the velocity-integrated values, and the others show the average weighted by \NHHavg\ along the \vlsr\ axis.}\label{FIG_RES1}
\end{figure*}

\begin{figure*}
\epsscale{.8}
\plotone{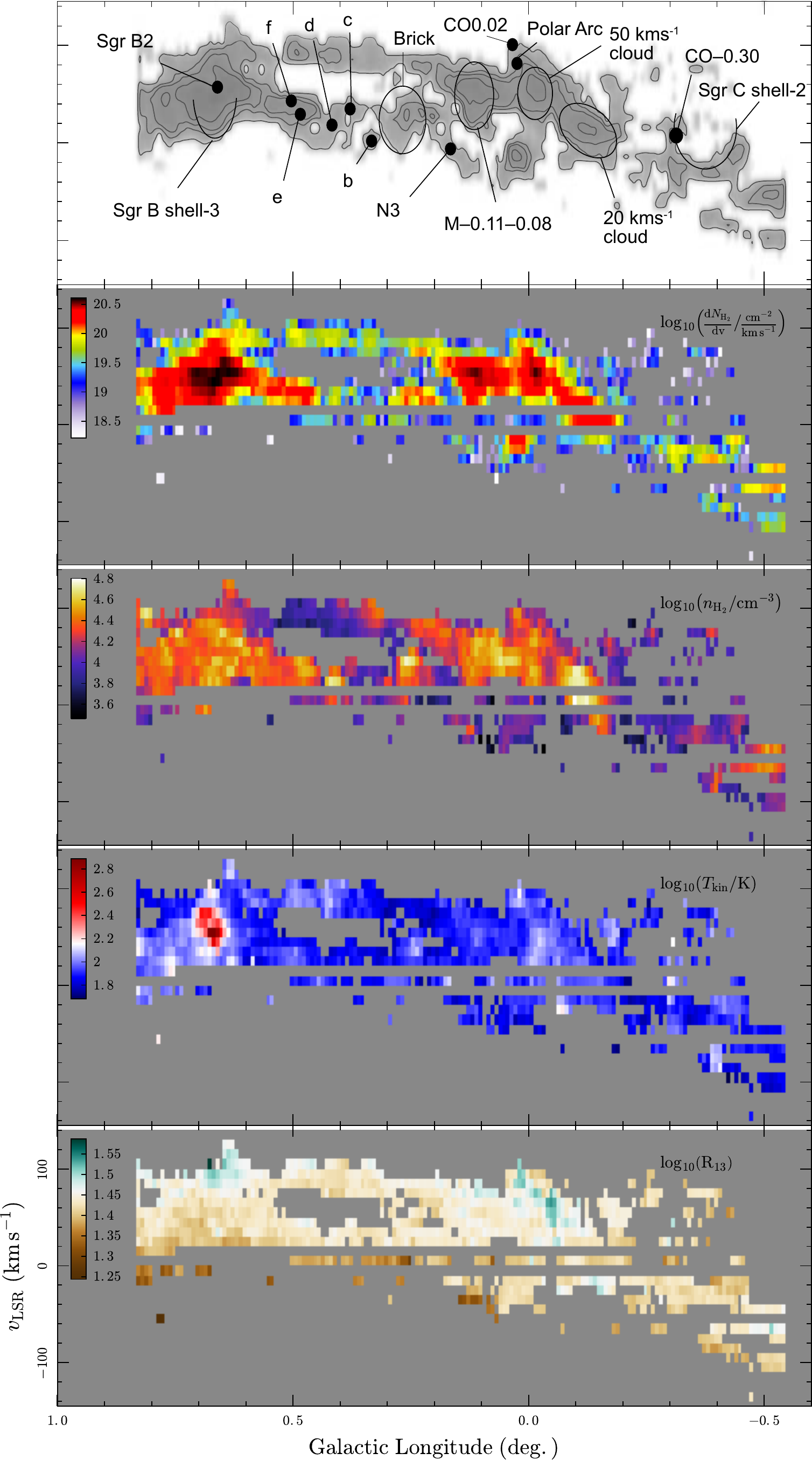}
\caption{Same as Figure \ref{FIG_RES1}\ but in the projection onto the $l$--$\vlsr$ plane.   The \NHHavg\ map shows the averages in the $\gb$ direction, and the others show the highest value along the $\gb$ axis at each $\gl$--$\vlsr$ coordinate. }\label{FIG_RES2}
\end{figure*}

Figure \ref{FIG_RES1} presents the 2-D distributions of \NHHavg, \Tkin, \nHH, and \Rt\ projected onto the $l$--$b$ plane.   The \NHHavg\ map shows the velocity-integrated values, and the others show the averages weighted by \NHHavg\ along the $\vlsr$ axis.
Their  $\gl$--$\vlsr$ diagrams are shown in Figure \ref{FIG_RES2}.   The averages along the $\gb$ axis are shown for \NHHavg, and the highest values along the $\gb$ axis at each $\gl$--$\vlsr$ coordinate are plotted for the other three parameters.

The \nHH\ map shows the physically natural outward decreasing gradients for the majority of the molecular clouds. 
The high \nHH\ regions are evenly distributed across the CMZ, and no systematic difference in \nHH\ is observed between the regions that host active SF regions or young massive clusters (Sgr A, Sgr B, and M$0.11$$-0.08$) and other inactive clouds;  
the peak \nHH\ values in the active SF regions are $10^{4.6}$--$10^{4.8}\ \pcc$, which is not particularly high compared with the peak densities of quiescent dense gas clumps such as the Brick cloud and cloud d in the dust-ridge region.

The spatial variation in \Tkin\ approximately follows the intensity variation in the HCN \JJ{4}{3}\ and \HCCCN\ \JJ{10}{9}\ lines.
It has a clear peak toward the Sgr B2 cluster-forming region, though the value of $10^{2.8}\ \kelvin$\ is unreliable because the intense continuum is not considered in our model. 
Similarly, the increase in \Tkin\ toward the Galactic Center ($\gl\sim -0^\circ.08$) and the center of the 50-\kmps\ cloud (M$-0.02$$-0.07$; $\gl\sim -0^\circ.02$) are considered to be a result of UV heating by the central cluster and \ion{H}{2} region G$-0.02$$-0.07$, respectively.
However, the correlation between the high-temperature regions and SF regions is weak almost everywhere else in the maps.  
The high-temperature spots in the Brick cloud, the N3 clump, the polar arc, CO$0.02$, CO$-0.30$, and the southeastern extension of the Sgr B2 complex lack known corresponding UV-heating sources, which would support the idea of mechanical heating by dissipating shocks \citep{Ao2013,Ginsburg2016}.

A few of the temperature peaks without UV sources are associated with broad-velocity features or expanding shells.  
Two clumps, i.e., CO$0.02$ and CO$-0.30$, are high-velocity(-width) compact clouds with atypical broad velocity widths of $\gtrsim 50\ \kmps$ at full width zero intensity \citep{Oka2012,Tanaka2015}.
The temperatures of both clumps are significantly higher than that of their surroundings. 
\cite{Tanaka2015}\ detected spots of high-excitation thermal methanol lines with $\Eu/\kB \gtrsim 100\ \kelvin$ inside CO$-0.30$ through interferometric observations, which is consistent with the high temperature measured in our analysis.
Two molecular shells, i.e., Sgr C shell 2 \citep{Tanaka2014} and  Sgr B shell 3 \citep{Tsuboi2015} exhibit \Tkin\ higher than 100 K in our map.

The variation in \Rt\ is small compared with the variations in \nHH\ or \Tkin; it is almost constant across the region with a slight increasing gradient toward the Galactic Center.
This is due to the hyperprior \PDF{S_{\Rt}}\ that explicitly imposes an upper limit on the dispersion of \Rt.
The 180-pc ring region, for which high values of \Rt\ are reported by \cite{Riquelme2013}, is not included in our analysis.
There is a subtle local spatial correlation between \Rt\ and \NHH\ in the Sgr A and Sgr B2 complexes, although the global correlation coefficient is slightly negative, $-0.1$.
If the local positive correlation between \NHH\ and \Rt\ is real, it is consistent with selective photodissociation, which predicts that $\COn$ is more stable against photodissociation than \COt\ in high-column-density regions.
However, the significance of the correlation would require further careful investigation; it could be an artifact introduced by an insufficient correction for the opacity effect for the main \HCOp\ isotopologue line.

The \gl--\vlsr\ plot for \Rt\ presented in Figure \ref{FIG_RES2}\ indicates the effect of contamination by the spiral arms on the analysis.
The velocity channels at $\vlsr = -30\ \kmps, -10\ \kmps$, and $10\ \kmps$ have larger fractions of low \Rt\ values than the other channels.
This can be explained by the self-absorption of HCN \JJ{1}{0}\ and \HCOp\ \JJ{1}{0}\ by the foreground low-density material;   as their \Ct\ isotopologue lines are less affected by self-absorption, \Rt\ is underestimated when contamination by the spiral arms is significant.
Although the obvious absorption dips at the velocities of the spiral arms have been removed in the initial filtering (\S\ref{SUBSECTION_DATA}), these results indicate that the velocity channels contiguous to the arms are also affected by self-absorption.
Therefore, we exclude these velocity channels from the analysis in the following sections.

\subsection{Molecular Abundances} \label{SECT_MOLMAPS}
\begin{figure*}
\epsscale{1}
\plotone{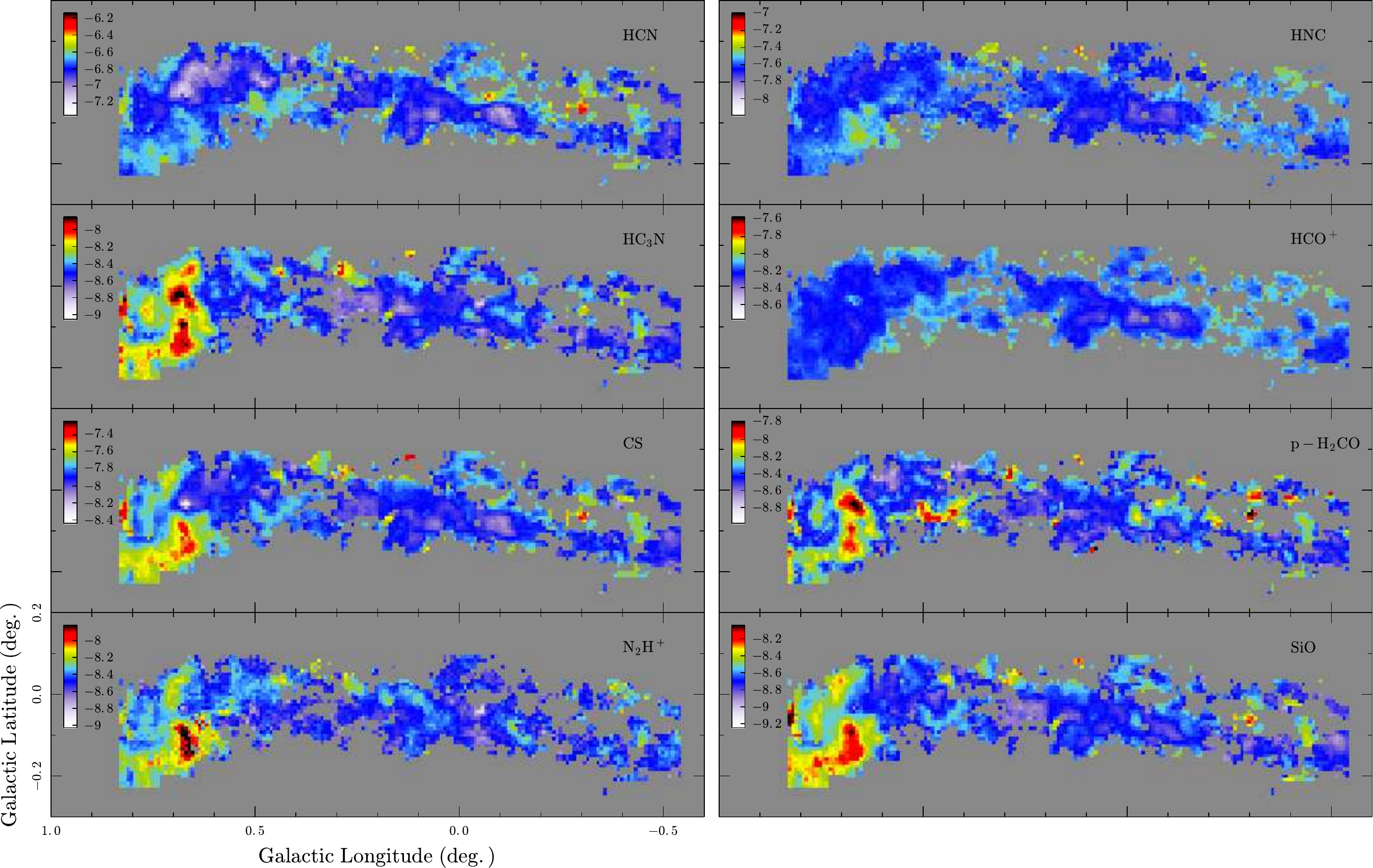}
\caption{Maps of the molecular fractional abundances calculated using the HB analysis.  The abundances are averaged over the full velocity range and weighted with \NHHavg\ for each line-of-sight. }\label{FIG_molx}
\end{figure*}

Figure \ref{FIG_molx} shows the molecular abundance maps projected onto the $\gl$--$\gb$ plane, where projection is performed by taking \NHHavg-weighted averages of the parameters. 
A trend common to several molecular species can be identified; that is, the abundances of SiO, \HCCCN, \formaldehyde, CS, and \NNHp are enhanced over a wide area of the Sgr B2 complex.  
This enhancement is not limited to the Sgr B2 cluster-forming region but also extends to the southeast extension, where active star formation is absent.
The other three species, i.e., HCN, HNC, and \HCOp, do not an exhibit enhancement in the Sgr B2 region; in particular, the latter two species have almost flat spatial distributions over the entire map.

\begin{figure}
\epsscale{.7}
\plotone{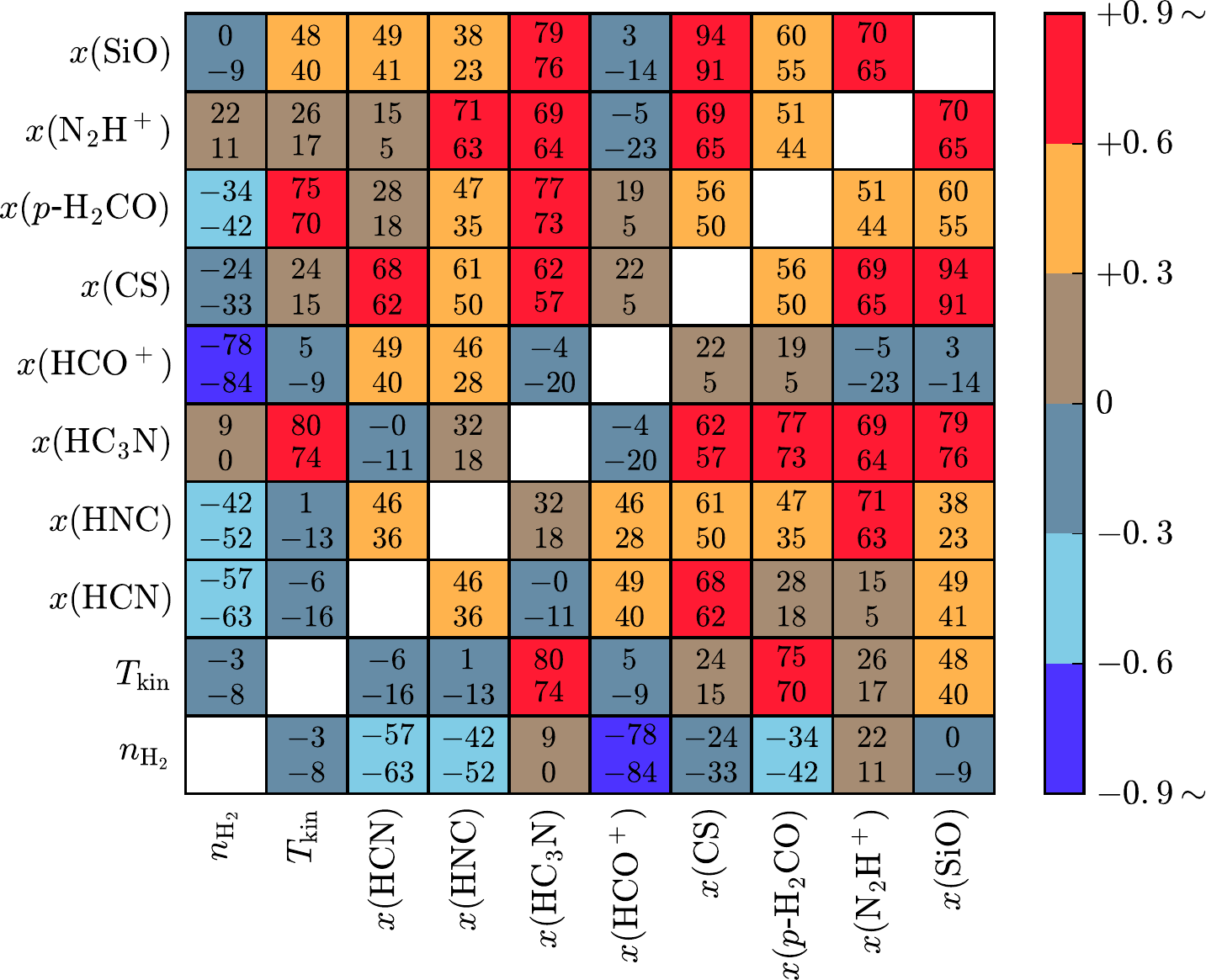}
\caption{Posterior correlation coefficient matrix of the logarithms of the physical condition parameters and molecular abundances.   The numbers in each cell show the \thn{25}--\thn{75}-percentile intervals multiplied by 100.  The cells are colored according to the medians.}\label{FIG_CORRCOEFF}
\end{figure}

Figure \ref{FIG_CORRCOEFF} shows the posterior correlation coefficient matrix for the logarithms of \Tkin, \nHH, and the molecular abundances. 
The species that have high abundances in the Sgr B2 complex have large positive correlation coefficients with each other.   
In addition, \HCCCN\ and \formaldehyde\ exhibit strong positive correlations with \Tkin.
The correlation coefficients with \nHH\ are negative or zero for all molecules except for \NNHp, which may imply that our analysis is affected by the degeneracy between \nHH\ and $\xmolp{\cdot}$ despite the use of the HB method.
Nevertheless, it is not likely that the negative correlation coefficients between the molecular abundances and \nHH\ are purely artificial, as the strength of the (anti-)correlation is not correlated with  \ncrit\ of the transitions used in the analysis.
For example, \HCOp\ and HCN have the largest and the second largest negative correlations with \nHH, respectively, but these species have the highest and lowest critical densities in our dataset.
The critical density of \HCOp \JJ{1}{0}\ is similar to that of \NNHp\ \JJ{1}{0}, which is the only species with a positive correlation with \nHH.

\subsection{{\it X}-factor and \Xdvdr}
\begin{figure*}
\epsscale{1.}
\plotone{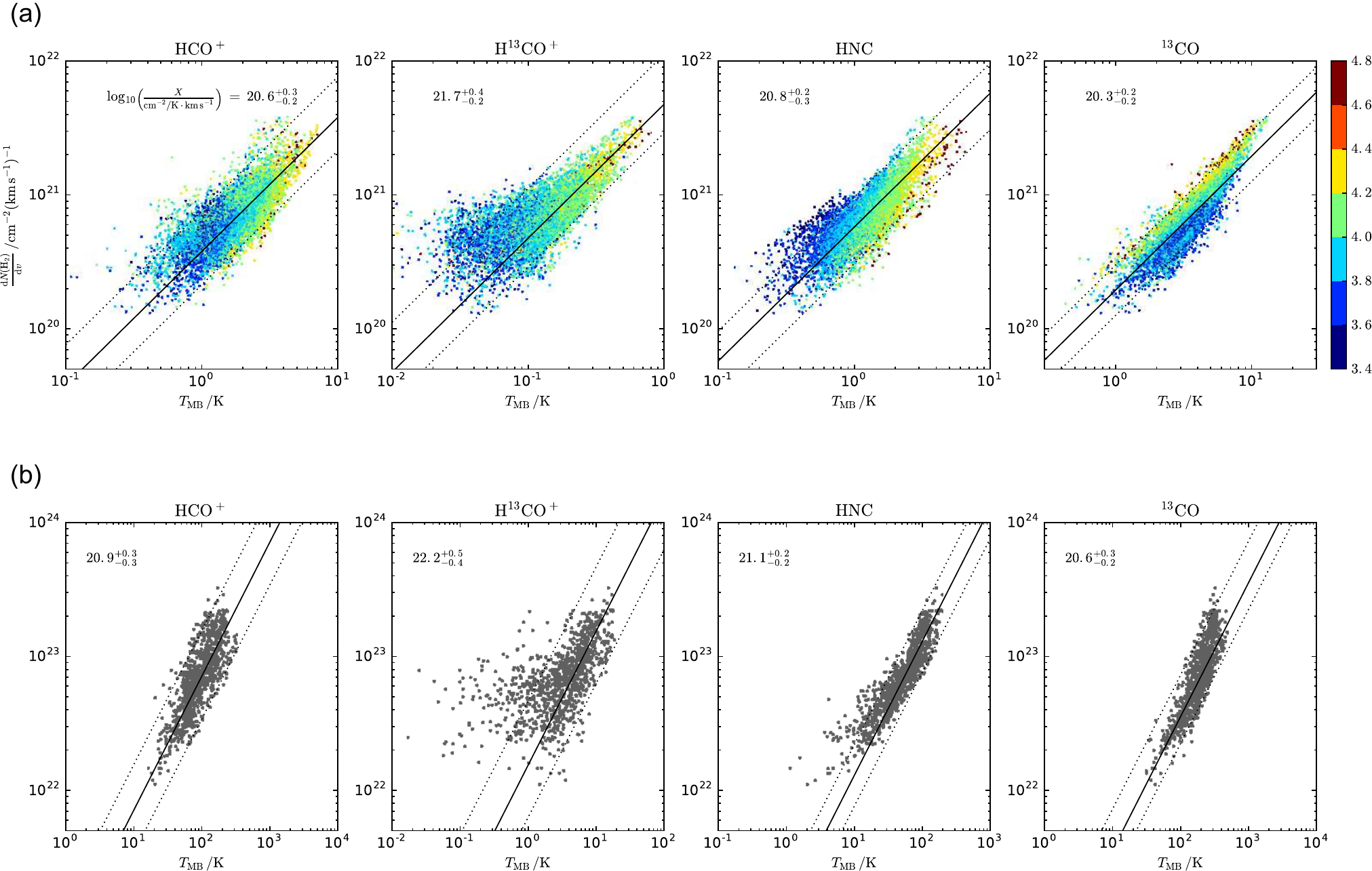}
\caption{(a) Voxel-by-voxel scatter plots of the column densities (per unit velocity), which are calculated using the HB analysis, against the brightness temperatures of \HCOp\ \JJ{1}{0}, \HCOpt\ \JJ{1}{0}, HNC\ \JJ{1}{0}, and \COt\ \JJ{2}{1}, with colors according to the density at each voxel. (b) Same as (a) but with velocity-integrated column densities estimated from the 500-\micron\ dust flux and the line integrated intensities.}\label{FIG_XFACTOR}
\end{figure*}

It is useful to obtain good gas mass tracer lines that are insensitive to the physical and chemical environments and to define their luminosity-to-mass conversion factors ({\it X}-factors).
In addition to the evident good tracer, i.e., \COt, we calculate the conversion factors for \HCOp, \HCOpt, and HNC, whose abundances show small spatial variations and weak dependencies on the physical conditions.
Figure \ref{FIG_XFACTOR}a\ shows the voxel-by-voxel scatter plots of the observed \HCOp, \HCOpt, and HNC intensities against the values of \NHHavg\ calculated using the HB analysis, with colors according to the ranges of \nHH.
The overlaid lines represent the medians and \thn{2.5}--\thn{97.5}-percentile intervals of the {\it X}-factors weighted by the intensities. 
The {\it X}-factors are consistent across the CMZ within an order of magnitude for all four analyzed lines.
The systematic variation over different \nHH\ ranges is approximately a factor of 4 from the lowest-density range ($\sim 10^3\ \pcc$) to the highest-density range ($\sim 10^{4.4}\ \pcc$) for \HCOpt\ and HNC. \HCOp\ is less dependent on \nHH\ because the \nHH\ dependencies of abundance and emissivity cancel each other. 

\begin{figure*}
\epsscale{0.6}
\plotone{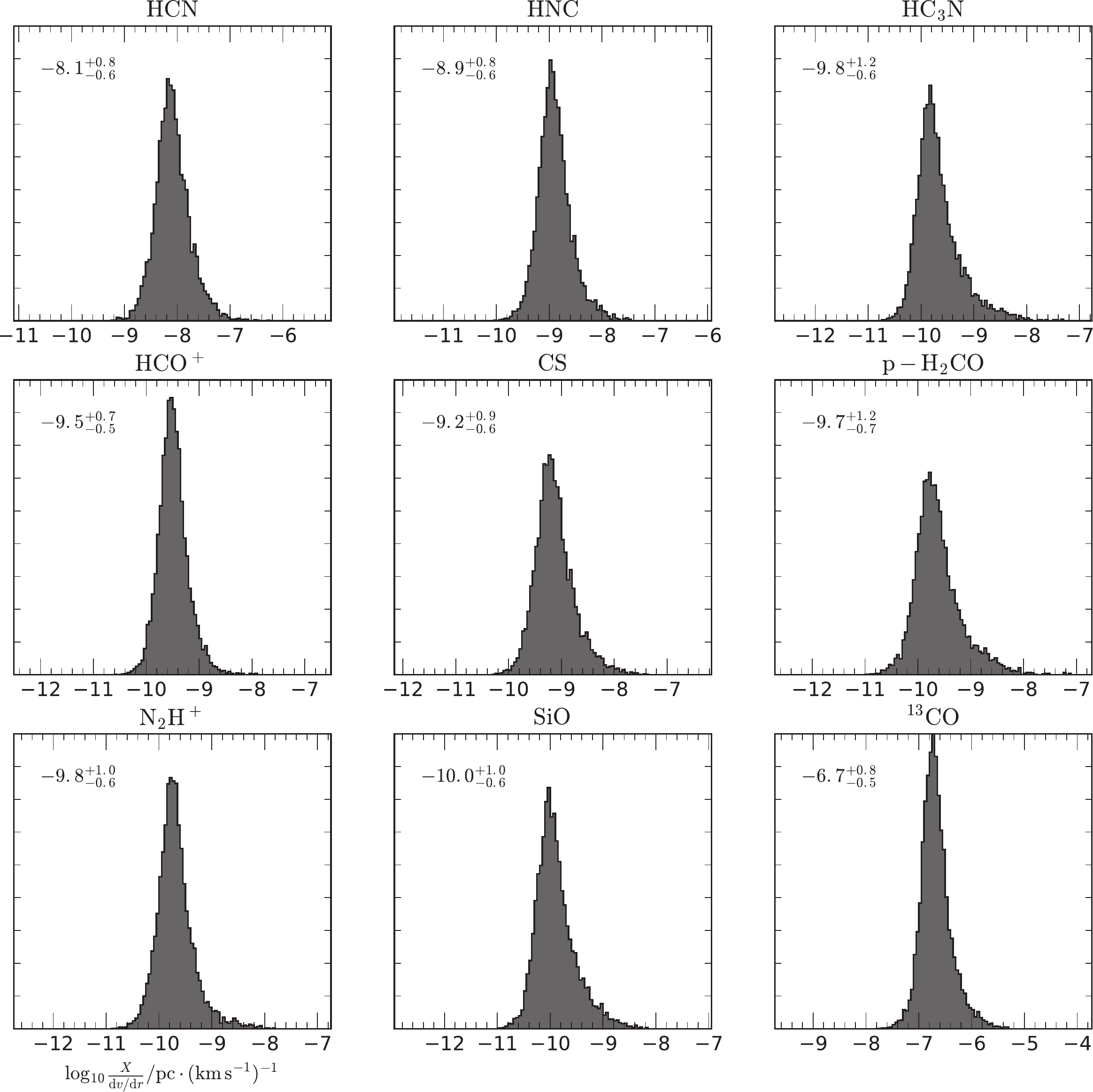}
\caption{Frequency histograms of the $\Xdvdr$ values for various molecules.  The frequencies are weighted by \NHHavg.  The median and \thn{2.5}--\thn{97.5}-percentile interval are displayed on each panel.  }\label{FIG_XDVDR}
\end{figure*}
 
We repeat the same analysis with the dust-based column densities instead of using the values of \NHHavg\ from the HB analysis results for the purpose of checking consistency.
These dust-based {\it X}-factors are equivalent to the intensity-to-dense-gas-mass conversion factors derived by \cite{Mills2017}. 
We calculate the column density from the {\it Herschel} 500-$\micron$ data \citep{Molinari2011} smoothed with a $60''$\ Gaussian kernel assuming a constant dust temperature of 20 K.
For the dust emissivity per unit gas mass, the formula in \cite{Pierce-Price2000}, $\kappa_{\nu} = 2.5\times10^{-4}\left(\frac{\lambda}{1\ \mathrm{mm}}\right)^{-\beta}\ \mathrm{cm}^{2}\,\mathrm{g}^{-1}$, is adopted.  
The dust emissivity index $\beta$ is assumed to be 2. 
Figure \ref{FIG_XFACTOR}b shows the scatter plots of the dust-based column densities against the velocity-integrated line intensities. 
Linearity with the dust-based column densities holds for the \HCOp, HNC, and \COt\ data, while \HCOpt\ appears more susceptible to instrumental noise owing to its weak intensities.
The dust-based scaling factors are systematically higher by a factor of $10^{0.3}$--$10^{0.5}$\ than those determined using the HB analysis.  
This difference is reasonable, considering the crudeness of the assumption for the dust temperature and  emissivity and the omission of the contribution from low-density gas in the HB analysis (see Subsection \ref{SECT_MULTCOMP}).

We evaluate the \Xdvdr\ parameter, i.e., the scaling factor that relates \nHH\ to the column densities per unit velocity. 
Figure \ref{FIG_XDVDR}\ shows the frequency histograms of $\Xdvdr\equiv \xmolp{\cdot}\dNHdv / \nHH$\ weighted by \NHHavg, which are calculated from the results of the HB analysis.
The \thup{2.5}--\thup{97.5}-percentile intervals of the parameter are less than two orders of magnitude for all analyzed molecules. 
HNC, \HCOp, and HCN exhibit the smallest dispersions in the values of $\Xdvdr$ among the analyzed molecules (except for \COt), which indicates their efficacy as physical condition probes.

%% file: table4.etex


\tablecolumns{3}
\small
\tablehead{\\ \colhead{molecule} & \colhead{transition} & \Em  }
\tabletypesize{\scriptsize}
\tablewidth{0pt}
\startdata
HCN    & \JJ{1}{0} & \textcolor{black}{ $0.97^{+0.11}_{-0.11}$}\phn \\
       & \JJ{4}{3} & \textcolor{black}{ $1.00^{+0.06}_{-0.07}$}\phn \\
\HCNt  & \JJ{1}{0} & \textcolor{black}{ $1.01^{+0.04}_{-0.04}$}\phn \\
\HCOp  & \JJ{1}{0} & \textcolor{black}{ $1.13^{+0.22}_{-0.23}$}\phn \\
\HCOpt & \JJ{1}{0} & \textcolor{black}{ $0.98^{+0.05}_{-0.05}$}\phn \\
\pformaldehyde & $3_{03}\mbox{--}2_{02}$ & \textcolor{black}{ $1.00^{+0.06}_{-0.07}$}\phn \\
               & $3_{21}\mbox{--}2_{20}$ & \textcolor{black}{ $1.00^{+0.02}_{-0.02}$}\phn \\
\COt   & \JJ{2}{1} & \textcolor{black}{ $1.00^{+0.03}_{-0.03}$}\phn \\
HNC    & \JJ{1}{0} & \textcolor{black}{ $1.00^{+0.03}_{-0.04}$}\phn \\
\HCCCN & \JJ{10}{9} & \textcolor{black}{ $1.00^{+0.02}_{-0.02}$}\phn \\
SiO    & \JJ{2}{1}  & \textcolor{black}{ $1.00^{+0.02}_{-0.02}$}\phn \\
\NNHp  & \JJ{1}{0}  & \textcolor{black}{ $1.00^{+0.02}_{-0.03}$}\phn \\
CS     & \JJ{1}{0}  & \textcolor{black}{ $1.00^{+0.03}_{-0.02}$}\phn
\enddata

%% file: discussion.etex
\subsection{Limitations of the HB Analysis}\label{SECTION_CAVEATS}
\begin{figure}
\ifdraft
\epsscale{0.7}
\else
\epsscale{0.7}
\fi
\plotone{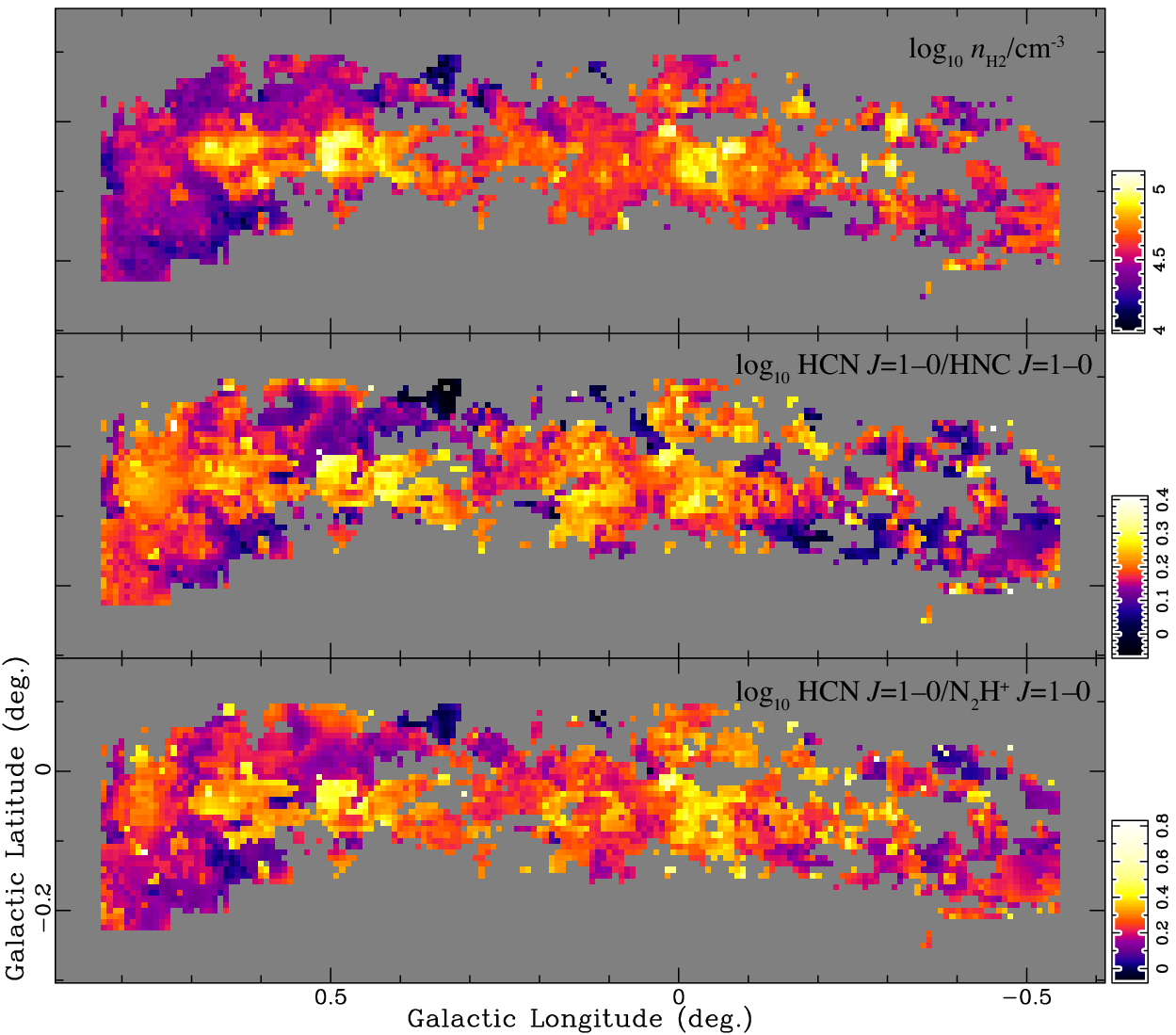}
\caption{(top panel) Volume density (\nHH) map obtained with the HB analysis where the weights of the input data $w_j$ are assumed to be 1 for all lines.  The highest values along the velocity axis are plotted on the $\gl$--$\gb$ plane.  Regions with irregularly high \nHH\ are present along $\gl \sim 0.5^\circ, -0.2^\circ$, and $-0.3^\circ$.   The correlation coefficient between \nHH\ and \NHH\ is $-0.1$.  (bottom and middle panels) The HCN \JJ{1}{0}/HNC \JJ{1}{0}\ and HCN \JJ{1}{0}/\NNHp\ \JJ{1}{0}\ maps, respectively.   Their similarity to the \nHH\ map indicates that \nHH\ is determined primarily by these intensity ratios in this analysis. } \label{FIG_WRONGANALY}
\end{figure}

The HB analysis successfully suppressed the artificial features that appeared in the results of the ML analysis and provided a more feasible 3-D distribution of the physical/chemical parameters. 
In this section, we summarize the limitations of our analysis, which potentially affect the validity of the results.
\paragraph{Prior and Hyperprior Functions}
First, the HB analysis is clearly dependent on the arbitrarily selected prior functions for the parameters and the systematic errors.  
We selected neutral priors, i.e., multivariate log-Student and log-normal functions, but the assumption of unimodal prior PDFs does not have observational or theoretical grounds except for \nHH\ and \NHH.  Considering the chemical and physical diversity among CMZ clouds, multimodal PDFs might be more realistic, even though they could materially complicate the model. 
In addition, the observation field selected for the analysis could alter the results; if we extract images of a certain region from the full CMZ map, the statistical properties of this subset could be different from those of the full dataset, which may lead to different values for the prior probability.   
\paragraph{Different Source Sizes and Velocity Widths for Different Lines}
Although our analysis assumes the same source sizes (or beam filling factors) and velocity widths for all lines, this assumption is generally not true;  
optically thin lines that trace high-temperature/density regions (such as the \HCNt\ and \formaldehyde\ lines) are likely to have compact spatial distributions and narrow spectral profiles compared with other lines and hence are likely to occupy a smaller volume in the position--position--velocity (PPV) space.
In other words, the grid size of our analysis ($2.4\ \pc\times 2.4\ \pc\times10\ \kmps$) may not be sufficiently small to allow the one-zone approximation within PPV volume elements.
This breakdown in the one-zone approximation can significantly affect the physical condition measurements.
An example of this situation might be observed for cloud CO$-0.30$, in which interferometric observations identified a few warm methanol spots inside a dense clump of HCN \JJ{3}{2}\  of approximately 1 pc in diameter \citep{Tanaka2015}.
If the \formaldehyde\ lines originate from the warm methanol spots, \Tkin\ probed by the \formaldehyde\ lines overestimates the temperature from the representative value for the volume element.
The overestimation of \Tkin\ leads to underestimation of \nHH, as shown in Figure \ref{LVG_SINGLE}.
Indeed, our analysis gives an unlikely low \nHH\ of $\sim 10^{3.5}\ \pcc$ at the voxels of CO$-0.30$, which does not agree with the presence of the intense \NNHp\ \JJ{3}{2}\ emission \citep{Tanaka2015}, whose \ncrit\ is $> 10^6\ \pcc$.   
This underestimation of \nHH\ might not be an exceptional case, considering the numerous class-I methanol sources found in the CMZ \citep{Yusef-Zadeh2013b}.
\paragraph{\HCNt\ Depletion}
Another important factor that is neglected in our ideal model is the isotopic fractionation in C-bearing species;  
even though we assume a common value of \Rt\ for HCN and \HCOp, theoretical works show that the [HCN]/[\HCNt] abundance ratio is significantly higher than the [C]/[\Ct] isotopic ratio in the chemical steady state, while the [\HCOpt]/[\HCOp] ratio is only slightly lower than it \citep{Roueff2015}.  
If isotopic fractionation is considered, it should affect the estimates of \Tkin\ and \nHH, on which the HCN \JJ{4}{3}/\HCNt\ \JJ{1}{0}\ and HCN \JJ{1}{0}/\HCNt\ \JJ{1}{0}\ ratios are dependent.
By performing tests with the manually given \HCNt\ depletion factor, we find that a depletion factor of $5$ for \HCNt\ causes a decrease of $\sim 0.2$ and an $\sim 0.1$ increase in $\log_{10}\ \nHH$ and $\log_{10} T$, respectively, in the HB analysis results.
\paragraph{Weights on the Observational Data} In the likelihood function (Equations \ref{EQ1} and \ref{EQ2}), we assumed different statistical weights for the molecules having multiple transitions in the dataset and for those that do not (i.e., the first and second groups in \S\ref{SUBSECTION_DATA}, respectively), in addition to the weights given by the RMS noise. 
These additional weights were introduced for a purely computational purpose;   without them, the latter group lines introduce a large amount of ``noise'' into the measurement of \nHH\ since they effectively decrease the weight of the HCN \JJ{4}{3}\ data, which are the most reliable probe of \nHH\ in our dataset. 
As an example, we show the \nHH\ map obtained with the HB analysis assuming $w_j = 1$ for all lines in Figure \ref{FIG_WRONGANALY}.   In this map, \nHH\ exhibits an irregular distribution that is not correlated with the \NHH\ distribution, 
where \nHH\ is determined primarily by the intensity ratios of HCN \JJ{1}{0}\ to HNC \JJ{1}{0}\ and \NNHp\ \JJ{1}{0}.
In the main analysis, we assumed that $w_j = 0.1$ for the lines that are not necessary for measuring the physical conditions; thus, we were able to suppress this unphysical behavior.
However, this also means that our present HB framework failed to determine the appropriate values for the systematic errors $\Emv$ that give a physically realistic solution without manually modifying the weights of the input data.
This can introduce arbitrariness in the analysis since we do not have a precise method to determine the appropriate weights prior to obtaining the results of the analysis.
\paragraph{Multiphase Gas}
Finally, a significant bias could be introduced by the high critical densities of the line used in our analysis if multiple density/temperature gas exists in the same beam.  This issue is discussed in more detail in Subsection \ref{SECT_MULTCOMP}.

\bigskip
Among these issues, the arbitrariness of the choice of the prior function is the largest problem with our HB analysis. 
However, we consider that the prior used in the HB analysis is at least more plausible than the flat prior used in the nonhierarchical analysis (and hence than the standard ML analysis) since \remove{the extremely high column density and} the anticorrelation between \NHH\ and \nHH\ in the latter analysis is unreasonable from a physical point of view.
This indicates that the errors due to the deficiencies of the simple one-zone non-LTE calculation have to be considered in the parameter inference; however, the standard method is not able to handle this problem, as discussed in Subsection \ref{SUBSECTION_METHOD}. 
An HB analysis with a log-normal or student prior is a method commonly used to solve such problems where the available information is limited without introducing too large biases.

\subsection{Multiphase Gas in the CMZ}\label{SECT_MULTCOMP}
\begin{figure*}
\ifdraft
\epsscale{0.7}
\else
\epsscale{0.7}
\fi
\plotone{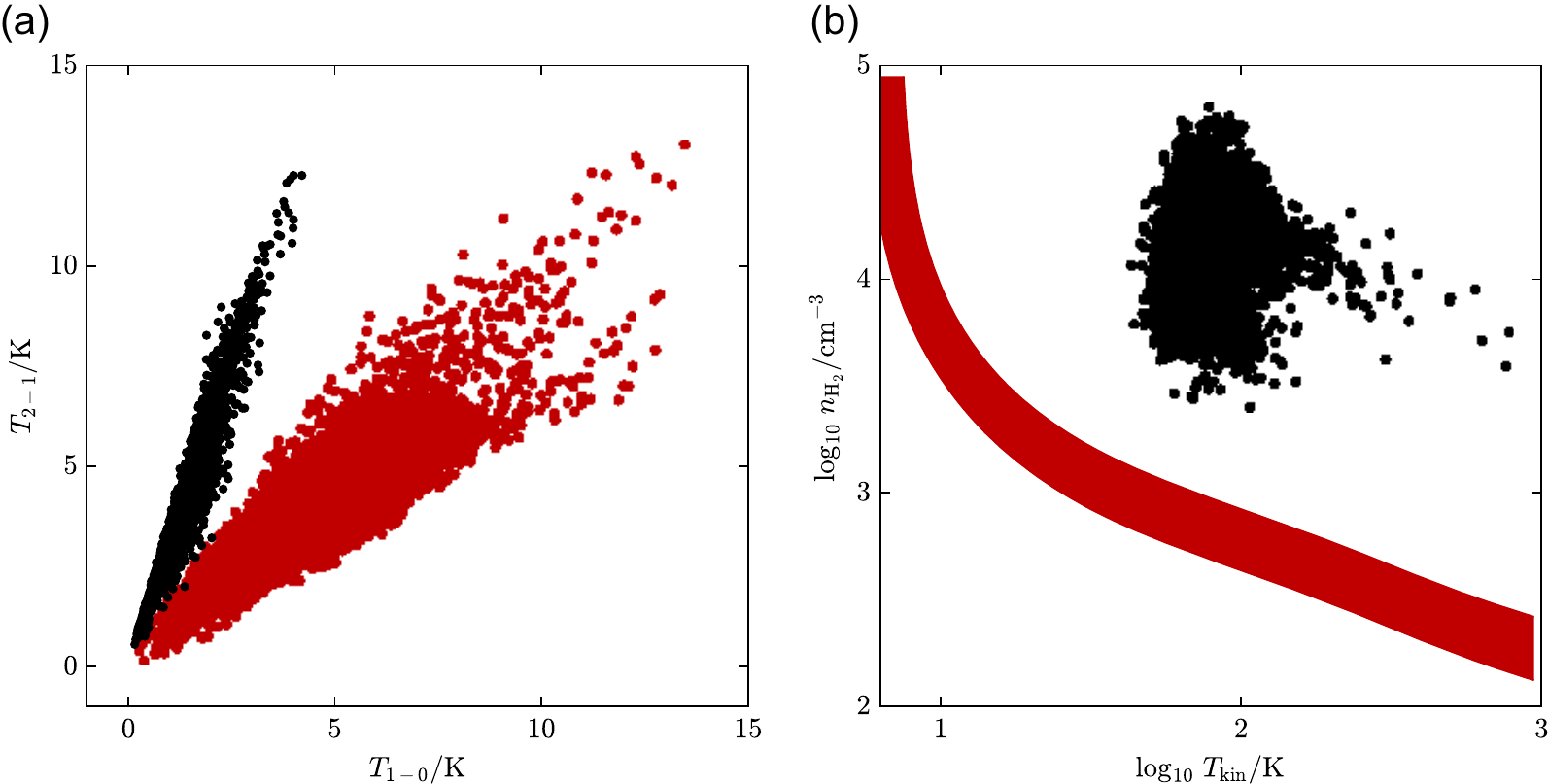}
\caption{(a) \textcolor{red}{Voxel-by-voxel \COt\ \JJ{1}{0}\ -- \JJ{2}{1}\ scatter plots from observations \citep[red;][]{Oka1999,Ginsburg2016} and for the values calculated using the parameters obtained through the HB analysis (black).  (b) \Tkin -- \nHH\ scatter plot for the HB results (black points).  The hatched region represents the parameter range corresponding to the \thn{25}--\thn{75}-percentile interval of the observed \COt\ \JJ{2}{1}/\JJ{1}{0}\ intensity ratio.}  } \label{FIG_MULTI}
\end{figure*}

\begin{deluxetable*}{lrccc}
\tabletypesize{\scriptsize}
\tablecaption{Cold-to-Warm Ratios for the Line Intensities and the Column Density\label{TABLE5}}
\input{table5_r.etex}
\end{deluxetable*}

Our analysis assumes a single temperature and density component for individual voxels, and the deviation from the one-zone model is regarded as systematic errors that are included in the \Em\ parameters. 
However, the presence of a multitemperature gas in the CMZ has been shown in several previous studies on the basis of the direct detection of the multicomponent ammonia rotation temperature \citep{Huettemeister1993,Arai2016} and indirectly from the inconsistency among temperatures measured using different probes \citep[e.g.,][]{Nagayama2007,Martin2004,Nagai2007,Ott2014,Ginsburg2016}.
The recent observations of multi-$J$ ammonia inversion lines by \cite{Arai2016}\ show that the ammonia rotation diagrams at a few representative voxels in the CMZ maps have a significant excess at the (1,1) level from the curve with a rotation temperature of $\sim$ 100--200 K, which otherwise fits the transitions up to $J$=6.  They estimate the typical temperature for the cold component to be $20\ \kelvin$ and the column density to be comparable to that of the warm component.  

The results of the HB analysis also indicate the presence of multicomponent gas;
they are inconsistent with the observations of the \COt\ \JJ{1}{0}\ line, which is not included in our dataset.
Figure \ref{FIG_MULTI}a compares the \COt\ \JJ{1}{0}\ intensity calculated using the parameters derived through the HB analysis with the data taken using the NRO 45-m telescope \citep{Oka1999}. 
The calculated luminosity accounts for only $\sim 30\%$ of the total observed luminosity.
Figure \ref{FIG_MULTI}b shows the \Tkin--\nHH\ parameter range calculated from the observed \COt\ \JJ{2}{1}/\JJ{1}{0}\ intensity ratio ($0.88\pm0.18$) on the assumption of optically thin conditions.  
These values of \Tkin\ and \nHH\ differ from those obtained through the HB analysis by approximately an order of magnitude when one of these parameters is fixed. 
If the \COt\ \JJ{1}{0}\ and \JJ{2}{1}\ emissions originate from the same cold gas with a temperature of 20 K that causes the (1,1) excess in the ammonia emissions \citep{Arai2016},  the best-fit value of $\nHH$ is obtained to be $10^{3.2}\ \pcc$.   
This is qualitatively consistent with the simultaneous \Tkin\ and \nHH\ measurements performed by \cite{Nagai2007}, who derived a combination of a low $\Tkin$  of $\sim 40\ \kelvin$ and a low $\nHH $ of $\sim 10^{3.5}\ \pcc$ from the \JJ{1}{0}\ and \JJ{3}{2}\ transitions of CO and \COt.
The actual density of the cold gas  could be lower than $10^{3.2}\ \pcc$, considering the contribution from the warm gas to the \COt\ \JJ{2}{1}\ luminosity. 
Thus, this cold gas component is likely to have a lower temperature and density than the values derived using the HB analysis.

We obtain an approximate estimate of the effect of the cold gas on our analysis using the above estimates.
Table \ref{TABLE5}\ summarizes the cold-to-warm ratios of the line intensities, defined as the ratios of the intensities calculated for the cold component to those for the warm component, i.e., the gas visible in the tracers used in our analysis.  
The intensities of the cold component are calculated on the bases of three assumptions for \Tkin\ and \nHH\ ($T_{\rm cold}$, and $n_{\rm cold}$), where the molecular fractional abundances are assumed to be the same as those obtained from the HB analysis.
The intensities of the warm component are calculated directly from the median parameter values obtained using the HB analysis.
The column density of the cold component is adjusted so that the cold-to-warm ratio of \COt\ \JJ{1}{0}\ is 2 for each parameter set of $T_{\rm cold}$ and $n_{\rm cold}$.   
The calculated cold-to-warm ratios of \NHH\ are also presented in Table \ref{TABLE5}.
The intensity ratios are $\lesssim 0.1$ for most of the lines when $T_{\rm cold} > 20\ \kelvin$, except for \COt\ \JJ{2}{1}.
The ratios may even decrease if the positive correlations between \Tkin\ and $\xmolp{\cdot}$ determined using the HB analysis hold for the two temperature components.
Therefore, it could be tentatively concluded that the effect of the multitemperature gas on the HB analysis is negligible in our HB analysis, and our results are safely biased to the warm-temperature component.

The multiphase gas could be more accurately modeled by using a multimodal prior instead of the unimodal prior used in this analysis.  
However, multiphase gas modeling would require correspondingly large datasets for the results to be significant;  at least three transitions are necessary for each molecule, and their upper state energies and critical densities should cover a sufficiently wide range of physical conditions, i.e., typically $\Tkin = 10$--100\ K and $\nHH = 10^{3-6}\ \pcc$.    Such large datasets are not available yet but may be provided by future wide-field and broadband observations using ALMA and other instruments.

\subsection{Principal Component Analysis}
We apply principal component analysis (PCA) to the 3-D data of \Tkin, \nH, and the molecular fractional abundances with a minor modification; we treat \Tkin\ and \nHH\ preferentially and subtract their covariant components from the molecular abundances in advance, thereby making the results easier to interpret than those of the analysis involving the full parameter space.

The parameters to be analyzed are represented by the matrix $Q \equiv \left({\qv_{T}}, {\qv_{n}}, {\qv_{x_1}}, {\qv_{x_2}}, \dots, {\qv_{x_\Nmol}} \right)$.
The parameter vectors $\qv_i$ are defined as $\pv_i - \left<p_{i,j}\right>_j$, where the notation $\left<\right>_j$ denotes the average for $j$, i.e., the spatial average.
The subscripts $T$, $n$, and $x_k$ are for  \Tkin, \nHH, and the fractional abundances of the \thup{k}\ species ($k$=1,2, ... \Nmol), respectively.

First, we decompose $Q$ into two components that are covariant and noncovariant with \Tkin\ and \nHH\ ($Q_{\mathrm{c}}$ and $Q_{\mathrm{nc}}$, respectively):
\begin{eqnarray}
Q &=& Q_{\mathrm{c}} + Q_{\mathrm{nc}}.
\end{eqnarray}
The condition $\transverse{\left(\qv_T, \qv_n\right)}\cdot Q_{\mathrm{nc}} = 0$ gives that $Q_{\rm c}  =  Q_{T,n} \cdot \left(\transverse{Q_{T,n}}\cdot Q_{T,n}\right)^{-1} \cdot \transverse{Q_{T,n}}\cdot Q$, where $Q_{T,n} \equiv \left(\qv_T, \qv_n\right)$. 

The component $Q_{\rm nc}$ ($= Q - Q_{\rm c}$) is further decomposed into the principal components $\myvector{w}_i$:
\begin{eqnarray}
Q_{\rm nc} &=&  \left(\myvector{w}_{1}, \myvector{w}_{2}, \dots, \myvector{w}_{\Nmol}\right) \cdot \transverse{V},
\end{eqnarray}
where $V$ is a matrix of the nonzero eigenvectors of $\transverse{\Qnc}\cdot\Qnc$, and $\left(\myvector{w}_{1}, \myvector{w}_{2}, \dots, \myvector{w}_{\Nmol}\right) \equiv\ Q_{\rm nc} \cdot V$.

Thus, the parameter space has been decomposed into linear combinations of $\qv_T$, $\qv_n$, $\myvector{w}_1$, $\myvector{w}_2$, \dots, $\myvector{w}_\Np$.  
The original parameters are expressed using the coefficient vectors in the new basis ($x_T\left(X\right) \equiv \transverse{\qv_X}\cdot\frac{\qv_T}{|\qv_T|}$ and so on).
The total variance of all parameters, $\Lambda\equiv\mathrm{tr}\left({\transverse{Q}\cdot Q}\right)$, is decomposed as 
\begin{eqnarray}
\Lambda & =& \mathrm{tr}\left(\transverse{Q_\mathrm{c}}\cdot Q_\mathrm{c}\right) + \mathrm{tr}\left(\transverse{Q_\mathrm{nc}}\cdot Q_\mathrm{nc}\right)  \nonumber \\
  & = & \mathrm{tr}\left(\transverse{Q_\mathrm{c}}\cdot Q_\mathrm{c}\right) + \sum_{i=1,2,\dots, \Nmol} \lambda_i , \label{EQ_CONTRIB}
\end{eqnarray}
where $\lambda_i$ denotes the non-zero eigenvalues of $S$ corresponding to the \thup{i}\ principal component (PC$i$) of $Q_\mathrm{nc}$.
The first term in Equation \ref{EQ_CONTRIB}\ is the contribution from \Qc, and the second is the sum of the variances of the PCs. 
The former cannot be decomposed further owing to the non-zero correlation between \Tkin\ and \nH.
We define the contribution ratios as the ratios of each term to the total variance, i.e., 
$\mathrm{tr}\left(\transverse{Q_\mathrm{c}}\cdot Q_\mathrm{c}\right)/\Lambda$ for the $Q_\mathrm{c}$ component and $\lambda_i/\Lambda$ for PC$i$, by extending the definition in the standard PCA.

\begin{figure*}
\epsscale{1.}
\plotone{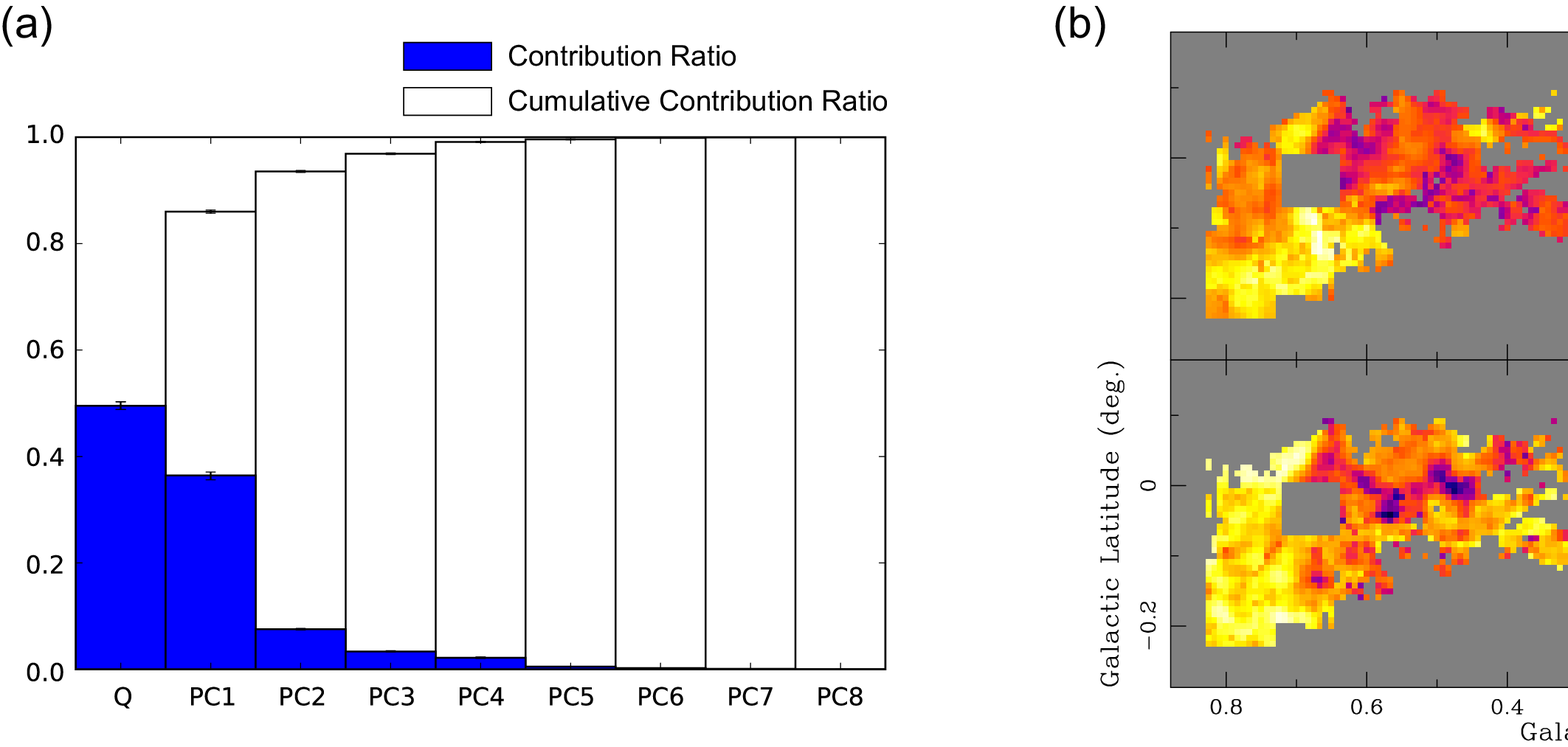}
\caption{Results of the PCA with the molecular abundances. (a) Contribution ratios and cumulative contribution ratios with error bars for the \thn{25}--\thn{75}-percentile intervals.  (b) \NHH-weighted average of the median values of PC1 and PC2 projected on the $l$--$b$ plane.   The $5'\times5'$ square region toward the Sgr B2 cluster-forming region, the $30''\times30''$ region toward \sgras, and the velocity channels of the spiral arms are not used for the PCA analysis, as the HB analysis is inaccurate for those regions owing to the omission of continuum emission in the excitation equations and self-absorption. } \label{RES_PCA1}
\end{figure*}

The marginal posterior PDFs of the PCs and their contribution ratios are calculated by repeating the analysis for values at every MCMC step. 
The $5'\times5'$ square region toward the Sgr B2 cluster-forming region, the $30''\times30''$ region toward \sgras, and the velocity channels of the spiral arms are not used for the calculations since these regions show obvious artificial features in the maps of \Rt\ and several molecular abundances, presumably due to the omission of the continuum in the excitation analysis and self-absorption (Subsections \ref{SECT_MAPS} and \ref{SECT_MOLMAPS}).
The individual and cumulative contribution ratios are shown in Figure \ref{RES_PCA1}a.
\Qc\ and the first PC explain more than 80\% of the total variance, and the contribution ratio of PC1 is more than twice the sum over the remaining PCs.
The cumulative contribution ratio up to PC2 is 93\%.
Thus, we are able to reduce the original 10 variables into the 4-D space defined by \Tkin, \nHH, PC1, and PC2 without losing important information.

Figure \ref{RES_PCA1}b shows the 2-D maps of PC1 and PC2 averaged along the velocity axis, weighted by \NHH.
Both PC1 and PC2 represent the enhancement in multiple molecules in the Sgr B2 complex (Figure \ref{FIG_molx}).  
These two PCs also show large positive values in the southwestern rim of Sgr C shell-1 ($\gl\sim -0.45^\circ$).
The difference between the two PCs appears most clearly in the 50-\kmps\ cloud \textcolor{red}{($\gl\sim 0^\circ$)}, where PC1 is negative but PC2 is positive.
In addition to these regions, PC2 has relatively large positive values in the polar arc and M$0.11$$-0.08$.

\begin{figure*}
\epsscale{.8}
\plotone{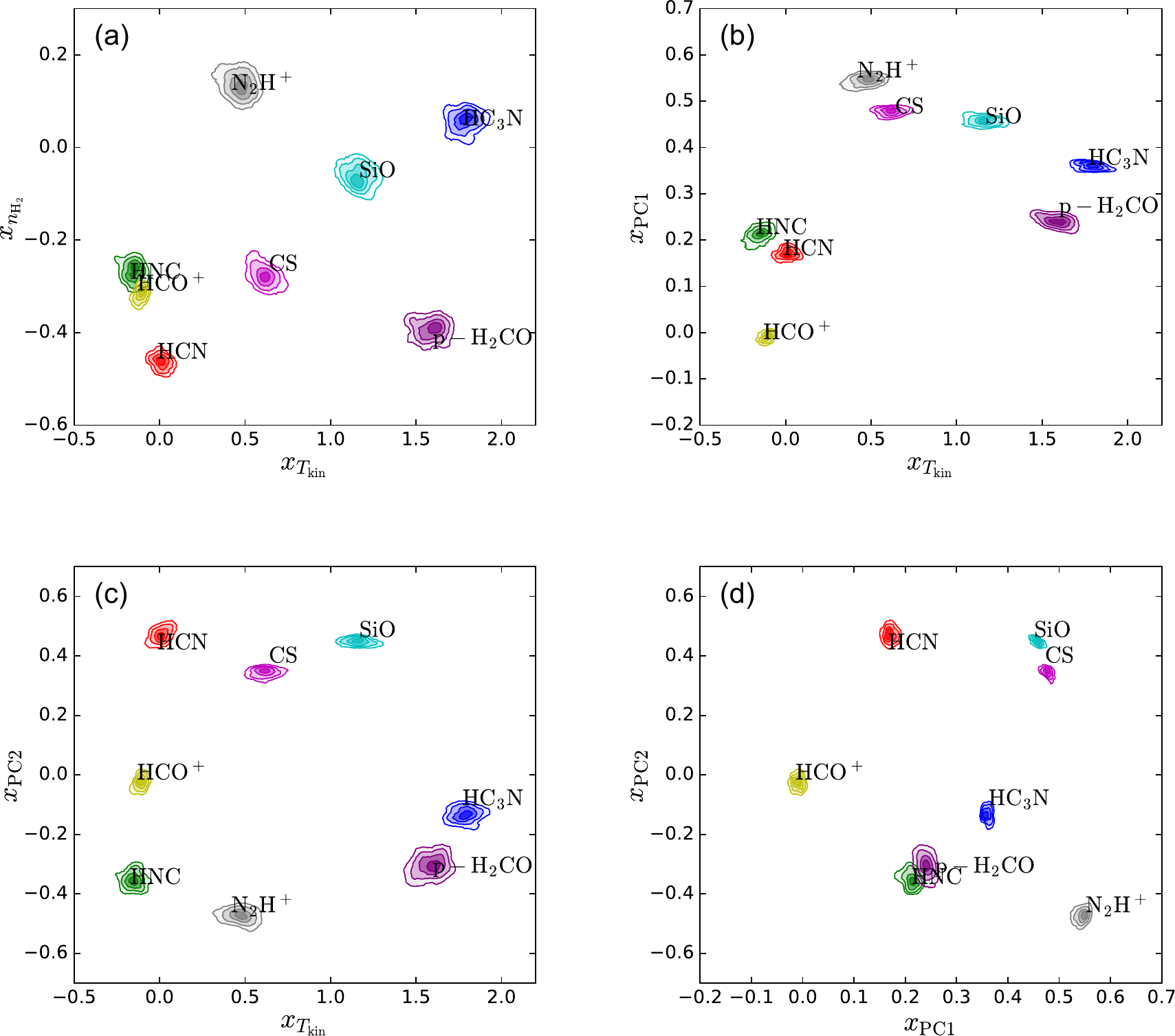}
\caption{Coefficient vector representation of the molecular abundances on the $x_{\Tkin}$--$x_{\nHH}$ (panel a), $x_{\Tkin}$--$x_{\mathrm{PC1}}$ (panel b), $x_{\Tkin}$--$x_{\mathrm{PC2}}$ (panel c), and $x_{\mathrm{PC1}}$--$x_{\mathrm{PC2}}$ (panel d)  planes.  Contours are drawn at the \thn{2.5}--\thn{97.5}-, \thn{10}--\thn{80}-, \thn{25}--\thn{75}-, \thn{40}--\thn{50}-percentile intervals of the marginal posterior distribution of each parameter pair.}\label{RES_PCA2}
\end{figure*}

The coefficient vector representations of the molecules in the $x_T$--$x_n$--$x_1$--$x_2$ space are shown in Figure \ref{RES_PCA2}. 
On the $x_T$--$x_n$ plane, we observe approximately the same trend as that found in the correlation coefficient matrix (Figure \ref{FIG_CORRCOEFF});  \HCCCN\ and \formaldehyde\ are distinguished by their large positive $x_T$, and \NNHp\ is the only molecule that has a significantly positive $x_n$.
The difference among the molecules becomes evident in their behaviors on the $x_T$--$x_1$ and $x_T$--$x_2$ planes.
Two groups are identified in the $x_T$--$x_1$ plot, i.e., species with both large $x_T$ and $x_1$ (CS, SiO, \HCCCN, \formaldehyde, and \NNHp) and those with both small $x_T$ and $x_1$ (HCN, HNC, and \HCOp);  they correspond to the groups classified according to whether they are enhanced in the Sgr B2 complex or not, which are immediately identified in the molecular abundance maps (Figure \ref{FIG_molx}).
The latter group is further classified into two subgroups according to the $x_2$ values;  SiO and CS have both large positive $x_1$ and $x_2$, whereas \HCCCN, \formaldehyde, and \NNHp\ have zero or negative $x_2$ values.

\begin{figure*}
\ifdraft
\epsscale{0.7}
\else
\epsscale{0.7}
\fi
\plotone{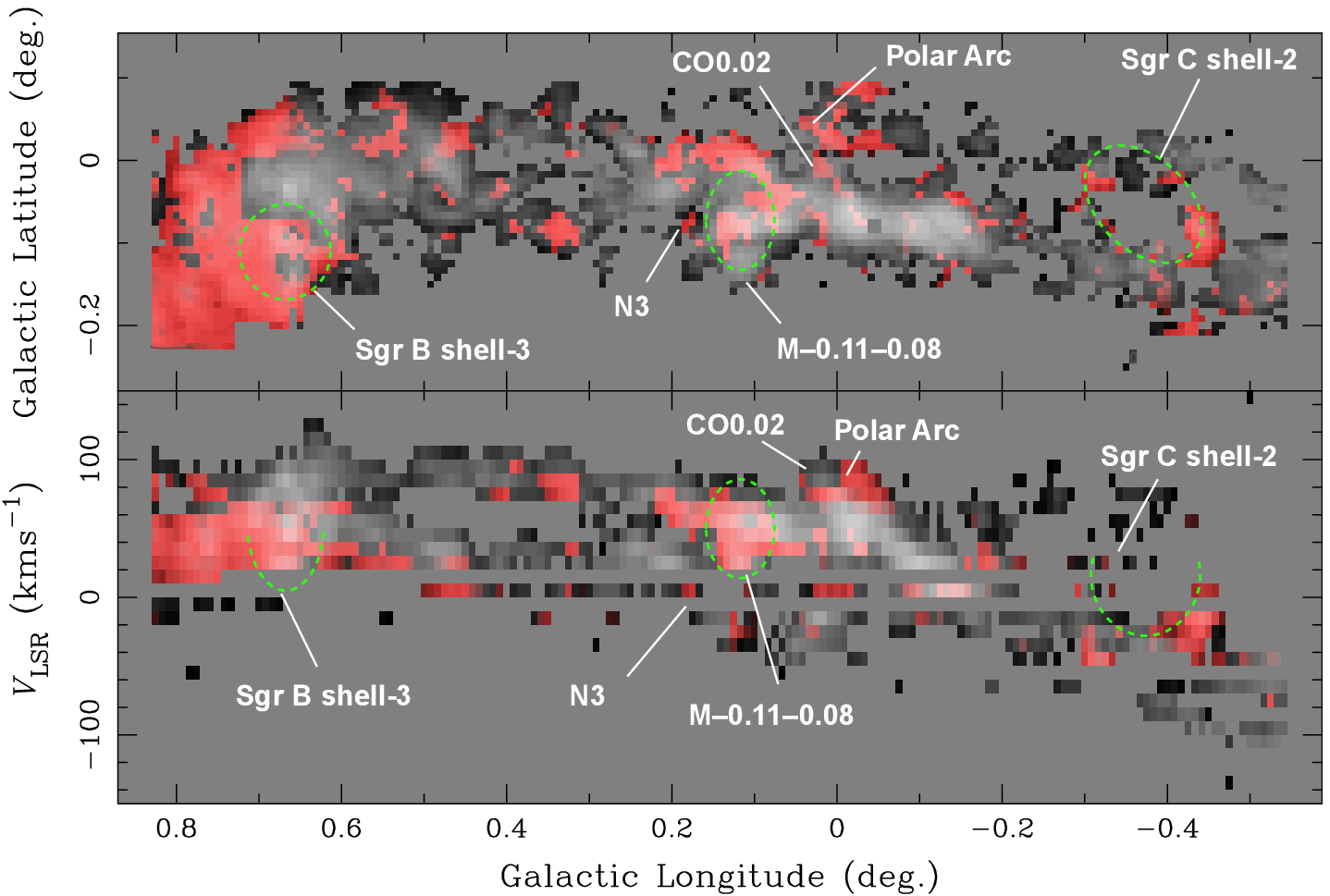}
\caption{\NHHavg\ map projected on the $\gl$--$\gb$ and $\gl$--$\vlsr$ planes (top and bottom panels, respectively), where the highest values along the projection axes are plotted.  The regions with positive PC1 and PC2 are highlighted in red.} \label{FIG_SHOCKEDREGION}
\end{figure*}

\subsubsection{PC1 and PC2: Regions with Fast Shocks}
SiO and CS are distinguished from other species by their  large positive $x_1$ and $x_2$ values.
PC1 and PC2 constitute approximately 80\% of the total variance of their abundance.
As SiO is a well-proven tracer of fast shocks, PC1 and PC2 are likely to be related to regions with shock velocities $\gtrsim 20\ \kmps$, where the sputtering of dust silicate cores occurs.
High CS abundances are also observed in the fast-velocity regions in molecular outflow sources \citep{Jorgensen2004,Tafalla2010}.
In addition, the large $x_T$ values for SiO and CS corroborate the interpretation that they represent shocked regions.

However, the interpretation of PC1 is somewhat ambiguous since a large value of $x_1$ is also measured for \NNHp.  
The molecule is an established tracer of cold quiescent gas, as it is destroyed in shocked gas via the reaction with molecules desorbed from dust.
Hence, the apparent spatial correlation between \NNHp\ and SiO in our data is unlikely to indicate the similarity between their chemical characteristics;  
it is more likely that our analysis is not sufficiently precise to separate the \NNHp-rich preshock gas from the post-shock turbulent gas, which frequently coexist in single-dish beams \citep[e.g.,][]{Tanaka2015}. 
Moreover, PC1 may contain an artificial correlation created by the uncertainty in \NHH\ due to the neglection of the spatial variation in \xmolp{CO}\ and the contribution from the low-density component discussed in Subsection \ref{SECT_MULTCOMP}.

Therefore, PC2 may better represent the fast-shock regions than PC1;  the molecules with large positive $x_2$ (HCN, CS, and SiO) and those with large negative $x_2$ (\NNHp\ and HNC) correspond well with the species enhanced in fast-shock regions and those abundant in cold quiescent gas, respectively \citep{Hirota1998,Tafalla2010}. 
Therefore, we could translate the combination of positive PC1 and PC2 values as a signature of fast-shock chemistry.   Figure \ref{FIG_SHOCKEDREGION}\ presents the $\gl$--$\gb$ and $\gl$--\vlsr\ distributions of \NHHavg, with the regions with positive PC1 and PC2 highlighted in red. 
In addition to the southeast extension of Sgr B2,  some other smaller regions such as Sgr C shell-2, Polar Arc, CO$-0.2$, M$-0.11$$-0.08$, and the clump N2 exhibit this signature of fast-shock regions.

\subsubsection{Large Molecules: Regions with Slow Shocks}
The coefficient vector diagram has another distinctive group constituted by two relatively large molecules, i.e., \HCCCN\ and \pformaldehyde, which are characterized by large $x_T$, moderate $x_1$, and negative $x_2$\ values.
These molecules could be classified as species enhanced in hot cores in the sense that their formation paths include dust-grain chemistry.  
The strong \Tkin\ dependence of their abundance may indicate that their origin is thermal desorption from heated dust; 
however, this hot-core explanation does not conform with the observed decoupling of the gas and dust temperatures.  
The {\it dust} temperature is uniformly low ($\sim 20\ \kelvin$) throughout the CMZ, and it is not correlated with the {\it gas} temperature \citep{Pierce-Price2000}.

The correlation of their abundances with \Tkin\ is better explained by the impacts of low-velocity shocks with velocities $\lesssim 20\ \kmps$.
The gas-phase abundances of large organic molecules can be enhanced by the mechanical sputtering of the icy grain mantle \citep{Requena-Torres2006,Harada2015}, as observed in molecular outflow sources in SF regions \citep{Jorgensen2004,Tafalla2010}.  
In addition, mechanical heating by decaying shocks is considered to be a dominant mechanism of gas heating in the CMZ \citep[e.g.,][]{Rodriguez-Fernandez2000,Tanaka2009,Ao2013,Ginsburg2016}.
The positive correlation between \Tkin\ and the \HCCCN\ and \formaldehyde\ abundances lends observational support for the theory of hot-core chemistry without hot cores induced by low-velocity shocks \citep{Requena-Torres2006}.

\subsubsection{Two Regimes of Shock Chemistry}
It is noteworthy that the above-mentioned two categories, which are distinguished by $x_T$, $x_1$, and $x_2$, correspond well to two regimes of  shock chemistry commonly observed in two well-studied molecular outflow sources, L1448-mm and IRAS 04166+2706 \citep{Tafalla2010}.  
The SiO and CS abundances increase in the fast-wing regime with velocities $\gtrsim 20\ \kmps$ where sputtering of the grain core is significant.
The slow-wing regime is characterized by high abundances of \formaldehyde\ and \methanol, which are caused by dust mantle sputtering, while these large molecules decrease in the fast-wind regime.
Although \HCCCN\ is ambiguous in the classification by \cite{Tafalla2010}, the morphological similarity of its \JJ{5}{4}\ line to the low-excitation \methanol\ line in the Mopra 7-mm survey of the CMZ data \citep{Jones2013} would justify classifying them in the same category, at least in the CMZ.

Thus, the spatial variations in SiO, CS, \HCCCN, \formaldehyde, and \Tkin\ in the CMZ could be consistently explained by a single parameter, i.e., the strength of the shock.
The correlation of \Tkin\ with the slow-wing species is better than that with the fast-wing species; this could be because slow shocks are simply more common than fast shocks.
Even though our analysis does not include the CND, our classification partly overlaps with that proposed for the CND by \cite{Takekawa2014}; the GMC-type species in the CND include \HCCCN\ and other hot-core-type species, while the CND-type and hybrid-type consist of small molecules including SiO and CS, which may indicate the significance of  fast-shock chemistry in the CND \citep{Harada2015}.

Several regions with high \Tkin\ values or those with both high \Tkin\ and a rich abundance of fast-shock tracers (i.e., high PC1 and PC2 values) exhibit kinematics that are indicative of the interaction with shocks, such as the expanding shells in  Sgr B2 and Sgr C \citep{Tsuboi2015,Tanaka2014}, the extremely broad emissions of CO$0.02$ and CO$-0.30$ \citep{Oka2011,Tanaka2014,Tanaka2015}, and the polar arc, which are argued to be a part of the molecular outflow driven by a past activity of \sgras \citep{Hsieh2016}.
On the contrary, the region with high PC1 and PC2 but low \Tkin, such as GMC complex M$0.11$$-0.08$, might require another explanation.
\cite{Amo-Baladron2009} proposed the hypothesis of nonthermal sputtering of small dust grains by hard X-ray photons for SiO enrichment in the region and obtained a  spatial correlation between the SiO abundance and the 6.4 keV fluorescent Fe line.  Their hypothesis has an advantage over the explanation with shock chemistry, as it is consistent with the absence of the temperature rise in M$0.11$$-0.08$ in the results of the HB analysis.

\subsubsection{Effect of Cosmic Rays}
The intense CR field in the CMZ may induce  hot-core-like chemistry through the CR-induced UV photodesorption of grain mantle species \citep{Yusef-Zadeh2013b}.
In addition, it can be a significant gas-heating mechanism when the CR ionization rate ($\zeta$) is higher than $\sim 10^{-14}\ \mathrm{s}^{-1}$ \citep{Ao2013}.
Therefore, the same arguments as those used for shock heating and shock chemistry could  apply to  CR heating and  CR-induced chemistry if the primary source of the CR acceleration is SNR--MC interactions.

We may be able to separate the effect of CRs from shocks by investigating the spatial variation in \HCOp, as \xmolp{\HCOp}\ is expected to be scaled by ${\zCR}\cdot{\nHH}^{-1}$ \citep{Papadopoulos2007}.
The observed strong negative correlation between \xmolp{\HCOp}\ and \nHH\ is consistent with this regime, whereas this could be alternatively understood as a decreasing inward gradient due to UV photodissociation. 
However, further positive evidence for effects of CRs on chemistry and thermal processes is not obtained with our data; \xmolp{\HCOp}\ exhibits a considerably flat spatial variation except for the \nHH\ dependence, and no significant correlation with \Tkin\ is observed.

\subsubsection{HCN and HNC}
The  HCN  \JJ{1}{0}\ line is frequently used to estimate the mass of dense gas (typically with $\nHH$ $\gtrsim 10^4\ \pcc$), which is supposed to be directly linked to star formation activities \citep[e.g.][]{Lada2012a}.  
Our analysis shows that the spatial variation in the HCN abundance is approximately correlated with that of the conversion factor from the HCN \JJ{1}{0}\ luminosity to dense gas mass that is found by \cite{Mills2017};
the abundance is relatively high for the clouds with over-luminous HCN emission (the 50-\kmps\ cloud) compared with the clouds with regions with under-luminous HCN emission (the Sgr B2 cluster-forming region and the 20-\kmps\ cloud).   
This supports the argument in \cite{Mills2017}\ that the large spatial variation in the HCN-to-dense-gas conversion factor is created primarily by the variation in the HCN abundance, though the opacity effect significantly masks the abundance variation.
One possible cause of the variation in the HCN abundance is shock chemistry, as this abundance is observed to increase in fast-shock regions in molecular outflow sources \citep{Tafalla2010,Mills2017}; 
in fact, the position of HCN in the $x_T$--$x_2$ vector component diagram (Figure \ref{RES_PCA2}c) is close to those of other fast-shock tracers such as SiO and CS.

The [HCN]/[HNC] isomer ratio increases sharply with \Tkin\ according to the systematic survey by \cite{Hirota1998}, who found that the ratio is $\sim 1$ for dark clouds with $\Tkin$ $\sim 10\ \kelvin$,  while it increases to $\sim 10$ for a high-mass SF region with $\Tkin \gtrsim 20\ \kelvin$. 
The mean [HCN]/[HNC] ratio for the CMZ is 10, which is close to the values for the latter regime; the CMZ is hot-core-like in terms of the [HNC]/[HCN] isomer ratio.
\Tkin\ dependence of the isomer ratio is absent within our data.

%% file: table5_r.etex


\tablecolumns{5}
\small
\tablehead{ \\ & & \colhead{$T_{\rm cold}$=40 K } & 
                \colhead{20 K } &               
                \colhead{10 K } \\
            & & \colhead{$n_{\rm cold}=10^{3.0}\ \pcc$ } &  
                \colhead{$10^{3.2}\ \pcc$ } &               
                \colhead{$10^{3.7}\ \pcc$ }  }
\tabletypesize{\scriptsize}
\tablewidth{0pt}
\startdata
HCN    & \JJ{1}{0}      & \textcolor{black}{0.09}  & \textcolor{black}{ 0.11}  & \textcolor{black}{ 0.20 } \\
       & \JJ{4}{3}      & \textcolor{black}{ 0.01} & \textcolor{black}{ 0.006} & \textcolor{black}{ 0.003 } \\
\HCNt  & \JJ{1}{0}      & \textcolor{black}{ 0.05 } & \textcolor{black}{ 0.05} & \textcolor{black}{ 0.10 } \\
\HCOp  & \JJ{1}{0}      & \textcolor{black}{ 0.09 } & \textcolor{black}{ 0.10} & \textcolor{black}{ 0.21 } \\
\HCOpt & \JJ{1}{0}      & \textcolor{black}{ 0.09 } & \textcolor{black}{ 0.07} & \textcolor{black}{ 0.18 } \\
\pformaldehyde & $3_{03}\mbox{--}2_{02}$        & \textcolor{black}{ 0.06} & \textcolor{black}{ 0.07} & \textcolor{black}{ 0.18 } \\
               & $3_{21}\mbox{--}2_{20}$        & \textcolor{black}{ 0.04} & \textcolor{black}{ 0.05} & \textcolor{black}{ 0.07 } \\
\COt   & \JJ{2}{1}      & \textcolor{black}{ 0.52} & \textcolor{black}{ 0.51} & \textcolor{black}{ 0.50 } \\
HNC    & \JJ{1}{0}      & \textcolor{black}{ 0.08} & \textcolor{black}{ 0.10} & \textcolor{black}{ 0.24 } \\
\HCCCN & \JJ{10}{9}     & \textcolor{black}{ 0.02} & \textcolor{black}{ 0.02} & \textcolor{black}{ 0.02 } \\ 
SiO    & \JJ{2}{1}      & \textcolor{black}{ 0.04} & \textcolor{black}{ 0.05} & \textcolor{black}{ 0.14 } \\
\NNHp  & \JJ{1}{0}      & \textcolor{black}{ 0.07} & \textcolor{black}{ 0.09} & \textcolor{black}{ 0.20 } \\
CS     & \JJ{1}{0}      & \textcolor{black}{ 0.11} & \textcolor{black}{ 0.13} & \textcolor{black}{ 0.31 } \\
\\
\textcolor{black}{\NHH}   &                & \textcolor{black}{0.84} & \textcolor{black}{0.88} & \textcolor{black}{1.2} 
\enddata

%% file: summary.etex
 We have presented the maps of HCN \JJ{4}{3}, \HCNt\ \JJ{1}{0}, HNC \JJ{1}{0}, and \HCCCN\ \JJ{10}{9}\ covering the major part of the CMZ with spatial and velocity resolutions of $\sim 20''\times 2\ \kmps$, which we obtained  using the ASTE 10-m telescope and NRO 45-m telescope.
Spatially resolved measurements of \Tkin, \nHH, and the fractional abundances of eight molecules have been conducted using our data and survey data taken from the literature by combining the LVG calculation with HB inference. 
The primary results obtained from the analysis are summarized below.
\begin{enumerate}
\item The HCN \JJ{4}{3}\ and \HCCCN\ \JJ{10}{9}\ maps exhibit similar spatial distributions, whose intense emissions are limited to high-temperature regions, while HCN \JJ{1}{0}\ and HNC \JJ{1}{0}\ trace the mass distribution of dense gas.
\item The 3-D (2-D in space and 1-D in velocity) distributions of \NHH, \Tkin, \nHH, the beam filling factor, the [$^{12}\mathrm{C}$]/[\Ct] isotopic abundance ratio, and the fractional abundances of HCN, \HCOp, HNC, \HCCCN, \pformaldehyde, SiO, \NNHp, and CS are calculated by employing the HB analysis.  The model includes parameters for systematic errors that represent the unmeasurable uncertainties from calibration errors and deviations from the ideal model, such as the one-zone LVG approximation.  A multivariate log-Student prior and log-normal prior are assumed for the primary parameters and errors, respectively. 
\item  The HB analysis is successful in suppressing strong artificial correlations among the parameters, while these artifacts strongly affect the results of the nonhierarchical method that is equivalent to the standard maximum likelihood analysis.  All line intensities are fitted with the HB analysis without assuming large systematic errors. 
\item The measured values of \nHH\ are distributed in the range of $10^{3.4}$--$10^{4.8}\ \pcc$\ with the highest frequency at \textcolor{black}{$10^{4.2}\ \pcc$}.  These results are consistent with previous measurements conducted using molecular lines with similar critical densities to those used by us.  No systematic difference is observed between \nHH\ for highly active cluster-forming regions (such as Sgr B2 and G$-0.02$$-0.07$) and quiescent massive clumps (such as the brick cloud). 
\item The typical value of \Tkin\ is $\sim 10^{1.8}\ \kelvin$, which is consistent with the previous LTE analysis performed using the same \formaldehyde\ data.    The distribution of the high-temperature gas is not correlated with the star-forming regions, except for two high-temperature spots toward the Sgr B2 cluster-forming region and G$-0.02$$-0.07$.  A few of the high-temperature regions without UV-heating sources are associated with broad-velocity features or clouds with the kinematics of expanding shells, such as CO0.02, CO$-0.30$, Sgr C shell 2, and Sgr B shell 3.
\item The fractional abundances of SiO, CS, \HCCCN, \NNHp, and \formaldehyde\ exhibit considerable enhancement in the Sgr B2 complex.  The HCN, \HCOp, and HNC abundances show small variations across the CMZ, except for a significant negative correlation of the former with \nHH, making them good probes of the gas mass and  physical conditions.   
\item The \COt\ \JJ{1}{0}\ intensity estimated from our analysis explains only $\sim 30\%$\ of the observed luminosity; this suggests the presence of a second component with a lower temperature and  density, which is not observable in the high-density tracers used in the analysis.
\item The spatial variations in \Tkin, \nHH, and the fractional abundances of the eight species can be reduced into four primary components, i.e., the covariant components with \Tkin\ and \nHH\ and the first and second principal components (PC1 and PC2) of the remaining variances, whose sum accounts for approximately $93\%$ of the total variance in the parameter space.   Two distinct  molecular species are identified according to their coefficients for the four components ($x_T$, $x_n$, $x_1$, and $x_2$): one category with large $x_1$, large $x_2$, and moderate-to-large $x_T$ (SiO and CS) and the other with large $x_T$, smaller $x_1$ than the first group, and negative $x_2$ (\HCCCN\ and \formaldehyde).   The former and latter groups approximately correspond to the molecules enhanced in the fast and slow shocks, respectively.  This indicates that the strength of the mechanical sputtering of dust grains is one of the primary determinants of  molecular chemistry in CMZ clouds.  This is corroborated by their positive dependence on \Tkin.  
\item The \HCOp\ abundance is anticorrelated with \nHH.  This is consistent with CR ionization or photodissociation theories, even though further evidence for the effect of the CRs on gas heating is not obtained.  The high [HCN]/[HNC] isomer ratio of $10$ is close to the value for high-mass SF regions.
\end{enumerate}

%% file: acknowledgements.etex
The authors are grateful to the staff of the National Astronomical Observatory Japan for their generous support during the observations.  
We also thank the anonymous referee and Dr. Feigelson, whose comments helped refine the analysis.
This work was supported by JSPS KAKENHI Grant Numbers 26800105 and 16K17666.
The data cubes and the source code of the analysis software presented in this paper will be available on the NRO website\footnote{\url{http://www.nro.nao.ac.jp/~nro45mrt/html/results/data.html}}.